\documentclass[english,11pt,aps,prd,a4paper,preprintnumbers,  floatfix,nofootinbib,showpacs,superscriptaddress,  notitlepage]{revtex4-1} 

 \pdfoutput=1
\usepackage[usenames,dvipsnames]{color}  
\usepackage{graphicx}
\usepackage{setspace}
\usepackage{upgreek}
\usepackage{braket}

\usepackage{microtype}
\usepackage{multirow}

\usepackage{bm}
\usepackage{bbm}
\usepackage{mathrsfs}
\usepackage[bbgreekl]{mathbbol}

\usepackage{caption}
\usepackage{stackengine}
\usepackage{subcaption}
\usepackage{amsmath}
\usepackage{amssymb}
\usepackage[colorlinks=true,citecolor=darkred,urlcolor=darkred, pdfborder={0 0 0}]{hyperref}
\usepackage[normalem]{ulem}
\usepackage{xcolor}
\usepackage{graphicx}
\makeatletter
\def\p@subsection{}
\makeatother
\usepackage{xcolor}
\usepackage{float}
\usepackage{placeins} 

\usepackage{array}
\definecolor{darkred}{rgb}{0.6,0,0}

\definecolor{linkcolor}{rgb}{0,0,0.5}

\def\gsim{\raise0.3ex\hbox{$\;>$\kern-0.75em\raise-1.1ex\hbox{$\sim\;$}}}
\def\lsim{\raise0.3ex\hbox{$\;<$\kern-0.75em\raise-1.1ex\hbox{$\sim\;$}}}

\def\beqn#1{\begin{equation}\label{#1}}
\def\eeqn{\end{equation}}

\def\beqa#1{\begin{eqnarray}\label{#1}}
\def\eeqa{\end{eqnarray}}

\usepackage{caption}
\usepackage{subcaption}
\usepackage{adjustbox}

\def\Z2{$\mathcal{Z_2}$}

\usepackage{bbold}
\usepackage{ragged2e}

\newcommand {\ignore}[1]{}

\def\cevns{CE$\nu$NS }
\def\eves{E$\nu$ES }
\usepackage{mathrsfs}

\def\321{$\mathrm{SU(3) \otimes SU(2) \otimes U(1)}$ }
\def\eves{E$\nu$ES~}
\definecolor{matplotlibBrown}{rgb}{0.65, 0.16, 0.16}

\usepackage{orcidlink}

\newcommand{\AddrIISERB}{Department of Physics, Indian Institute of Science Education and Research - Bhopal, \\ 
Bhopal Bypass Road, Bhauri, Bhopal 462066, India}

\newcommand{\AddrAHEP}{%
  AHEP Group, Institut de F\'{i}sica Corpuscular--CSIC/Universitat de Val\`{e}ncia \\ %
  C/Catedr\'atico Jos\'e Beltr\'an, 2 E-46980 Paterna, Spain}

\begin{document}

\title{\textcolor{BrickRed}{Constraining  low scale dark hypercharge symmetry at spallation, reactor and dark matter direct detection experiments }}

\author{Anirban Majumdar~\orcidlink{0000-0002-1229-7951}}\email{anirban19@iiserb.ac.in}
\affiliation{\AddrIISERB}
\author{Dimitrios K. Papoulias~\orcidlink{0000-0003-0453-8492}}\email{dipapou@ific.uv.es}
\affiliation{\AddrAHEP}
\author{Hemant Prajapati~\orcidlink{0000-0001-5104-9427}}\email{hemant19@iiserb.ac.in}
\affiliation{\AddrIISERB}
\author{Rahul Srivastava~\orcidlink{0000-0001-7023-5727}}\email{rahul@iiserb.ac.in}
\affiliation{\AddrIISERB}

\begin{abstract}
\vspace{0.5cm}
Coherent elastic neutrino-nucleus (CE$\nu$NS) and elastic neutrino-electron scattering (E$\nu$ES) data are exploited to constrain ``chiral'' $U(1)_{X}$ gauged models with light vector mediator mass. 
These models fall under a distinct class of new symmetries called dark hypercharge symmetries. A key feature is the fact that the $Z'$ boson can couple to all Standard Model fermions at tree level, with the $U(1)_X$ charges determined by the requirement of anomaly cancellation.
Notably, the charges of leptons and quarks can differ significantly depending on the specific anomaly cancellation solution.  As a result, different models exhibit distinct phenomenological signatures and can be constrained through various experiments.
In this work, we analyze the recent data from the COHERENT experiment, along with results from dark matter (DM) direct detection experiments such as XENONnT, LUX-ZEPLIN, and PandaX-4T, and place new constraints on three benchmark models. Additionally, we set constraints from a performed analysis of TEXONO data and discuss the prospects of improvement in view of the next-generation DM direct detection DARWIN experiment.
\end{abstract}

\maketitle  

\section{\label{sec:Intro}Introduction}

The Standard Model (SM) is a robust and extensively validated framework that accurately describes fundamental particles and their interactions at the current experimental energy scale. However, the discovery of neutrino oscillations~\cite{Super-Kamiokande:1998kpq, SNO:2002ziz} combined with the tantalizing evidence for the existence of nonluminous objects---dubbed Dark Matter (DM)---from cosmological observations~\cite{Zwicky:1933gu,Sofue:2000jx, Bertone:2004pz,Planck:2018vyg}, provide a strong motivation to look for new particles and interactions beyond the Standard Model (BSM). To address these shortcomings, the SM can be extended in various ways. One of the simplest and most well-motivated examples is the introduction of an additional Abelian $U(1)_{X}$ gauge symmetry, along with the required
new fermions for canceling the gauge anomalies. The latter could participate in SM neutrino mass mechanisms or constitute  potential DM candidates. A common feature of $U(1)_{X}$ symmetry is the introduction of a novel neutral gauge boson ($Z'$), usually assumed to be heavier than the electroweak scale. This choice is highly inspired by Grand Unified Theories~\cite{Langacker:1981hs}, as they could break into 
$SU(3)_{C} \otimes SU(2)_{L} \otimes U(1)_{Y} \otimes U(1)_{X} $ at lower energies. The phenomenological aspects of such heavy bosons have been extensively studied at various colliders, including the LHC~\cite{ATLAS:2019erb,CMS:2021ctt}, LEP~\cite{Electroweak:2003ram,Essig:2009nc}, and other fixed-target experiments \cite{Bross:1989mp,CHARM:1985anb,Gninenko:2012eq,NA64:2016oww,NA64:2019auh}.

However, there exist numerous phenomenological and theoretical motivations to study new Abelian extensions at a lower scale. For example, some cosmological observations suggest that  DM might be self-interacting~\cite{Spergel:1999mh} with the new light $Z'$  acting as a mediator for such interactions~\cite{Duerr:2018mbd}. Known anomalies such as the muon magnetic moment could also be explained in the presence of a light mediator~\cite{AtzoriCorona:2022moj}. Light mediators are phenomenologically very interesting; indeed, due to few parameters of the underlying theory, $U(1)_{X}$-based models are highly predictive. In general, the key parameters characterizing the $Z'$ boson interactions are the gauge coupling $\textsl{g}_{_x}$, mass $M_{Z'}$, $Z-Z'$ mixing angle, and fermion charges under $U(1)_X$. For an underlying  symmetry, the charges of SM and BSM fermions are fixed by gauge anomaly cancellation, while the rest parameters can be constrained by various terrestrial experiments. Relevant experimental probes include DM direct detection~\cite{AristizabalSierra:2020edu, Majumdar:2021vdw, A:2022acy, DeRomeri:2024dbv}, neutrino scattering~\cite{Lindner:2018kjo, AtzoriCorona:2022moj, AristizabalSierra:2022axl, Coloma:2022umy, DeRomeri:2022twg, Melas:2023olz}, beam dumps~\cite{Bross:1989mp,CHARM:1985anb,Gninenko:2012eq,NA64:2016oww} and colliders~\cite{ATLAS:2016bps}. Further constraints can be set from deep-sky investigations. Typical such examples include  astrophysical observations such as supernova cooling~\cite{Cerdeno:2021cdz}, and through the recent Planck measurement of the comic microwave background~\cite{Planck:2018vyg}, e.g., from constraining $N_\text{eff}$~\cite{Li:2023puz, Esseili:2023ldf, Ghosh:2023ilw, Ghosh:2024cxi}. 

In the literature, various $U(1)_{X}$ gauge symmetries have been proposed, while the limits placed on their relevant parameters are usually probed with varying significance.
For instance, $B-L$ is a well-studied symmetry where all generations of leptons and quarks are charged under the $U(1)_{X}$ symmetry, resulting in tree-level couplings between different fermions and the $Z'$. Therefore, this model is severely constrained by different scattering experiments and cosmological observations~\cite{AtzoriCorona:2022moj, DeRomeri:2024dbv, Ghosh:2024cxi}. 
On the other hand, the case of $L_{i}-L_{j}$~\cite{He:1990pn}, $B-3L_{i}$~\cite{Ma:1997nq,Lee:2010hf}, $B_{i}-3L_{j}$~\cite{Bonilla:2017lsq,Alonso:2017uky} etc., where $i,j=1,2,3$ denote the generations of SM fermions which are charged under the corresponding symmetry, constitutes a class of less constrained symmetries~\cite{Majumdar:2021vdw, AtzoriCorona:2022moj, DeRomeri:2024dbv}. This is because for these symmetries not all quarks and leptons couple to $Z'$ boson at the tree level. 
Note that the aforementioned symmetries are called ``vector" $U(1)_X$ symmetries. This is because in these cases, the gauge anomalies are canceled by assigning both the left and right components of a given fermion the same  $U(1)_X$ charge.

On the other hand, recently, a class of symmetries called ``dark hypercharge symmetry'' (DHC) has been proposed, with the charges of SM and BSM fermions under the $U(1)_{X}$ being nontrivial \cite{Prajapati:2024wuu}. In contrast to all previous cases, the SM fermions within DHC are assumed to be chiral under $U(1)_{X}$, i.e., the left and right components of a given fermion are assigned different charges under $U(1)_{X}$. Therefore, in this case, gauge anomaly cancellation is achieved by introducing three right-handed SM singlet fermions. An interesting feature of such symmetries is the fact that the $Z'$ may couple to all SM fermions at tree level, with the corresponding $U(1)_{X}$ charges being relatively large. Also under such symmetries the relative charges of leptons and quarks can be varied and each specific choice of charge corresponds to a different model (see Table \ref{tab:Final_ano_con}). Consequently, different models exhibit a distinct phenomenological signature and can be constrained through various experiments.

The recent observation of coherent elastic neutrino-nucleus scattering (CE$\nu$NS) on CsI~\cite{COHERENT:2017ipa, COHERENT:2021xmm} and liquid argon (LAr)~\cite{COHERENT:2020iec} reported by the COHERENT Collaboration, has revolutionized the searches for novel light mediators providing stringent constraints, especially in the few MeV mass regime~\cite{Abdullah:2022zue}. An exhaustive number of related works have placed new bounds on novel light  mediators~\cite{Dent:2016wcr,Papoulias:2017qdn, Khan:2019cvi, AristizabalSierra:2019ykk, Miranda:2020tif, Cadeddu:2020nbr}, 
predicted in the context of various motivated models~\cite{Abdullah:2018ykz, Flores:2020lji, Banerjee:2020zvi, delaVega:2021wpx, AtzoriCorona:2022moj, Coloma:2022avw, DeRomeri:2023cjt,Bernal:2022qba}.  
Recent studies have also explored the possibility of generalized interactions~\cite{AristizabalSierra:2018eqm, Flores:2021kzl, DeRomeri:2022twg} 
as well as the production of new particles via upscattering~\cite{Brdar:2018qqj,Candela:2023rvt, Candela:2024ljb}. 
Improved constraints are expected from the next generation \cevns experiments at the European Spallation Source (ESS)~\cite{Baxter:2019mcx, Chatterjee:2022mmu} and from reactor-based \cevns experiments~\cite{CONUS:2021dwh, CONNIE:2019xid, Alfonso-Pita:2022eli, Lindner:2024eng}. 
Moreover, with the current advancements in multiton DM direct detection experiments such as as XENONnT~\cite{XENON:2022ltv}, LUX-ZEPLIN (LZ)~\cite{LZ:2022lsv}, and PandaX-4T~\cite{PandaX:2024cic}, the detection of solar neutrino-induced low-energy electron recoils through elastic neutrino-electron scattering (E$\nu$ES) events is now possible. These experiments have achieved an electron recoil  energy threshold as low as $\sim 1\mathrm{~keV}_{ee}$, making them excellent facilities to probe new physics phenomena by searching for spectral distortions at low energies~\cite{Cerdeno:2016sfi, Majumdar:2021vdw, Schwemberger:2022fjl}. 
Constraints set from the analysis of XENONnT, LZ and PandaX-4T, indicate stringent limits, especially in the keV mass regime, see, e.g., Refs.~\cite{Khan:2022bel, A:2022acy, DeRomeri:2024dbv} 

In this work, our goal is to put constraints on various chiral symmetric models predicted in the context of DHC. The present study focuses on three benchmark chiral models which involve large fermion charges, resulting to stringent constraints in comparison to established models such as $B-L$. After presenting the theoretical framework of the considered chiral models, our analysis strategy is to perform a timely analysis utilizing available data from  recent \cevns and E$\nu$ES measurements, induced by neutrinos originating from pion decay at rest, the Sun and reactor facilities.  Concerning CE$\nu$NS, we exploit the full COHERENT dataset  considering both the CsI and LAr measurements, and  present individual as well as combined analysis constraints. Then, for our E$\nu$ES-related analyses we employ the recent data from XENONnT, LZ and PandaX-4T experiments, and again we place new bounds obtained from each experiment separately and also present the combined limits. We furthermore derive the projected sensitivity from the next generation DM direct detection DARWIN~\cite{DARWIN:2020bnc} experiment, for which we conclude that an order of magnitude improvement with respect to current constraints is expected. Moreover, we utilize the available data from the reactor \eves experiment TEXONO~\cite{TEXONO:2009knm} which, however, results to subdominant constraints compared to  COHERENT and DM direct detection experiments.  Finally, we  discuss the complementarity of our present results with those obtained by beam-dumps, collider and astrophysical searches, and conclude that the present analysis offers the dominant constraints in the keV--MeV range.

The remainder of this paper is organized as follows. In Sec.~\ref{Sec:Theory}, we outline the theoretical framework of our present study. This, includes a discussion of gauge anomaly cancellation after introducing the new $U(1)_X$ gauge group, the mechanism of mass generation for the gauge mediators, and the interactions between fermions and gauge bosons within and beyond the SM. Additionally, we provide a brief theoretical description of \eves and \cevns within and beyond the SM, highlighting the implications of the extra $U(1)_X$ symmetry on these processes. Section~\ref{Sec:Data_Analysis} describes the details of the performed data analysis procedures for the different experiments considered. In Sec.~\ref{Sec:Results}, we present the results of our sensitivity analysis, and finally, we conclude with a summary of our findings in Sec.~\ref{Sec:Conclusions}.

\section{\label{Sec:Theory}Theory}
\subsection{Gauge anomalies}
In quantum field theory, symmetries play a fundamental role, with classical action invariance leading to conserved quantities. However, the quantization of the theory may lead to a violation of classically conserved currents which signifies a lack of invariance in the quantum effective action, hence indicating an anomaly. Anomalies can affect both global and gauge symmetries. In global symmetries, it merely means that classical selection rules are not applicable in the quantum domain and processes that are classically forbidden may still occur in the quantum realm. Moreover,  anomalies within gauge symmetry, termed gauge anomalies, can have profound implications. Specifically, gauge symmetry is pivotal in establishing the unitarity and renormalizability of gauge theories.  In a consistent gauge theory, it is imperative that these anomalies should mutually cancel out when contributions from multiple chiral fermions are considered~\cite{Witten:1982fp,Adler:1969gk,Bell:1969ts,Alvarez-Gaume:1983ihn,Delbourgo:1972xb}. For instance, within the SM framework $U(1)$, hypercharge serves as a potential origin for anomaly generation. 
Therefore, the SM hypercharges must obey four distinct anomaly cancellation conditions, as discussed in Ref. \cite{Prajapati:2024wuu}.
However, as it is well-known, the hypercharges of all the SM fermions add up in a way such that these anomalies are trivially canceled, making it a mathematically consistent theory and free from gauge anomalies~\cite{Geng:1989tcu,Minahan:1989vd}. 

The incorporation of new symmetries in BSM scenarios introduces further anomaly cancellation conditions and imposes additional constraints on possible charges of SM and BSM fermions. For example, the new anomaly cancellation conditions for a $SU(3)_C \otimes SU(2)_L \otimes U(1)_Y \otimes U(1)_X$ theory are shown below, which includes all gauge and mixed gauge-gravitational anomalies~\cite{Prajapati:2024wuu}. 
\begin{subequations}
\label{U1x anomaly cancellation}
\begin{align}
&[SU(3)_{C}]^2[U(1)_{X}]=  \sum\limits_{i} X_{Q'^{^{i}}} -  \sum\limits_{j} X_{q'^{^{j}}}.\label{ano1} 
\\& [SU(2)_{\mathtt{L}}]^2[U(1)_{X}]=  \sum\limits_{i} X_{L'^{^{i}}} + 3\sum\limits_{j} X_{Q'^{^{j}}}. \label{ano2}
\\& [U(1)_{Y}]^2 [U(1)_{X}] = \sum\limits_{i,j} ( Y_{L'^{^{i}}}^2   X_{L'^{^{i}}} + 3 Y_{Q'^{^{j}}}^2   X_{Q'^{^{j}}}  )  - \sum\limits_{i,j} ( Y_{l'^{^{i}}}^2 X_{l'^{^{i}}} + 3 Y_{q'^{^{j}}}^2 X_{q'^{^{j}}}  ). \label{ano3}
\\& [U(1)_{Y}] [U(1)_{X}]^2 = \sum\limits_{i,j} ( Y_{L'^{^{i}}}   X_{L'^{^{i}}}^2 + 3 Y_{Q'^{^{j}}}   X_{Q'^{^{j}}}^2  )  - \sum\limits_{i,j} ( Y_{l'^{^{i}}} X_{l'^{^{i}}}^2 + 3 Y_{q'^{^{j}}} X_{q'^{^{j}}}^2  ). \label{ano4}
\\&  [U(1)_{X}]^3= \sum\limits_{i,j} ( X_{L'^{^{i}}}^{3} + 3 X_{Q'^{^{j}}}^{3}  ) - \sum\limits_{i,j} ( X_{l'^{^{i}}}^{3} + 3 X_{q'^{^{j}}}^{3}  ). \label{ano5}
\\& [\text{Gravity}]^2[U(1)_{X}]= \sum\limits_{i,j} ( X_{L'^{^{i}}} + 3 X_{Q'^{^{j}}}  ) - \sum\limits_{i,j} ( X_{l'^{^{i}}} + 3 X_{q'^{^{j}}} ). \label{ano6}
\end{align}
\end{subequations}
Here, $Y_{\psi}$ represents the hypercharge of fermion $\psi$, with $X_{\psi}$ being the corresponding $U(1)_{X}$ charge, while the indices $i$,$j$ run over the fermion generations. We use $L' = \{L, \Psi\}$ to represent the collection of all SM \{$L= (\nu_\mathtt{L}, e_\mathtt{L})^{T}$\} and BSM ($\Psi$) fermions which are doublets under $SU(2)_{L}$ and singlets of $SU(3)_{C}$.
Moreover, SM ($l = e_\mathtt{R}$) and BSM ($\uppsi$) fermions  which are singlets under both $SU(2)_{L}$ and $SU(3)_{C}$, are denoted as $l' = \{l, \uppsi\}$. For fermions charged under $SU(3)_{C}$, $Q' = \{Q, \Psi_{c}\}$ denotes the set of all SM \{$Q=(u_\mathtt{L}, d_\mathtt{L})^{T}$\} and BSM ($\Psi_{c}$) $SU(2)_{L}$ doublets. Similarly, $q' = \{q, \uppsi_{c}\}$ includes all the SM \{$q = (u_\mathtt{L}, d_\mathtt{L})^{T}$ \} and BSM ($\uppsi_{c}$) $SU(2)_{L}$ singlet fermions, charged under $SU(3)_{C}$.

To cancel gauge anomalies, all six conditions shown in Eq.~\eqref{U1x anomaly cancellation} along with the anomaly involving SM gauge symmetries should vanish. To this purpose, three new right handed fermions ($f^{k},~k=1,2,3$ ) are introduced, acting as singlets within the SM, and also being charged under $U(1)_{X}$ with the corresponding charge denoted as $X_{f^{k}}$, where the index $k=1,2,3$. Furthermore, in addition to gauging anomalies, invariance of the SM Yukawa structure under $U(1)_{X}$ is also demanded. We thus consider the following Lagrangian 
\begin{equation}
\label{yukawa terms of e u d}
-\mathscr{L}_\text{Yukawa} \supset Y_{e}^{ij}\overline{L}^{i} \phi e_{\mathtt{R}}^{j} +Y_{u}^{ij} \overline{Q}^{i} \tilde{\phi} u_{\mathtt{R}}^{j}  + Y_{d}^{ij}\overline{Q}^{i} \phi d_{\mathtt{R}}^{^{j}}  +  \text{H.c.}\, ,
\end{equation}
where $\phi$ is the SM Higgs doublet with $\tilde{\phi}=i\sigma_{2}\phi^{*}$, while $\sigma_{2}$ is the second Pauli matrix. To fully preserve the SM Yukawa structure, identical charges for all generations of SM fermions are assumed, i.e., $X_{\psi}^{i} = X_{\psi}^{j}= X_{\psi}$. This choice allows to generate masses for the SM fermions from a single Higgs doublet, as done in the SM framework. Under this assumption, constraints on the $U(1)_{X}$ charges of SM fermions arising from the Yukawa sector can be expressed through the Higgs $U(1)_{X}$ charge ($X_{\phi}$), as
\begin{equation} 
\label{Higgs Charge Final}
X_{_{\phi}}=X_{L}-X_{e_{_{\mathtt{R}}}}=X_{Q}-X_{d_{_{\mathtt{R}}}}=X_{u_{_{\mathtt{R}}}}-X_{Q}.
\end{equation}
Then, by solving the anomaly cancellation conditions of Eq.~(\ref{ano1}-\ref{ano6}), using Eq.~\eqref{Higgs Charge Final},  the following constraints on fermion charges are implied~\cite{Prajapati:2024wuu}.  
\begin{subequations}
\label{Eq:Final_ano_con}
\begin{align}
&X_{Q}=-\frac{X_{L}}{3},~X_{u_{_{\mathtt{R}}}}=\frac{2X_{L}}{3}-X_{e_{_{\mathtt{R}}}},~X_{d_{_{\mathtt{R}}}}= -\frac{4X_{L}}{3}+X_{e_{_{\mathtt{R}}}},~X_{_{\phi}}=X_{L}-X_{e_{_{\mathtt{R}}}},\\
& \sum_{k=1}^{3} (X_{f^{k}})^3=  3 (2X_{L} - X_{e_{_{\mathtt{R}}}})^{3},~~~ \sum_{k=1}^{3} X_{f^{k}}= 3 (2X_{L} - X_{e_{_{\mathtt{R}}}})\, .
\end{align}
\end{subequations}

\begin{table*}[ht]
\begin{center}
\begin{adjustbox}{width=1\textwidth}
\renewcommand{\arraystretch}{1.8}
\begin{tabular}{|@{\hspace{2.5pt}} c  @{\hspace{2.5pt}}|@{\hspace{2.5pt}} c @{\hspace{2.5pt}}|@{\hspace{2.5pt}} c@{\hspace{2.5pt}}|@{\hspace{2.5pt}} c@{\hspace{2.5pt}}| @{\hspace{2.5pt}} c @{\hspace{2.5pt}}@{\hspace{2.5pt}}|}
 \hline 
 Fields & Sol 1 & Sol 2 & Sol 3 & Sol 4 \\ 
 \hline
 \hline 
 $Q$ & $\frac{-X_{L}}{3}$ & $-\frac{X_{L}}{3}$ & $\frac{1}{s}$ & $-\frac{X_{L}}{3}$  \\ 
 $u_{_{\mathtt{R}}}$ & $\frac{-4X_{L}}{3}$ & $-\frac{4X_{L}}{3}+\kappa$ & $-(\kappa-\frac{4}{s})$ & $\frac{-4X_{L}}{3}-\frac{s^{2}-\kappa^{2}}{8}$  \\ 
 $d_{_{\mathtt{R}}}$ & $\frac{2X_{L}}{3}$ & $\frac{2X_{L}}{3}-\kappa$ & $\kappa-\frac{2}{s}$ & $\frac{2X_{L}}{3}+\frac{s^{2}-\kappa^{2}}{8}$ \\ 
 $L$ & $X_{L}$ & $X_{L}$ & $-\frac{3}{s}$ & $X_{L}$  \\ 
 $e_{_{\mathtt{R}}}$ & $2X_{L}$ & $2X_{L}-\kappa$ & $\kappa-\frac{6}{s}$ & $2X_{L}+\frac{s^{2}-\kappa^{2}}{8}$  \\
 \hline
 \hline
 $f^{1}$ & $0$ & $\kappa$ & $5\kappa$ & $\frac{1}{8}(5s^{2}+3 \kappa^{2})$ \\ 
 $f^{2}$ & $\kappa$ & $\kappa$ & $-4\kappa$ &  $-\frac{1}{8}(4s^{2}+ 3 \kappa s +\frac{\kappa^{3}}{s})$ \\ 
 $f^{3}$ & $-\kappa$ & $\kappa$ & $-4\kappa$ &  $\frac{1}{8}(-4s^{2}+ 3 \kappa s +\frac{\kappa^{3}}{s}) $   \\ 
 \hline
 \hline
 $\phi$ & $-X_{L}$ & $\kappa - X_{L}$ & $-(\kappa-\frac{3}{s})$ & $-(X_{L}+\frac{s^{2}-\kappa^{2}}{8})$  \\ 
 \hline
\end{tabular}
\end{adjustbox}  
\end{center}
\caption{Fermion charges under $U(1)_{X}$ symmetry satisfying gauge anomaly cancellation conditions and Higgs charge for expressing SM invariant Yukawas, see Ref. \cite{Prajapati:2024wuu}. Each column corresponds to a different class of solutions, and each specific set of values for  $X_{L}, \kappa, s$ represents a distinct model.}
\label{tab:Final_ano_con}
\end{table*}

Table~\ref{tab:Final_ano_con}, shows possible solutions of Eq.~\eqref{Eq:Final_ano_con} with each column representing a separate solution, see Ref. \cite{Prajapati:2024wuu} for more details. Note that each set of values for $X_{L}, \kappa, s$ represents a distinct model, and the free parameters can be varied to make relative changes in fermionic $U(1)_{X}$ charges. As a result, each model exhibits a different phenomenological signature and they can be constrained through various experiments.

\subsection{Mass spectrum of gauge mediators} \label{Sub_Sec:Mass_spectrum_of_gauge_mediators}

 One of the most prominent features of extra $U(1)_{X}$ symmetries is the prediction of a novel electrically neutral gauge boson ($Z'$). As discussed earlier, the masses of all SM fermions can be generated by the SM Higgs boson. Notice that, this holds true for the full set of possible solutions listed in Table \ref{tab:Final_ano_con}. In this section, we will focus our attention on the masses of SM ($A,\,Z,\,W^{\pm}$) and BSM ($Z'$) bosons. The mass spectrum of  gauge bosons arises from the expansion of the kinetic terms of the scalar fields after  spontaneous symmetry breaking (SSB) of both the hypercharge  and  \( U(1)_{X} \) symmetries. The kinetic terms of scalars are written as
\begin{equation}\label{Eq:Kinetic_term_scalar}
(D_{\rho}\phi)^{\dagger}D^{\rho}\phi+(D_{\rho}\chi_{i})^{\dagger}D^{\rho}\chi_{i}~.
\end{equation}
Since the SM Higgs alone cannot account for the masses of all vector bosons, additional scalar fields, denoted as $\chi_{i}$, are introduced alongside it. These new scalars are singlets of the SM but carry a charge under the $U(1)_{X}$ symmetry. They could also participate in the mass generation of SM neutrinos and new fermions. Since we will not focus on a specific model, in general we can consider any number of $\chi_{i}$'s. The pseudoscalar element of one of these complex scalars serves as the Goldstone boson for the $Z'$ boson. The covariant derivative, $D_{\rho}$, in that case is defined as
\begin{equation}\label{codrivative}
D_{\rho}= \partial_{\rho} +igT^{a}W^{a}_{\rho} + ig'\frac{Y}{2}B_{\rho}+i\textsl{g}_{_{x}}XC_{\rho}\,.
\end{equation}
In this context, $g'$, $g$, and $\textsl{g}_{_{x}}$ denote the gauge couplings corresponding to  $U(1)_{Y}$, $SU(2)_{L}$, and $U(1)_{X}$, respectively, while the charges associated to $U(1)_{Y}$ and $U(1)_{X}$ are represented by $Y$ and $X$, respectively. The generators of $SU(2)_{L}$ are given by $T^{a} = \sigma^{a}/2$, where $\sigma^{a}$ represents the Pauli matrices.
The SM Higgs doublet and $\chi_{i}$'s, acquire a vacuum expectation value (VEV), to break both the electroweak and $U(1)_{X}$ symmetries. The corresponding VEVs can be expressed as follows:
\begin{equation}\label{VEV}
\langle \phi \rangle = \frac{1}{\sqrt{2}}\begin{bmatrix}
0 \\
v
\end{bmatrix},~~~~ \langle \chi_{i} \rangle = \frac{v_{i}}{\sqrt{2}}~.
\end{equation} 
By substituting the covariant derivative and fields with the expression defined in Eq.~\eqref{codrivative} and~\eqref{VEV}, the mass matrix of neutral bosons in the basis $(B_{\rho},W_{\rho}^{3},C_{\rho})$ takes the form 
\begin{equation}\label{Eq:Gauge_Boson_Mass_mat}
\mathcal{M}^2_{_{V}}= \frac{v^{2}}{4}\begin{pmatrix}
g'^{2} & -gg' & 2g'X_{_{\phi}}\textsl{g}_{_{x}}\\
-gg'   &  g^{2} & -2gX_{_{\phi}}\textsl{g}_{_{x}}\\
2g'X_{_{\phi}}\textsl{g}_{_{x}} & -2gX_{_{\phi}}\textsl{g}_{_{x}} & 4u^{2}\textsl{g}_{_{x}}^{2}
\end{pmatrix}\,,
\end{equation}
where $u^{2}=X^{^{2}}_{_{\phi}}+u_{\chi}^{2}/v^{2}$, with $u_{\chi}$  defined as $u_{\chi}=\sqrt{\sum_{i}(X^{^{2}}_{\chi_{_{i}}}v_{i}^{2})}$.  Here, $X_{\chi_{_{i}}}$ represents the $U(1)_{X}$ charge of the scalar fields $\chi_{i}$, and $X_{_{\phi}}$ denotes the $U(1)_{X}$ charge of the SM Higgs doublet. Note that, the Higgs field is generally charged under the new $U(1)_{X}$ symmetry, as shown in Table~\ref{tab:Final_ano_con}. This charge leads to mass mixing between the SM ($B_{\rho},W_{\rho}^{3}$) and BSM ($C_{\rho}$) neutral bosons in the gauge basis, as shown in Eq.~\eqref{Eq:Gauge_Boson_Mass_mat}. Let us also stress that, although these bosons can also mix through the kinetic terms of gauge bosons
(kinetic mixing), we assume this kinetic mixing to be negligible in our context.  

{We proceed by diagonalizing the mass matrix given in Eq.~\eqref{Eq:Gauge_Boson_Mass_mat} with the aid of an orthogonal matrix, that relates the gauge ($B^{\rho}, W_{3}^{\rho}, C^{\rho}$) and mass ($A^\mu, Z^\mu, Z^{\prime \mu}$) basis. The mass and gauge eigenstates are related as
\begin{equation}
\label{unitary matrix}
\begin{bmatrix}
A^{\rho} \\
Z^{\rho} \\
Z^{\prime\rho}
\end{bmatrix} =
\begin{bmatrix}
\cos\theta_{W} &~ \sin\theta_{W} &~0\\
-\cos\alpha \sin\theta_{W} & \cos\alpha \cos\theta_{W}
&~ -\sin \alpha\\
-\sin\alpha \sin\theta_{W} &~  \sin\alpha\cos\theta_{W} &~ \cos\alpha 
\end{bmatrix}  \begin{bmatrix}
B^{\rho} \\
W_{3}^{\rho}\\
C^{\rho}
\end{bmatrix}.
\end{equation}
Following this, one mass eigenstate becomes zero which is identified as the photon, while the remaining two mass eigenstates read
\begin{equation}\label{mass}
\begin{aligned}
M_{Z'}^{2}= \frac{v^{2}}{8}(A_{0}-\sqrt{B_{0}^{2}+C_{0}^{2}}),&&&&&&\,M_{Z}^{2}=\frac{v^{2}}{8}(A_{0}+\sqrt{B_{0}^{2}+C_{0}^{2}})\,,
\end{aligned}
\end{equation}
with $A_0$, $B_0$ and $C_0$ defined as  
\begin{equation}
\begin{aligned}
      A_{0}=&g^{2}+{g'}^{2}+4u^{2}\textsl{g}_{_{x}}^{2} \, ,\\   B_{0}=&4X_{_{\phi}}\textsl{g}_{_{x}}\sqrt{g^{2}+{g'}^{2}} \, ,\\
      C_{0}=& g^{2} + {g'}^{2} - 4u^{2}\textsl{g}_{_{x}}^{2}\,.
      \end{aligned}
\end{equation}
The rotation angles are defined as
\begin{equation}
\label{Eq:angle}
\tan\theta_{W} = \frac{g'}{g},~~~ \tan2\alpha = \frac{4X_{_{\phi}}\textsl{g}_{_{x}}\sqrt{g^{2}+{g'}^{2}}}{| g^{2} + {g'}^{2} - 4u^{2}\textsl{g}_{_{x}}^{2} |} = \frac{2X_{\phi} g_{_x} v M_{Z}^{\text{SM}}}{| (M_{Z}^{\text{SM}})^{2} - [v^{2}X_{\phi}^{2} + u_{\chi}^{2}]g_{_x}^{2}  |} \, ,
\end{equation}
while the $u_{\chi}$ parameter can be expressed as a function of $M_{Z'}$ and $\textsl{g}_{_{x}}$,
\begin{equation}
\label{Eq:VEV_Paramter_ux}
u_{\chi} = \frac{M_{Z'}}{\textsl{g}_{_{x}}}\sqrt{ \left |  \frac{4M_{Z'}^{2} -v^{2} (g^{2} + g'^{2} + 4 \textsl{g}_{_{x}}^{2}X_{\phi}^{2})}{4M_{Z'}^{2} -v^{2} (g^{2} + g'^{2} )} \right | } =  \frac{M_{Z'}}{\textsl{g}_{_{x}}}\sqrt{ \left |  \frac{M_{Z'}^{2} - [(M_{Z}^{\text{SM}})^{2} + v^{2}\textsl{g}_{_{x}}^{2}X_{\phi}^{2}]}{M_{Z'}^{2} -(M_{Z}^{\text{SM}})^{2}} \right | } 
\end{equation}
Before closing, the following comments are in order. From the right-hand side (rhs) of Eq.~(\ref{mass}), one can notice that in the context of the considered chiral model the mass of the SM $Z$ boson receives contributions from $Z-Z'$ mixing, as opposed to the SM case where it reads $M_{Z}^{\text{SM}} = v\sqrt{g^{2}+g'^{2}}/2$. On the other hand, the $W$ boson mass remains identical to the SM expression given by $M_{W}^{2} = v^{2} g^{2}/{4}$.  

%
\subsection{\label{SubSec:SM_BSM_interaction}Interactions between fermions and gauge bosons within and beyond SM}
In this section, we study the gauge boson interactions with SM and BSM fermions. We first focus on the $W^{\pm}$ boson and the photon.
As already mentioned above, since there is no additional mixing in the charged sector, the $W$ mass and its interactions with other fermions remain the same as in the SM. Additionally, by choosing an appropriate rotation matrix, we prevent mixing between the photon and the new boson, ensuring that the QED current remains unaltered. Hence, regarding the charged current and QED sectors, one has the same relations as in the SM case
\begin{equation}\label{Eq:Fermi_constant_and_fine_structre_constant}
    \frac{G_F}{\sqrt{2}} = \frac{g^{2}}{8M_{W}^{2}}\, , ~~~ \alpha_{\text{EM}} = \frac{e^{2}}{4 \pi} = \frac{g^{2}\sin^{2}\theta_{W}}{4 \pi} \, . 
\end{equation}
Here, $G_F$ and $\alpha_{\text{EM}}$ are the well-known Fermi and fine structure constants, respectively, and we have used the identity, $g \sin{\theta_{W}} = g' \cos{\theta_{W}} = e$. Since, $G_F$ and $\alpha_{\text{EM}}$ are measured with higher precision compared to the gauge couplings, it is convenient to define Weinberg angle $\theta_{W}$, as~\footnote{The two definitions of the Weinberg angle $\theta_{W}$, given in Eq.~\eqref{Eq:angle} and Eq.~\eqref{Eq:Winberg_angle_def} are equivalent.} 
\begin{equation}\label{Eq:Winberg_angle_def}
    \sin^{2}\theta_{W} = \frac{\pi \alpha_{\text{EM}}}{\sqrt{2}G_FM_{W}^{2}}.
\end{equation}

We now proceed by focusing our attention on another precisely measured parameter of the SM, namely the electroweak parameter $\rho$, defined as
\begin{equation}\label{rho}
\rho=\frac{M_{W}^{^{2}}}{M_{Z}^{^{2}} {\cos}^{2}{\theta_{W}}}\, .
\end{equation}
The latter plays a crucial role in the study of electroweak interactions~\cite{ROSS1975135} and in the SM it is equal to 1 assuming tree-level interactions. On the other hand, due to $Z-Z'$ mixing through the angle $\alpha$, within the chiral model the $Z$ mass exceeds its SM predicted value~\footnote{This assumes  $Z'$ being lighter than $Z$.} as implied by Eq.~\eqref{mass}, and hence the $\rho$ parameter in this case becomes less than $1$. 
Let us also remind that in such scenarios a new parameter $\rho'$ can be introduced which is equal to 1 at the tree level~\cite{Bento:2023weq, Bento:2023flt} 
\begin{equation}
    \rho' = \frac{M_{W}^{2}}{(M_{Z}^{2} \cos^{2}\alpha + M_{Z'}^{2}\sin^{2}\alpha ) \cos^{2} \theta_{W}} = 1 \, .
\end{equation}
Using the latter definition, the $\rho$ parameter can be rewritten, as
\begin{equation}
\label{Eq.:rho_param_BSM}
    \rho -1 = \left[ \left( \frac{M_{Z'}}{M_{Z}} \right)^{2} - 1 \right] \sin^{2} \alpha.
\end{equation}

Figure~\ref{Fig:rho_vs_gx} illustrates the $\rho$ parameter as a function of $\textsl{g}_{_{x}}$, by considering various $M_{Z'}$ and $X_{\phi}$ values. The most recent global fit to electroweak precision data provides a $3\sigma$ allowed range for the $\rho$ parameter that has been measured to be $1.00038 \pm 0.00060$~\cite{ParticleDataGroup:2020ssz}. The red-shaded area in the plot represents the exclusion region for the $\rho$ parameter at the $3\sigma$ confidence level. It follows  that  $\textsl{g}_{_{x}}X_{\phi}\lesssim 5.5 \times 10^{-3}$ is adequate to satisfy the constraint on the $\rho$ parameter. In this particular limit Eq.~\eqref{Eq:VEV_Paramter_ux} simplifies to $u_{\chi} \approx M_{Z'} / g_x$. As a result, the $\rho$ parameter could be approximated as  

\begin{equation}
\label{Eq.Rho_Param}
 \rho \approx \frac{1}{1 + \frac{4X_{\phi}^{2}\textsl{g}_{_{x}}^{2}}{g^{2}+g'^{2}}},  
\end{equation}
indicating that  effectively $\rho$  is independent of $M_{Z'}$, as corroborated by Fig.~\ref{Fig:rho_vs_gx}.
\begin{figure}[t]
   \centering
   \captionsetup{justification=raggedright}
  \includegraphics[width=0.49\textwidth]{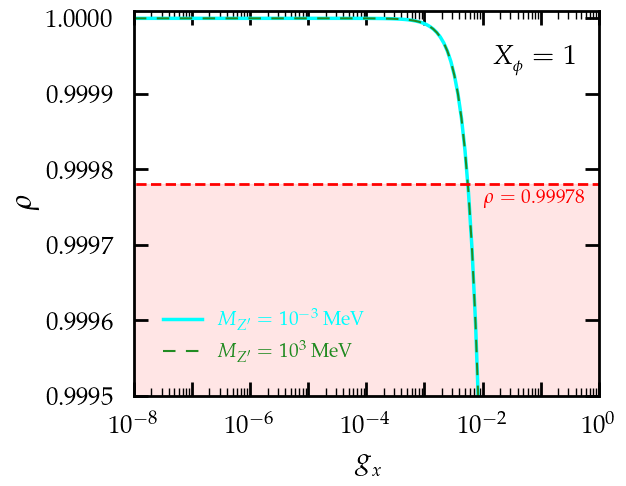}
   \includegraphics[width=0.49\textwidth]{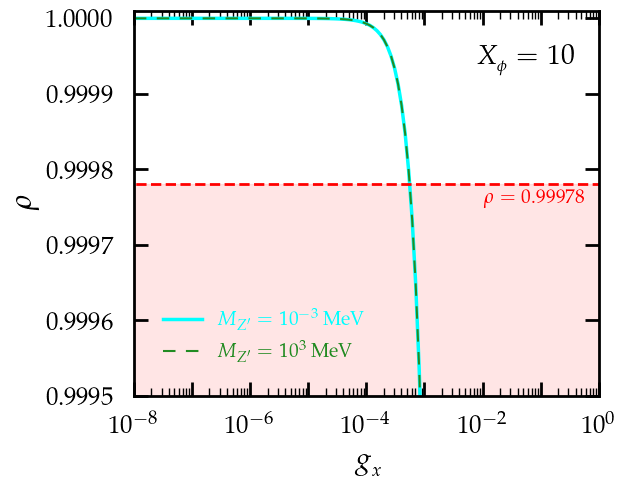}
  \caption{Variation of $\rho$ parameter as a function of $\textsl{g}_{_{x}}$ for  $M_{Z'}=10^{-3}$ MeV and $M_{Z'}= 1$~GeV. The $\rho$ parameter is found to be independent of $M_{Z'}$ within the parameter's range of interest. Left (right) panels correspond to $X_\phi=1$ ($X_\phi=10$).}
  \label{Fig:rho_vs_gx}
\end{figure}
\begin{table*}[h]
\begin{center}
\begin{adjustbox}{max width=\textwidth}
\renewcommand{\arraystretch}{2}
\begin{tabular}
{|@{\hspace{3pt}} c  @{\hspace{3pt}}|@{\hspace{3pt}} c  @{\hspace{3pt}}|@{\hspace{3pt}} c @{\hspace{3pt}}|@{\hspace{0.5pt}} c@{\hspace{3pt}}|@{\hspace{3pt}} c@{\hspace{3pt}}| @{\hspace{3pt}} c @{\hspace{3pt}}|@{\hspace{3pt}} c @{\hspace{3pt}}|}
\hline
\centering \textbf{Fields}& \centering \textbf{Hypercharge} & \centering \textbf{Weak Isospin} & \multicolumn{4}{@{\hspace{10pt}} c @{\hspace{10pt}}|}{$\mathbf{U(1)_{X}}$ \textbf{Charge}}\\
\cline{4-7}
 & \centering$(\mathbf{Y})$ &\centering ($\mathbf{T^{3}}$) & \hspace{.2cm} \textbf{BM 1} \hspace{.2cm}& \hspace{.2cm}\textbf{BM 2} \hspace{.2cm}& \hspace{.2cm}\textbf{BM 3}\hspace{.2cm} &\hspace{.2cm} \textbf{B - L}~~\hspace{0.cm}\\
\hline
$u_{_{\mathtt{L}}}$ & 1/3 & 1/2 & -13/3 & -3 & -1/3 & 1/3 \\
\hline
$u_{_{\mathtt{R}}}$ & 4/3 & 0 & -4/3 & 10 & 5/3 & 1/3\\
\hline
$d_{_{\mathtt{L}}}$ & 1/3 & -1/2 & -13/3 & -3 & -1/3 & 1/3\\
\hline
$d_{_{\mathtt{R}}}$ & -2/3 & 0 & -22/3 & -16 & -7/3 & 1/3\\
\hline
$e_{_{\mathtt{L}}}$ & -1 & -1/2 & 13 & 9 & 1 & -1\\
\hline
$e_{_{\mathtt{R}}}$ & -2 & 0 & 10 & -4 & -1 & -1\\
\hline
$\nu_{_{\mathtt{L}}}$ & -1 & 1/2 & 13 & 9 & 1 & -1\\
\hline
\hline
$f^{1}$ & 0& 0 & 16& -110 & 10 & -1\\
\hline
$f^{2}$ & 0& 0 & 16 & 88 & -18 & -1\\
\hline
$f^{3}$ & 0 & 0 & 16 & 88 & 17& -1\\
\hline
\hline
$\phi$ & 1 & - & 3 & 13 & 2 & 0\\
\hline
\end{tabular}
\end{adjustbox}
\end{center}
\caption{SM and $U(1)_{X}$ charges of various fields relevant to this study. The $U(1)_{X}$ charges for different fields are provided for four benchmark model (BM) choices. The first three benchmark $U(1)_{X}$ charge combinations represent chiral solutions to the anomaly cancellation conditions, while the last one corresponds to the vector $B-L$ model. It is noteworthy that for each benchmark model choice, identical charges have been assigned to all generations of a given fermion.}
\label{Tab:BP_Charges}
\end{table*}

We now turn our attention on  neutral current processes mediated by $Z$ and $Z'$ bosons. Let us first start with the interactions mediated by the SM gauge boson $Z$ within the chiral model. In this case the relevant Lagrangian density  can be expressed as
 %
\begin{equation} \label{SM_lagrangian}
\mathscr{L}_{\text{int}} \subset - \frac{g}{\cos\theta_{W}} \overline{\psi} \gamma^{\rho}\left(g_{\psi_{_{\mathtt{L}}}}^{z}\mathtt{L} + g_{\psi_{_{\mathtt{R}}}}^{z}\mathtt{R}\right)\psi Z_{\rho}\,.
\end{equation} 
Here, $\psi$ can be any SM or BSM fermion, $\mathtt{L}$ and $\mathtt{R}$ are left and right-handed chiral projection operators, respectively. The left- and right-handed couplings $g_{\psi_{\mathtt{L}}}^{z}$ and $g_{\psi_{\mathtt{R}}}^{z}$ are  expressed as
\begin{subequations}\label{SM Couplings}
\begin{align}
&g_{\psi_{_{\mathtt{L}}}}^{z} = \Big(T_{\psi_{_{\mathtt{L}}}}^3 - Q_{\psi_{_{\mathtt{L}}}}\sin^{2}\theta_{W}\Big)\cos{\alpha} - \frac{X_{\psi_{_{\mathtt{L}}}}\textsl{g}_{_{x}}}{g}\sin{\alpha}\cos{\theta_{W}}\,, \\
&g_{\psi_{_{\mathtt{R}}}}^{z} = - Q_{\psi_{_{\mathtt{R}}}}\sin^{2}\theta_{W}\cos{\alpha} - \frac{X_{\psi_{_{\mathtt{R}}}}\textsl{g}_{_{x}}}{g}\sin{\alpha}\cos{\theta_{W}}\, ,
\end{align}
\end{subequations}
where, $T_{\psi}^3 $ and $Y_{\psi}$ represent the third component of the weak isospin and the hypercharge of the fermion  $\psi$, respectively. The QED charge of $\psi$ is defined as $Q_{\psi} = T_{\psi}^3 + Y_{\psi}/2$, while  $X_{\psi}$ represents its  $U(1)_X$ charge, as given in Table~\ref{Tab:BP_Charges}. 

At this point, it is worth noticing the presence of BSM contributions in the fermion couplings with the SM $Z$ boson. These, arise due to $Z-Z'$ mixing (see Appendix~\ref{Appendix:1} for more details), and are characterized  by the gauge coupling $\textsl{g}_{_{x}}$, mixing angle $\alpha$ and fermion charge under $U(1)_{X}$. To satisfy the constraint on the $\rho$ parameter discussed previously, the mixing angle $\alpha$ can be approximated as $\alpha \approx 2.67X_{\phi}\textsl{g}_{_{x}}$, and hence the BSM contribution becomes essentially proportional to $\textsl{g}_{_{x}}^{2}$. Therefore, within the $\textsl{g}_{_{x}}$ limit of interest in this study, one can safely approximate $\cos{\alpha}\sim 1$ and completely disregard the BSM contributions to the SM $Z$ boson couplings shown in Eq.~\eqref{SM Couplings}. Subsequently,  the $Z$ boson couplings within the chiral model reduce to the purely SM couplings  $g_{\psi_{\mathtt{L}}}^{z}$ and $g_{\psi_{\mathtt{R}}}^{z}$. For completeness, the latter are listed in Table~\ref{Tab:SM_Charges} for the various relevant fermions considered in this study.
\begin{table}[t]
    \centering
    \begin{adjustbox}{max width=\textwidth}
    \begin{tabular}{|c|c|c|}
        \hline
        \hline
        Fermions $(\psi)$ & $g_{\psi_{_\mathtt{L}}}^z$ & $g_{\psi_{_\mathtt{R}}}^z$ \\[0.2cm]
        \hline
        \hline
        $\nu^{i}$ & $+\frac{1}{2}$ & 0 \\
        $e^{i}$ & $-\frac{1}{2} + \sin^2{\theta_W}$ & $\sin^2{\theta_W}$ \\
        $u^{i}$ & $+\frac{1}{2} - \frac{2}{3}\sin^2{\theta_W}$ & $-\frac{2}{3}\sin^2{\theta_W}$ \\
        $d^{i}$ & $-\frac{1}{2} + \frac{1}{3}\sin^2{\theta_W}$ & $\frac{1}{3}\sin^2{\theta_W}$ \\
        \hline
    \end{tabular}
    \end{adjustbox}
    \caption{Couplings between left-handed ($g_{\psi_{_\mathtt{L}}}^z$) and right-handed ($g_{\psi_{_\mathtt{R}}}^z$) fermions with the SM $Z$ boson,  where $i=1,2,3$ is the flavor index.}
    \label{Tab:SM_Charges}
\end{table}

We now turn to $Z'$-mediated interactions. In this case, the relevant Lagrangian density takes the form
\begin{equation} 
\label{BSM_lagrangian}
\mathscr{L}_{\text{int}} \subset - \overline{\psi} \gamma^{\rho}\left(g_{\psi_{_{\mathtt{L}}}}^{z'}\mathtt{L}+g_{\psi_{_{\mathtt{R}}}}^{z'}\mathtt{R}\right)\psi ~Z'_{\rho} \, ,
\end{equation} 
while the resulting BSM couplings can be expressed as (see Appendix~\ref{Appendix:1})
\begin{subequations}
\label{Eq.BSM_Charge}
\begin{align}
& g_{\psi_{_{\mathtt{L}}}}^{z'} = \Big( gT_{\psi_{_{\mathtt{L}}}}^{3}\cos{\theta_{W}} - \frac{g'Y_{\psi_{_{\mathtt{L}}}}}{2}\sin{\theta_{W}} \Big)\sin{\alpha} + X_{\psi_{_{\mathtt{L}}}}\textsl{g}_{_{x}}\cos{\alpha}\,, \\
& g_{\psi_{_{\mathtt{R}}}}^{z'} =   - \frac{g'Y_{\psi_{_{\mathtt{R}}}}}{2}\sin{\theta_{W}} \sin{\alpha} + X_{\psi_{_{\mathtt{R}}}}\textsl{g}_{_{x}}\cos{\alpha}\,.
\end{align}
\end{subequations}
It is interesting to notice that the above BSM couplings include a component similar to the SM one but suppressed by $\sin{\alpha}$. By recalling that $\sin{\alpha}$ is of the same order as $\textsl{g}_{_{x}}$, the SM contribution to the BSM couplings cannot be disregarded, unlike in the case of the $Z$ boson discussed previously where the BSM contribution was neglected.

The present study considers three different benchmark models (BM 1 -- BM 3), see e.g. Table~\ref{Tab:BP_Charges}, which satisfy the anomaly cancellation conditions presented in Eq.~\eqref{U1x anomaly cancellation}. The first model, BM 1, corresponds to Sol 2, with $X_{L} = 13$ and $\kappa = 16$. The second model, BM 2, is based on Sol 3, where $s = -1/3$ and $\kappa = -22$. The third model, BM 3, comes from Sol 4, with $X_{L} = 1$, $s = 1$ and $\kappa = 5 $. BM 1 is chosen because it predicts large lepton charges. In BM 2, the charges of all fermions are integer, and the charges of BSM fermions are quite large. BM 3  is the same benchmark model as considered in Ref. \cite{Prajapati:2024wuu}. Finally, we will also consider the well-studied vector $B-L$ model, for comparison and completeness.

\subsection{Elastic neutrino-electron scattering within and beyond the SM}
\label{SubSec:EvES}
 E$\nu$ES is a well-studied weak interaction process in which neutrinos or antineutrinos of all flavors ($\alpha = \{e, \mu, \tau\}$) scatter elastically on electrons. In what follows, we present briefly the relevant formalism exploring E$\nu$ES interactions within and beyond the SM.
\subsubsection{\eves within SM}
\label{SubSec:EvES_SM}
The effective interaction Lagrangian relevant for the \eves within SM can be expressed as follows~\cite{Giunti:2007ry}
\begin{equation}
\label{Eq:EvES_SM_Lagrangian}
\begin{aligned}
    \mathscr{L}_\mathrm{SM}^{\nu e} \subset - 2\sqrt{2}G_F~\Big\{ \Big[\bar{\nu}_\alpha\gamma^\rho \mathtt{L} \nu_\alpha \Big] \left[\bar{e}\gamma_\rho \left(g_{e_{_\mathtt{L}}}^z \mathtt{L} + g_{e_{_\mathtt{R}}}^z \mathtt{R}\right)e \right] ~+~ \Big[\bar{\nu}_e\gamma^\rho \mathtt{L} e \Big] \Big[\bar{e}\gamma_\rho \mathtt{L} \nu_e \Big] \Big\},
\end{aligned}
\end{equation}
where the couplings $g_{e_{_\mathtt{L}}}^z$ and $g_{e_{_\mathtt{R}}}^z$ are provided in Table~\ref{Tab:SM_Charges}. It should be noted that Eq.~\eqref{Eq:EvES_SM_Lagrangian} includes both neutral- and charged-current contributions for $\nu_e-e^-$ scattering, while for $\nu_{\mu,\tau}-e^-$ scattering only neutral-current contributions are relevant.

The corresponding tree-level differential \eves cross section with respect to the electron recoil energy $T_{er}$ has been previously written as~\cite{Kayser:1979mj}
\begin{equation}
\left[\frac{d\sigma_{\nu_\alpha}}{d T_{er}}\right]_\text{SM}^{\nu e}=  \frac{G_F^2m_e}{2\pi}\left[(\textsl{g}_V + \textsl{g}_A)^2 + (\textsl{g}_V - \textsl{g}_A)^2\left(1-\frac{T_{er}}{E_\nu}\right)^2  -(\textsl{g}_V^2-\textsl{g}_A^2)\frac{m_e T_{er}}{E_\nu^2}\right] \, ,
\label{EqEvesSM}
\end{equation}
where $E_\nu$ is the incoming neutrino energy and $m_e$ the electron mass. The vector and axial vector couplings are defined as
\begin{equation}
 \label{table:EveS_SM_couplings}
 \textsl{g}_V= g_{e_{_{\mathtt{L}}}}^{z} +g_{e_{_{\mathtt{R}}}}^{z} + \delta_{\alpha e} = -\frac{1}{2}+2\sin^2\theta_w + \delta_{\alpha e}, \qquad  \textsl{g}_A = g_{e_{_{\mathtt{L}}}}^{z} - g_{e_{_{\mathtt{R}}}}^{z} + \delta_{\alpha e}=-\frac{1}{2}+ \delta_{\alpha e} \,, 
\end{equation}
with the Kronecker delta, $\delta_{\alpha e}$, term accounting for the charged-current contributions to the differential cross section, which are only present for $\nu_e$--$e^-$ and $\bar{\nu}_e$--$e^-$ scattering. For antineutrino scattering $(\bar{\nu}_\alpha e^- \rightarrow \bar{\nu}_\alpha e^-)$, the \eves differential cross section is given by Eq.~\eqref{EqEvesSM} with the substitution $\textsl{g}_{A}\rightarrow -\textsl{g}_{A}$.

\subsubsection{\eves within BSM}
In the presence of a new $U(1)_X$ gauge symmetry, BSM contributions to the \eves arise from $Z'$-mediated interactions. The Lagrangian density for this interaction can be expressed as
\begin{equation}
    \label{Eq:EvES_BSM_Lagrangian}
    \mathscr{L}_\mathrm{BSM}^{\nu e}\subset -\Big[g_{\nu_{\alpha}}^{z'}\bar{\nu}_\alpha\gamma^\rho\mathtt{L}\nu_\alpha+\bar{e}\gamma^\rho\left(g_{e_{_{\mathtt{L}}}}^{z'}\mathtt{L}+g_{e_{_{\mathtt{R}}}}^{z'}\mathtt{R}\right)e\Big]Z'_{\rho}\,.
\end{equation}
Here, the BSM couplings $g_{\nu_{\alpha}}^{z'}$, $g_{e_{_{\mathtt{L}}}}^{z'}$, and $g_{e_{_{\mathtt{R}}}}^{z'}$ can be derived from Eq.~\eqref{Eq.BSM_Charge}. It should also be noted that, as discussed in Sec.~\ref{SubSec:SM_BSM_interaction}, due to the mixing between $Z$ and $Z'$, the BSM couplings include components from the SM, which are not negligible. Let us stress that in Eq.~\eqref{Eq:EvES_BSM_Lagrangian} the chirality indices for $\nu_\alpha$ are dropped for simplicity. This is because all the models discussed here are accommodating left-handed neutrinos only.
Finally, the total differential cross section corresponding to SM plus BSM contributions is obtained by replacing the SM couplings $\textsl{g}_V$ and $\textsl{g}_A$, defined in Eq.~\eqref{table:EveS_SM_couplings}, with
\begin{equation}
\label{Eq.EvES_BSM_Couplings}
\textsl{g}_{V/A} \rightarrow \textsl{g}_{V/A} + \frac{(g_{e_{_{\mathtt{L}}}}^{z'} \pm g_{e_{_{\mathtt{R}}}}^{z'})~ g_{\nu_{\alpha}}^{z'}}{2\sqrt{2}G_{F}(2m_{e}T_{er}+M_{Z'}^{2})}\,.
\end{equation} 

Before closing this discussion, from Eq.~\eqref{Eq.EvES_BSM_Couplings}, readers should note that if $Z-Z'$ mixing is switched off (i.e. $\alpha=0$), the BM 3 model contribution becomes purely axial vector with the same strength as the $B-L$ model, which is purely vectorial. On the other hand, contributions from the other two chiral models (BM 1 and BM 2) include both vector and axial vector components, even for $\alpha=0$.

\subsection{Coherent elastic neutrino-nucleus scattering within and beyond the SM}
\label{SubSec:CEvNS}

CE$\nu$NS is a low-energy neutral current process in which  incident neutrinos scatter off an entire nucleus, involving nuclear ground-state $\to$ ground-state transitions. Here, we review \cevns and its interactions within and beyond the SM and discuss the relevant cross sections.

\subsubsection{\cevns within SM}
Within the framework of the SM, at low and intermediate neutrino energies $(E_\nu \ll M_{Z})$, \cevns is accurately described by an effective four-fermion Fermi interaction Lagrangian~\cite{Barranco:2005yy}
\begin{equation}
\label{equn:CEvNS_SM_Lagrangian}
\mathscr{L}_\mathrm{SM}^{\nu N} \subset -2\sqrt{2}G_F \sum_{\substack{q=u,d \\ \alpha=e,\mu,\tau}} \Big[\bar{\nu}_\alpha \gamma^\rho \mathtt{L} \nu_\alpha \Big] \Big[\bar{q} \gamma_\rho (g_{q_{_\mathtt{L}}}^z \mathtt{L} + g_{q_{_\mathtt{R}}}^z \mathtt{R}) q \Big] \, .
\end{equation}
The SM couplings $g_{q_{_\mathtt{L}}}^z$ and $g_{q_{_\mathtt{R}}}^z$ can be obtained from Eq.~\eqref{SM Couplings} and Table~\ref{Tab:SM_Charges}, with the relevant fermions, $q \equiv \{u, d\}$. As discussed in Sec.~\ref{SubSec:SM_BSM_interaction}, within the $\textsl{g}_{_{x}}$ limit of interest in this study, we can approximately neglect the BSM contributions to the SM couplings.
At the tree level, the SM differential \cevns cross section with respect to the nuclear recoil energy $T_{nr}$ is given by~\cite{Freedman:1973yd}
\begin{equation}
\label{equn:CEvNS_SM_xsec}
\left[\frac{d\sigma}{dT_{nr}}\right]_\mathrm{SM}^{\nu N} = \frac{G_F^2 m_N}{\pi} \left(Q_V^\text{SM}\right)^2 \left(1 - \frac{m_N T_{nr}}{2E_\nu^2}\right) \, ,
\end{equation}
where $m_N$ is the nuclear mass, while the SM vector weak charge $Q_V^\text{SM}$ takes the form~\cite{Papoulias:2018uzy}
\begin{equation}
\label{equn:CEvNS_SM_charges}
Q_V^\text{SM} = \left[g_p^V \mathbb{Z} + g_n^V \mathbb{N}\right] F_W(\mathtt{q}^2) \, .
\end{equation}
Here, $\mathbb{Z}$ and $\mathbb{N}$ denote the number of protons and neutrons in the nucleus, respectively, while the corresponding vector couplings for protons $(g_p^V)$ and neutrons $(g_n^V)$ are given by~\cite{Papoulias:2015vxa}
\begin{equation}
\begin{aligned}
g_{p}^V &= 2\left(g_{u_{_\mathtt{L}}}^z + g_{u_{_\mathtt{R}}}^z\right) + \left(g_{d_{_\mathtt{L}}}^z + g_{d_{_\mathtt{R}}}^z\right) = \frac{1}{2} - 2\sin^2{\theta_W} \, , \\
g_n^V &= \left(g_{u_{_\mathtt{L}}}^z + g_{u_{_\mathtt{R}}}^z\right) + 2\left(g_{d_{_\mathtt{L}}}^z + g_{d_{_\mathtt{R}}}^z\right) = -\frac{1}{2} \, .
\end{aligned}
\label{equn:CEvNS_SM_nuclear_couplings}
\end{equation}
It is important to note that the axial vector contribution to the \cevns cross section can be safely neglected since it is suppressed by the nuclear spin sum compared to the vector contribution~\cite{Barranco:2005yy}.

Equation~\eqref{equn:CEvNS_SM_xsec} is valid for sufficiently low momentum transfer to satisfy the coherency condition $\mathtt{q} \leq 1/R$~\cite{Papoulias:2018uzy}, with $R$ being the nuclear rms radius and $\mathtt{q}^2 = 2m_N T_{nr}$ denoting the magnitude of the three-momentum transfer. Moreover, to account for the finite nuclear spatial distribution, nuclear physics corrections are incorporated in Eq.~\eqref{equn:CEvNS_SM_charges} through the weak nuclear form factor $F_W(\mathtt{q}^2)$ for which the Klein-Nystrand parametrization is adopted~\cite{Klein:1999qj}
\begin{equation}
\label{equn:kl_ff}
F_W(\mathtt{q}^2)=\frac{3j_1(\mathtt{q}R)}{\mathtt{q} R} \,  \left(\frac{1}{1+\mathtt{q}^2a_k^2}\right) \, .
\end{equation}
Here, $j_1(x)$ stands for the first-order spherical Bessel function, with $a_k=0.7$ fm being the range of the Yukawa potential, while the nuclear rms radius is taken to be $R=1.23\, \mathbb{A}^{1/3}$~fm with $\mathbb{A}=\mathbb{N}+\mathbb{Z}$ being the atomic mass number.

\subsubsection{\cevns within BSM}
BSM contributions to  \cevns mediated by a $Z'$ boson can arise from the following interaction Lagrangian
\begin{equation}
        \mathscr{L}_\mathrm{BSM}^{\nu N}\subset -\Big[g_{\nu_{\alpha}}^{z'}\bar{\nu}_\alpha\gamma^\rho\mathtt{L}\nu_\alpha+\bar{q}\gamma^\rho\left(g_{q_{_{\mathtt{L}}}}^{z'}\mathtt{L}+g_{q_{_{\mathtt{R}}}}^{z'}\mathtt{R}\right)q\Big]Z'_{\rho}\,,
\end{equation}
where the  quark level BSM couplings  with the relevant fermions $\psi \equiv \{\nu, u, d\}$, are adopted from Eq.~\eqref{Eq.BSM_Charge}. In this study we use several benchmark $U(1)_{X}$ charges corresponding to relevant fields, as detailed in Table~\ref{Tab:BP_Charges}. The differential cross section taking into account SM and BSM contributions can be expressed as
\begin{equation}
\left[\frac{d\sigma}{dT_{nr}}\right]_\mathrm{BSM}^{\nu N} =  \left(1+\frac{Q_V^\mathrm{BSM}~ g_{\nu_{{\alpha}}}^{z'}}{2\sqrt{2}G_{F}Q_V^\mathrm{SM}(2m_{N}T_{nr}+M_{Z'}^{2})}\right)^2\left[\frac{d\sigma}{dT_{nr}}\right]_\mathrm{SM}^{\nu N}\,.
\end{equation} 
Similarly to the \eves case, the chirality indices for $\nu_\alpha$ are omitted again, while the corresponding BSM weak charge $Q_V^\mathrm{BSM}$  reads
\begin{equation}
Q_V^\mathrm{BSM} = \left[\mathbb{Z}\left\{2\left(g_{u_{_{\mathtt{L}}}}^{z'}+g_{u_{_{\mathtt{R}}}}^{z'}\right)+\left(g_{d_{_{\mathtt{L}}}}^{z'}+g_{d_{_{\mathtt{R}}}}^{z'}\right)\right\}+\mathbb{N}\left\{\left(g_{u_{_{\mathtt{L}}}}^{z'}+g_{u_{_{\mathtt{R}}}}^{z'}\right)+2\left(g_{d_{_{\mathtt{L}}}}^{z'}+g_{d_{_{\mathtt{R}}}}^{z'}\right)\right\}\right]F_W(\mathtt{q}^2) \, .
\end{equation}

\section{\label{Sec:Data_Analysis}Data Analysis}
In this section, we provide the necessary details regarding the detector simulation procedure and data analysis treatment of the recorded \cevns and \eves signals in current experiments such as COHERENT, DM detectors (XENONnT, LZ, PandaX-4T, and DARWIN), and TEXONO. Additionally, we discuss the methodology of the statistical analysis framework followed for the different experiments analyzed.

\subsection{\label{subsec:COHERENT}COHERENT}
 We consider the full data reported by the COHERENT collaboration from the CsI~\cite{COHERENT:2021xmm} and LAr~\cite{COHERENT:2020iec} measurements~\footnote{We do not consider the recent measurement on Ge due to the lack of sufficient statistics~\cite{Adamski:2024yqt}.}. For the two experiments, a detailed analysis utilizing  both energy and time available is conducted by following the methodology outlined in Ref.~\cite{DeRomeri:2022twg}.

First, we describe the procedure followed to estimate the expected number of \cevns events at the CsI detector, which has a fiducial mass $m_{\mathrm{det}} = 14.6~\mathrm{kg}$ and a  distance $L = 19.3~\mathrm{m}$ from the Spallation Neutron Source (SNS) facility at Oak Ridge National Laboratory (ORNL). The expected \cevns events, in each nuclear recoil energy bin, $i$,  for the different interaction channels $\kappa \equiv\{\mathrm{SM},\mathrm{~BSM}\}$ is evaluated by
\begin{equation}
\label{eq:Nevents_CEvNS}
\begin{split}
R^{i,\alpha}_{\mathrm{CE}\nu\mathrm{NS}} &= N_T \int_{T_{nr}^i}^{T_{nr}^{i+1}} \hspace{-0.3cm} dT_{nr}^{\mathrm{reco}}\, \epsilon_E(T_{nr}^{\mathrm{reco}}) \int_0^{T_{nr}^{\mathrm{max}}} dT_{nr} \, \mathcal{P}(T_{nr}^{\mathrm{reco}}, T_{nr}) \\
& \times \int_{E_\nu^{\mathrm{min}}(T_{nr})}^{E_\nu^{\mathrm{max}}} dE_\nu \, \frac{d \Phi_{\nu_\alpha}(E_\nu)}{d E_\nu} \left[\frac{d\sigma}{dT_{nr}}\right]_\kappa^{\nu N}\, ,
\end{split}
\end{equation}
where $N_T = m_\mathrm{det} N_A / M$ represents the number of target nuclei (Cs or I) in the detector with $N_A$ being the Avogadro number, and $M$  the molar mass of the target. Further, $T_{nr}$ and $T_{nr}^\mathrm{reco}$ represent true and measured reconstructed nuclear recoil energy, respectively. The minimum neutrino energy necessary to induce a nuclear recoil of energy $T_{nr}$ can be calculated from the kinematics of the process and to an excellent approximation takes the form $E_\nu^\mathrm{min}=\sqrt{m_NT_{nr}/2}$, while the maximum recoil energy is obtained by inverting the last expression and reads $T_{nr}^\text{max}= 2(E_\nu^\text{max})^2 / ( m_N)$. The neutrino fluxes from the SNS consist of a prompt ($\nu_\mu$ beam from $\pi^+$ decay at rest) followed by a delayed ($\bar{\nu}_\mu$ and $\nu_e$ beam from $\mu^+$ decay at rest)  neutrino beam. The differential neutrino energy spectra can be estimated by employing the Michel spectra~\cite{Michel:1949qe, Bouchiat:1957zz}
\begin{equation}
\begin{aligned} 
\frac{d \Phi_{\nu_\mu}(E_\nu)}{d E_\nu} & = \eta \, \delta\left(E_\nu-\frac{m_{\pi}^{2}-m_{\mu}^{2}}{2 m_{\pi}}\right) \quad &(\text{prompt})\, , \\ 
\frac{d \Phi_{\bar{\nu}_\mu}(E_\nu)}{d E_\nu} & = \eta \frac{64 E^{2}_\nu}{m_{\mu}^{3}}\left(\frac{3}{4}-\frac{E_\nu}{m_{\mu}}\right) \quad &(\text{delayed})\, ,\\ 
\frac{d \Phi_{\nu_e}(E_\nu)}{d E_\nu} & = \eta \frac{192 E^{2}_\nu}{m_{\mu}^{3}}\left(\frac{1}{2}-\frac{E_\nu}{m_{\mu}}\right) \quad &(\text{delayed}) \, ,
\end{aligned}
\label{Eq.:SNS_Flux}
\end{equation}
where the flux normalization is given by $\eta=rN_\mathrm{POT}/4\pi L^2$, with  $r = 0.0848$ being the neutrino yield per flavor produced for each proton on target (POT), while $N_\mathrm{POT}=3.198\times 10^{23}$ is the number of POT accumulated during the running period of the experiment. The energy-dependent detector efficiency is modeled by
\begin{equation}
\label{eq:CsI_E_efficiency}
\epsilon_E(x) = \frac{a}{1+e^{-b(x-c)}}+d \, , 
\end{equation}
with $x=\mathrm{PE}(T_{nr}^\mathrm{reco})+\alpha_7$ and $a = 1.32045$, $b = 0.285979$, $c = 10.8646$, $d = -0.333322$~\cite{COHERENT:2021xmm}. At this point, it should be stressed that the nuisance parameter \(\alpha_7\) has been introduced in order to incorporate the 1\(\sigma\) uncertainty on \(\epsilon_E\). In the performed $\chi^2$ analysis (see below), the latter parameter is allowed to vary freely within the range \([-1, +1] \times \text{PE}\) with no prior, for details see Ref.~\cite{DeRomeri:2022twg}. The number of photoelectrons is given by $\mathrm{PE}(T_{nr}^\mathrm{reco})=\mathrm{LY}\times T_{er}^\mathrm{reco}(T_{nr}^\mathrm{reco})$, with the light yield $\mathrm{LY}=13.35 \mathrm{~PE}/\mathrm{keV}_{ee}$~\cite{COHERENT:2021xmm}, while the electron-equivalent energy is considered as $T_{er}^\mathrm{(reco)}\left(T_{nr}^\mathrm{(reco)}\right)=x_1T_{nr}^\mathrm{(reco)}+x_2\left( T_{nr}^\mathrm{(reco)}\right)^2+x_3\left(T_{nr}^\mathrm{(reco)}\right)^3+x_4\left(T_{nr}^\mathrm{(reco)}\right)^4$. The values of these parameters are taken from the supplementary document of Ref.~\cite{COHERENT:2021xmm}, $x_1=0.0554628$, $x_2=4.30681~\mathrm{MeV}^{-1}$, $x_3=-111.707\mathrm{~MeV}^{-2}$, $x_4=840.384\mathrm{~MeV}^{-3}$.  Further, the true nuclear recoil spectra are smeared using the resolution function
\begin{equation}
\mathcal{P}(T_{nr}^\mathrm{reco},T_{nr}) =  \frac{(\mathfrak{a} (1 + \mathfrak{b}))^{1 + \mathfrak{b}}}{\Gamma(1 + \mathfrak{b})} \cdot x^\mathfrak{b} \cdot e^{-\mathfrak{a}(1 +\mathfrak{b})x}\, ,
\end{equation}
where $x=\mathrm{PE}(T_{nr}^\mathrm{reco})$, while $\mathfrak{a}=0.0749/T_{er}(T_{nr})$, and $\mathfrak{b} = 9.56 \times T_{er}(T_{nr})$~\cite{COHERENT:2021xmm}. Subsequently, the nuclear-recoil spectrum from Eq.~\eqref{eq:Nevents_CEvNS} is converted to an electron-equivalent spectrum using the quenching factor (QF) measurements in~\cite{COHERENT:2021pcd}. Finally, for direct comparison with experimental data, we further convert the electron-recoil spectrum into a photoelectron (PE) spectrum based on the light yield information from Ref.~\cite{COHERENT:2021xmm}.

Since the CsI detector in COHERENT cannot distinguish between nuclear and electron recoils, our calculations also consider the expected number of \eves~events
\begin{equation}
\label{eq:Nevents_EvES}
\begin{split}
R^{i,\alpha}_{\mathrm{E}\nu\mathrm{ES}} &= N_T \int_{T_{er}^i}^{T_{er}^{i+1}} \hspace{-0.3cm} dT_{er}^{\mathrm{reco}}\, \epsilon_E(T_{er}^{\mathrm{reco}}) \int_0^{T_{er}^\text{max}} dT_{er} \, \mathcal{P}(T_{er}^{\mathrm{reco}}, T_{er}) \\
& \times \mathbb{Z}_\mathrm{eff}(T_{er})\int_{E_\nu^{\mathrm{min}}(T_{er})}^{E_\nu^{\mathrm{max}}} dE_\nu \, \frac{d \Phi_{\nu_\alpha}(E_\nu)}{d E_\nu} \left[\frac{d\sigma_{\nu_\alpha}}{dT_{er}}\right]_\kappa^{\nu e}\, ,
\end{split}
\end{equation}
$E_\nu^\mathrm{min}=1/2\left[T_{er}+\sqrt{T_{er}^2+2m_eT_{er}}\right]$ is the minimum neutrino energy essential to generate an electron recoil with energy $T_{er}$, while the maximum true recoil energy reads $T_{er}^\text{max} = 2 (E_\nu^\text{max})^2/(2 E_\nu^\text{max}+ m_e)$. The quantity $\mathbb{Z}_\mathrm{eff}(T_{er})$ takes into account the atomic binding effect and represents the effective number of electrons that can be ionized for an energy deposition $T_{er}$.  For the Cs and I atoms, we consider the distribution of $\mathbb{Z}_\mathrm{eff}(T_{er})$ from Ref.~\cite{Thompson2009}. The computed \eves~spectrum is eventually converted into a PE-spectrum in a similar manner to  the \cevns case.

Due to the pulsed nature of the SNS neutrino spectra, timing information is also incorporated  by distributing the predicted number of both \cevns and \eves energy spectra $R^i$ into each time bin $j$. The time distributions $\mathbb{f}^{\nu_\alpha}_T(t_{\mathrm{rec}})$ for $\nu_\alpha = \nu_\mu,\, \overline{\nu}_\mu,\, \nu_e$ are given by~\cite{Picciau:2022xzi,COHERENT:2021xmm}
\begin{subequations}
\begin{align}
&\mathbb{f}^{\nu_\mu}_T(t_{\mathrm{rec}}) = \frac{1}{\sqrt{2\pi} b_{POT} \tau_{\pi^+}} \int_0^{t_{\mathrm{rec}}} \exp \left( -\frac{(t' - a_{POT})^2}{2b_{POT}^2} \right) \exp \left( -\frac{(t_{\mathrm{rec}} - t')}{T_{\pi^+}} \right) dt'\,, \\
&\mathbb{f}^{\nu_e, \overline{\nu}_\mu}_T(t_{\mathrm{rec}}) = \frac{1}{\tau_{\mu^+}} \int_0^{t_{\mathrm{rec}}} \mathbb{f}^{\nu_\mu}_T(t') \exp \left( -\frac{(t_{\mathrm{rec}} - t')}{T_\mu} \right) dt'\,,
\end{align}
\end{subequations}
where $a_\mathrm{POT} = 0.44~\mu s$ and $b_\mathrm{POT} = 0.15~\mu s$ are the protons-on-target trace parameters from the SNS, and  $\tau_{\pi^+}$, $\tau_{\mu^+}$ are the pion and muon lifetime, respectively.  The expected number of events in each PE bin $i$ and time bin $j$ is finally given by
\begin{equation}
\label{eq:R_ij}
 R^{ij}= \sum_{\alpha=\mathrm{SNS}}\, \int_{t_\mathrm{rec}^j}^{t_\mathrm{rec}^{j+1}}dt_\mathrm{rec}\mathbb{f}^{\nu_\alpha}_T(t_\mathrm{rec}, \alpha_6)\epsilon_T(t_\mathrm{rec})R^{i,\alpha}\,,
\end{equation}
where $t_\mathrm{rec}$ is the reconstructed time and $\epsilon_T(t_\mathrm{rec})$ is the time-dependent efficiency taken from the supplementary document of Ref.~\cite{COHERENT:2021xmm}. An additional nuisance parameter $\alpha_6$ is included to account for the uncertainty in the beam arrival time, allowing   the timing distribution to vary freely in the range of $\pm 250~\mu s$. In the final step, a weighted sum for the Cs and I nuclei is performed to obtain the total expected events, i.e. $ R^{ij}_{\mathrm{CE\nu NS}}$ and $R^{ij}_{\mathrm{E \nu ES}}$ for \cevns and \eves, respectively, considering the 9 PE and 11 time bins as reported in Ref.~\cite{COHERENT:2021xmm}.

For the statistical analysis of the COHERENT-CsI-2021 data, we use the following Poissonian least-squares function \cite{DeRomeri:2022twg}
\begin{equation}\label{eq:chi2CsI}
	\chi^2_{\mathrm{CsI}}\Big|_{\mathrm{CE\nu NS} + \mathrm{E \nu ES}} =2\sum_{i=1}^{9}\sum_{j=1}^{11}\left[ R_\mathrm{th}^{ij}  -  R^{ij}_{\mathrm{exp}} +  R^{ij}_{\mathrm{exp}} \ln\left(\frac{R^{ij}_{\mathrm{exp}}}{ R_\mathrm{th}^{ij}} \right)\right]\\+ \sum_{k=0}^{5}\left(\dfrac{ \alpha_{k} }{ \sigma_{k} }\right)^2  \,,
\end{equation}
where $R_{ij}^{\mathrm{exp}}$ is the binned coincidence data from \cite{COHERENT:2021xmm} and $R^\mathrm{th}_{ij}$ is the predicted number of events, given by
\begin{equation}
\begin{aligned}
R_\mathrm{th}^{ij} &= (1 + \alpha_{0} + \alpha_{5})  R^{ij}_{\mathrm{CE\nu NS}} (\alpha_{4}, \alpha_{6}, \alpha_{7})  + (1 + \alpha_{0} )  R^{ij}_{\mathrm{E \nu E S}} (\alpha_{6}, \alpha_{7})\\[4pt]
 &+ (1 + \alpha_{1}) R^{ij}_\mathrm{BRN}(\alpha_{6}) + (1 + \alpha_{2}) R^{ij}_\mathrm{NIN}(\alpha_{6}) + (1 + \alpha_{3}) R^{ij}_\mathrm{SSB}  \,.
	\label{eq:Nth_CsI_chi2}
 \end{aligned}
\end{equation}
Here, $R^{ij}_\mathrm{BRN}$, $R^{ij}_\mathrm{NIN}$, and $R^{ij}_\mathrm{SSB}$ represent the background events from beam-related neutrons (BRN), neutrino-induced neutrons (NIN), and steady-state backgrounds (SSB), respectively, as reported in Ref.~\cite{COHERENT:2021xmm}. Equations~\eqref{eq:chi2CsI} and \eqref{eq:Nth_CsI_chi2} involve several nuisance parameters ($\alpha_i$) and their corresponding uncertainties ($\sigma_i$). Specifically, $\sigma_{0} = 11$\% accounts for the SNS flux normalization uncertainty,  $\sigma_{1} = 25\%$, $\sigma_{2} = 35\%$, and $\sigma_{3} = 2.1\%$ correspond to the uncertainties assigned for BRN, NIN, and SSB backgrounds, respectively, while $\sigma_{5} = 3.8\%$ is related to the QF. The number of events $R^{ij}_{\mathrm{CE\nu NS}}$ and $R^{ij}_{\mathrm{E}\nu\mathrm{ES}}$ also incorporate: $\alpha_{4}$, affecting only the CE$\nu$NS events due to the uncertainty arising from nuclear physics~\footnote{This is incorporated by adjusting the nuclear radius in Eq.~\eqref{equn:kl_ff} with $R = 1.23 \, \mathbb{A}^{1/3} (1+\alpha_4)$.}, with $\sigma_{4} = 5\%$, $\alpha_{6}$ which accounts for the beam timing uncertainty with no prior assigned as discussed above, and $\alpha_{7}$ which allows for $[-1,1]~\text{PE}$ deviations in the energy efficiency. Further details on the statistical analysis can be found in Ref.~\cite{DeRomeri:2022twg}.

For the analysis of COHERENT-LAr data, the \cevns event rate can be evaluated using Eq.~(\ref{eq:Nevents_CEvNS}), with $r=0.09$,  $N_\mathrm{POT}=1.38 \times10^{23}$,   $m_\text{det}=24$~kg and  $L=27.5$~m. 
The  energy efficiency function is provided in the data release~\cite{COHERENT:2020ybo}, while the reported quenching factor for the LAr target is
$\mathrm{QF}(T_{nr}) = 0.246 + 7.8 \times 10^{-4} T_{nr} (\mathrm{keV_{nr}})$~\cite{COHERENT:2020iec}.
The energy resolution effect is taken into account by smearing the true energy spectrum  with a normalized Gaussian function $\mathcal{P}(T_{er}^\mathrm{reco},T_{er})$ with resolution power given by 
$\frac{\sigma_{T_{er}^\text{reco}}}{ T_{er}^\text{reco}}= \frac{0.58}{\sqrt{T_{er}^\text{reco}(\mathrm{keV_{ee}})}}$~\cite{COHERENT:2020ybo}.
Notably,  thanks to the $F_{90}$ measurement performed by the COHERENT Collaboration, the LAr detector is capable of rejecting electron recoils and hence in the present analysis possible contributions from E$\nu$ES are ignored (see also Ref.~\cite{DeRomeri:2022twg}). Furthermore, unlike the COHERENT-CsI detector, for the case of LAr the time efficiency is not taken into account, i.e., $\epsilon_T=1$, while the uncertainty of the beam arrival time is also neglected. The neutrino time distributions considered are the same as in the CsI case, but the accepted spectra account for trigger times with $t_\text{trig} \leq 5~\mathrm{\mu s}$.
Following the COHERENT Collaboration, we consider 12 (index $i$)  bins of reconstructed recoil energy and 10 time-bins (index $j$)~\cite{COHERENT:2020iec}.

The statistical analysis performed for the LAr detector relies on the $\chi^2$ function~\cite{AtzoriCorona:2022moj}
\begin{equation}
    \begin{aligned}
	\left.\chi^2_{\mathrm{LAr}}\right|_{\mathrm{CE\nu NS}} = 
	\sum_{i=1}^{12}
	\sum_{j=1}^{10}
	\frac{1}{\sigma_{ij}^2}&\Bigl[ (1 + \beta_0 +  \beta_1 \Delta_\mathrm{CE\nu NS}^{F_{90+}} + \beta_1 \Delta_\mathrm{CE\nu NS}^{F_{90-}} + \beta_2 \Delta_\mathrm{CE\nu NS}^\mathrm{t_{trig}}) R^{ij}_{\mathrm{CE\nu NS}}  \\
    & + (1 + \beta_3) R^{ij}_\mathrm{SSB}  \\
    & + (1 + \beta_4 + \beta_5 \Delta_\mathrm{pBRN}^{E_+} + \beta_5 \Delta_\mathrm{pBRN}^{E_-}
    + \beta_6 \Delta_\mathrm{pBRN}^{t_\text{trig}^+} + \beta_6 \Delta_\mathrm{pBRN}^{t_\text{trig}^-} + \beta_7 \Delta_\mathrm{pBRN}^{t_\text{trig}^\text{w}}) R^{ij}_\mathrm{pBRN}\\
    & + (1 + \beta_8) R^{ij}_\mathrm{dBRN}  -  R^{ij}_{\text{exp}} \Bigr]^2  \\ 
     &+ \sum_{k=0,3,4,8}
	\left( 	\dfrac{ \beta_{k} }{ \sigma_{k} } 	\right)^2  +  \sum_{k=1,2,5,6,7} \left(\beta_{k}  	\right)^2 \,  ,
\end{aligned}
\label{chi2LAr}
\end{equation}
where $\sigma_{ij}^2 = R^{ij}_\text{exp} +  R^{ij}_\mathrm{SSB}/5$.  
The introduced nuisance parameters $\beta_0,~\beta_3,~\beta_4$ and $\beta_8$ take into account the normalization uncertainties of CE$\nu$NS, SSB, prompt BRN and delayed BRN,  with $\sigma_{0} = 13\%$, $\sigma_{3}=0.79\%$, $\sigma_{4}=32\%$ and $\sigma_{8} = 100\%$~\cite{COHERENT:2020iec}. 
Shape uncertainties are accounted for through   the nuisance parameters $\beta_1, \beta_2, \beta_5, \beta_6$ and $\beta_7$.
As explained in Ref.~\cite{DeRomeri:2022twg} and the data release~\cite{COHERENT:2020ybo}, the parameters $\beta_1$ and $\beta_2$  control the shape of the predicted CE$\nu$NS rates, with the relevant sources of systematics being  the $\pm 1\sigma$ energy distributions of the $F_{90}$ parameter, given by $\Delta_\mathrm{CE\nu NS}^{F_{90+}}$ and $\Delta_\mathrm{CE\nu NS}^{F_{90-}}$, and
 the mean time to trigger distribution, $\Delta_\mathrm{CE\nu NS}^\mathrm{t_{trig}}$. On the other hand, the nuisances  $\beta_5$, $\beta_6$ and $\beta_7$ are introduced for adjusting the shape of the prompt BRN background, with the relevant uncertainties being the  $\pm 1\sigma$ energy distributions ($\Delta_\mathrm{pBRN}^{E_+}$ and  $\Delta_\mathrm{pBRN}^{E_-}$), the $\pm 1 \sigma$ mean time to trigger  distributions ($\Delta_\mathrm{pBRN}^{t_\text{trig}^+}$ and $\Delta_\mathrm{pBRN}^{t_\text{trig}^-}$) and the 
trigger width distribution ($\Delta_\mathrm{pBRN}^{t_\text{trig}^\text{w}}$). 
The aforementioned $\Delta$-distributions are expressed as~\cite{AtzoriCorona:2022moj}
\begin{equation}
    \Delta_{\lambda}^{\xi_\lambda} = \frac{R^{ij}_{\lambda,\xi_\lambda} - R^{ij}_{\lambda,\mathrm{CV}}}{R^{ij}_{\lambda,\mathrm{CV}}} \, ,
\end{equation}
where $\lambda=$ \{CE$\nu$NS or  prompt BRN\},  while $\xi_\lambda$ enumerates the different sources relevant to each   $\lambda$. Finally, the central values (CV) of the \cevns ~or
prompt BRN rates are taken from ~\cite{COHERENT:2020ybo}. 

In the upper panel of Fig.~\ref{Fig:COHERENT_CsI_Events}, we present the fitted signals and experimental data, residual over SSB at the COHERENT CsI experiment. The upper-left panel shows the projection of fitted events onto the PE axis, while the upper-right panel displays the projection onto the $t_\mathrm{rec}$ axis. In the lower panel of Fig.~\ref{Fig:COHERENT_CsI_Events} we show  the corresponding  signals for the COHERENT-LAr detector, again residual over SSB. The lower-left panel shows the fitted spectra projected onto the reconstructed electron recoil energy axis, while the lower-right panel shows the projection onto the trigger time axis. The colored histograms represent the fitted signals for SM and different $U(1)_X$ models, calculated for a benchmark mediator mass of $M_{Z'} = 100$ MeV and several coupling constants $\textsl{g}_{_x}$. The black points with error bars indicate the experimental data residuals over the SSB. The displayed signal events incorporate fitted BRN (for both the CsI and LAr detectors) as well as  NIN (only for the CsI detector) background components.

\begin{figure}[t]
   \centering
   \captionsetup{justification=raggedright}
  \includegraphics[width=0.49\textwidth]{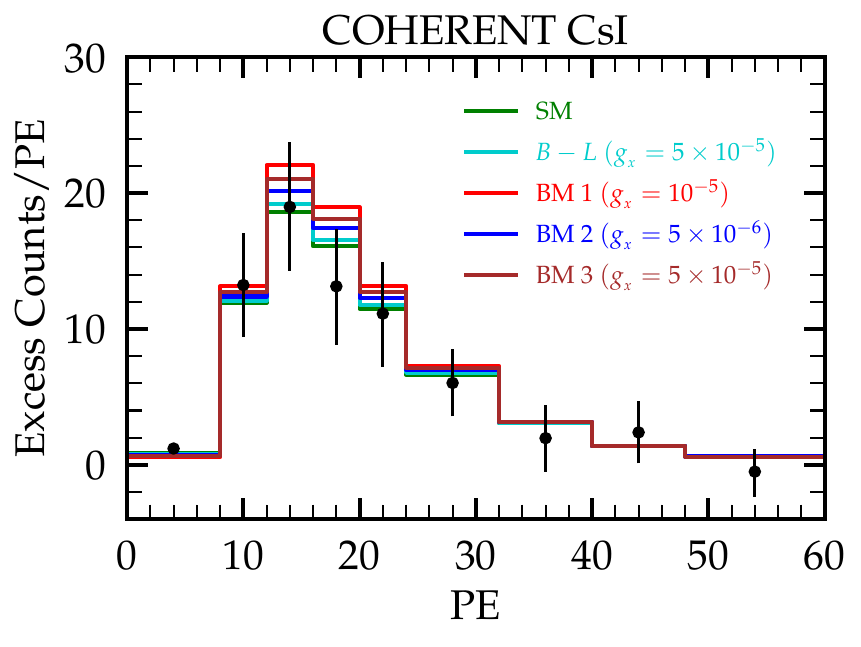}
  \includegraphics[width=0.49\textwidth]{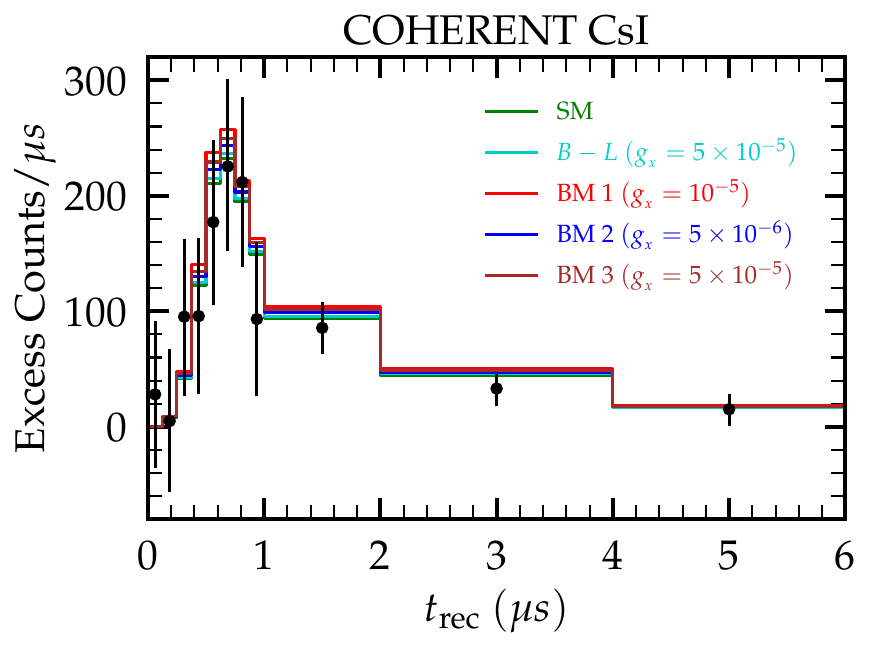}
   \includegraphics[width=0.49\textwidth]{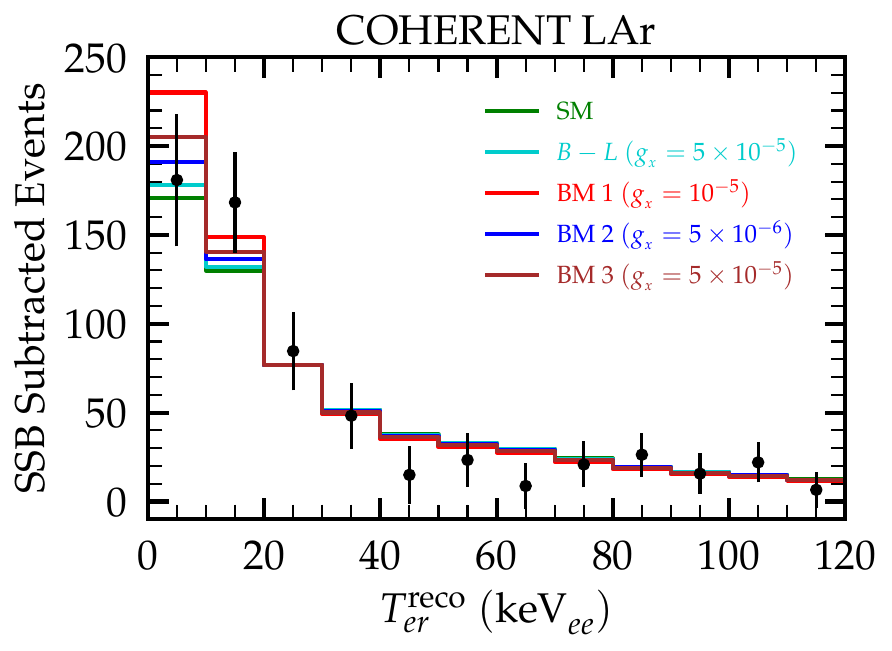}
   \includegraphics[width=0.49\textwidth]{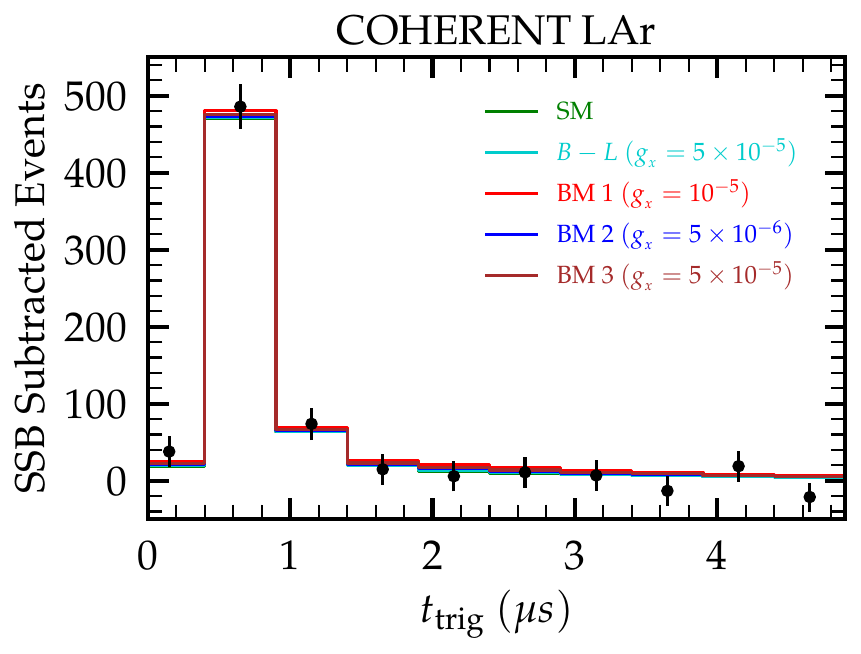}
  \caption{Fitted signals (colored histograms) and experimental data residual over SSB (black points with error bars) at  COHERENT experiment. Upper panel: fitted events projected onto PE (upper left) and $t_\mathrm{rec}$ (upper right) axes, at the COHERENT-CsI detector.  Lower panel: fitted events projected onto reconstructed electron recoil energy axis (lower left) and trigger time axis $t_\mathrm{trig}$ (lower right), for the case of  COHERENT-LAr detector. The fitted BSM events are evaluated for a benchmark mediator mass $M_{Z'} = 100$ MeV and various couplings $\textsl{g}_{_x}$, corresponding to different $U(1)_X$ models, as indicated in the legend. The displayed signal events also include fitted BRN and NIN background components (for details see the text).}\label{Fig:COHERENT_CsI_Events}
\end{figure}

\subsection{\label{subsec:Direct_Detection}DM direct detection experiments: XENONnT, LZ, PandaX-4T, and DARWIN}

DM direct detection experiments are sensitive to electron recoils induced by solar neutrinos. The electron recoil spectra arising from \eves in each electron recoil bin $i$, for the interaction channel $\kappa \equiv\{\mathrm{SM},\mathrm{~BSM}\}$ can be simulated as
\begin{equation}
\label{Eq:Events_Rate}
\begin{aligned}
R^i_{\mathrm{E}\nu\mathrm{ES}} = t_\mathrm{run} N_T &\int_{T_{er}^i}^{T_{er}^{i+1}}\hspace{-0.3cm} ~ dT_{er}^\mathrm{reco}\mathcal{A}(T_{er}^\mathrm{reco})\sum_{k=\mathrm{Solar}}\int_0^{T_{er}^{\mathrm{max,}k}} \hspace{-0.3cm}dT_{er}\mathrm{~}\mathcal{G}(T_{er}^\mathrm{reco},T_{er})\\
&\times \mathbb{Z}_\text{eff}(T_{er}) \int_{E_\nu^\mathrm{min}}^{E_{\nu,k}^\mathrm{max}}dE_\nu \frac{d\Phi_{\nu_{_{e}}}^k(E_\nu)}{dE_\nu} {\left[\frac{d\sigma}{dT_{er}}\right]}_\kappa^{\nu e}\,,
\end{aligned}
\end{equation}
where $T_{er}$ and $T_{er}^\mathrm{reco}$ represent the true and reconstructed electron recoil energies, respectively, while $t_\mathrm{run}$ is the experiment's running time. The function accounting for the binding effect is approximated through a series of step functions, introduced in Ref.~\cite{Chen:2016eab}, and defined as $\mathbb{Z}_\text{eff}(T_{er}) = \sum_{i=1}^{54}\Theta(T_{er}-\mathscr{B}_i)$. Here, $\Theta(x)$ is a Heaviside step function, and $\mathscr{B}_i$ denotes the single-particle atomic level binding energy of the $i$th electron in the Xenon atom, evaluated from Hartree-Fock calculations~\cite{Chen:2016eab}.

DM detectors are exposed to various astrophysical neutrinos. In the region of interest (ROI) of present and  next-generation DM detection experiments, the detectable \eves populations are predominantly triggered by the $\nu_e$ fluxes that emerge from various thermonuclear reactions occurring within the Sun's core, which in Eq.~(\ref{Eq:Events_Rate}) are denoted by $k\equiv\{\mathrm{pp},\mathrm{~} ^7\mathrm{Be},\mathrm{~} \mathrm{pep},\mathrm{~} ^{13}\mathrm{N},\mathrm{~} ^{15}\mathrm{O},\mathrm{~} ^{17}\mathrm{F},\mathrm{~} ^8\mathrm{B},\mathrm{~} \mathrm{hep}\}$ to  signify the reaction responsible for their production. We utilize the spectra from  Ref.~\cite{BahcallSNData, Bahcall:1987jc, Bahcall:1994cf, Bahcall:1995bt, Bahcall:1996qv, Haxton:2012wfz} with the normalizations and uncertainties corresponding to the high metallicity model from Ref.~\cite{Baxter:2021pqo}. However, within the ROI, current DM direct detection experiments such as XENONnT, LZ, and PandaX-4T are primarily sensitive to pp neutrinos, with a subdominant contribution from $^7\mathrm{Be}_{0.861}$ neutrinos, while the rest of the solar neutrino spectrum contributes negligibly. Therefore, for these experiments, we simulate the \eves signal considering only these two relevant flux components ($\mathrm{pp},\mathrm{~} ^7\mathrm{Be}_{0.861}$), while for the next-generation experiment DARWIN, we consider all components of the solar neutrino flux as done in Ref.~\cite{DeRomeri:2024dbv}. 
In our calculations, the maximum neutrino energy, $E_{\nu,k}^\mathrm{max}$, corresponds to the energy endpoint of the neutrino production process within the Sun, denoted by the index $k$. 
To account for the effect of neutrino oscillations in propagation, we have weighted the corresponding \eves cross section with the respective neutrino oscillation probability
\begin{equation}
{\left[\frac{d\sigma}{dT_{er}}\right]}_\kappa^{\nu e}=P_{ee}(E_\nu){\left[\frac{d\sigma_{\nu_e}}{dT_{er}}\right]}_\kappa^{\nu e}+P_{ef}(E_\nu){\left[\frac{d\sigma_{\nu_f}}{dT_{er}}\right]}_\kappa^{\nu e}\,,
\end{equation} 
here, $f\equiv\{\mu, \tau\}$. The survival probability of electron type neutrinos, $P_{ee}(E_\nu)$, is evaluated under the two-flavor approximation~\cite{Escrihuela:2009up}, and the oscillation probability $P_{ef}(E_\nu)=1-P_{ee}(E_\nu)$.

For accurately simulating the signal, we have also taken into account  detector-specific quantities such as, the efficiency, $\mathcal{A}(T_{er}^\mathrm{reco})$, and resolution, $\mathcal{G}(T_{er}^\mathrm{reco},T_{er})$, functions into Eq.~\eqref{Eq:Events_Rate}, following the recommendations of the various experimental Collaborations. For XENONnT, LZ, and PandaX-4T, the detection efficiencies have been adopted from Refs.~\cite{XENON:2022ltv}, \cite{LZ:2023poo}, and \cite{PandaX:2024cic}, respectively. Notice that, unlike XENONnT and PandaX-4T, the LZ collaboration has provided the detection efficiency in terms of the true recoil energy, $T_{er}$. The differential event spectra is further smeared with the Gaussian resolution function
\begin{equation}
\mathcal{G}(T_{er}^\mathrm{reco},T_{er}) = \frac{1}{\sqrt{2\pi}\sigma}\exp{\left[-\frac{(T_{er}-T_{er}^\mathrm{reco})^2}{2\sigma^2}\right]}\,.
\end{equation}
The various Collaborations have provided the resolution power, $\sigma$, in terms of either $T_{er}^\mathrm{reco}$ or $T_{er}$. The detector resolutions for different experiments are given as $\sigma_\text{XENONnT}(T_{er}^\mathrm{reco}) = 0.31 ~\sqrt{T_{er}^\mathrm{reco}} + 0.0037 ~T_{er}^\mathrm{reco}$\,\cite{XENON:2020iwh, XENON:2020rca}, $\sigma_\text{LZ}(T_{er}^\mathrm{reco}) = 0.323 ~\sqrt{T_{er}^\mathrm{reco}}$\,\cite{Pereira:2023rte}, and $\sigma_\text{PandaX-4T}(T_{er}) = 0.073 + 0.173 ~T_{er} - 6.5\times 10^{-3} ~T_{er}^2 + 1.1\times 10^{-4} ~T_{er}^3$\,\cite{PandaX:2022ood}. It should be noted that all values of \( T_{er} \) and \( T_{er}^{\mathrm{reco}} \) are denoted here in units of \( \mathrm{keV}_{ee} \). 

Concerning the forthcoming DARWIN experiment, the calculated spectrum comprises $51$ logarithmically distributed bins in the range  $(1,1500) \, \text{keV}_{ee}$~\cite{DARWIN:2020bnc}. For the evaluation of the expected event rate at DARWIN, a similar resolution function and detector efficiency as for the XENONnT case has been incorporated, while for  ${T_{er}^\text{reco}\Big|}_\text{DARWIN} > {T_{er}^\text{reco}\Big|}_\text{XENONnT,~max}$ a flat efficiency is assumed.

 The simulated signals for the XENONnT, LZ, PandaX-4T (run 1), and DARWIN (with $300 \mathrm{~ton}\times\mathrm{years}$ exposure) experiments as a function of  $T_{er}^\mathrm{reco}$ are presented in Fig.~\ref{Fig:DM_Detector_Events}. The colored histograms indicate the simulated signals for SM and different $U(1)_X$ models, calculated under the assumption of a benchmark mediator mass of $M_{Z'} = 100$ MeV and various coupling constants $\textsl{g}_{_x}$. For XENONnT, LZ, and PandaX-4T, the signal and total background contributions are also shown, all taken from the Collaborations~\footnote{For LZ, the $^{37}\mathrm{Ar}$ background component is separately shown with the orange histogram. This background arises from the cosmogenic activation of xenon prior to its underground deployment. The process results in the production of short-lived $^{37}\mathrm{Ar}$ isotopes, which decay during the first run \cite{LZ:2022lsv}.}. In contrast, the results for DARWIN demonstrate only the predicted signals. Note also that, the depicted total background spectra for XENONnT, LZ, and PandaX-4T experiments are also including the contribution from SM \eves rates.

\begin{figure}[t]
   \centering
   \captionsetup{justification=raggedright}
  \includegraphics[width=0.49\textwidth]{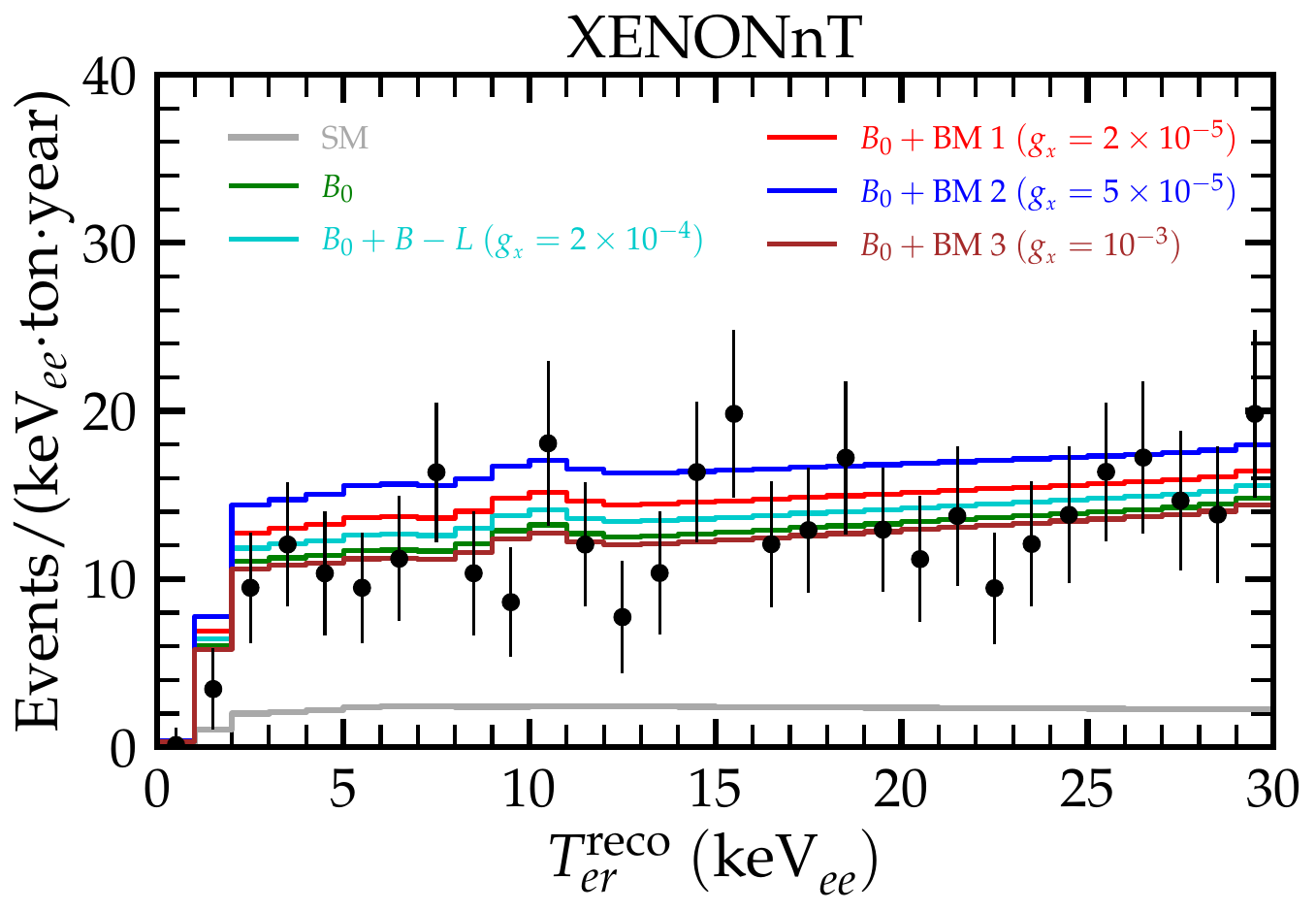}
   \includegraphics[width=0.49\textwidth]{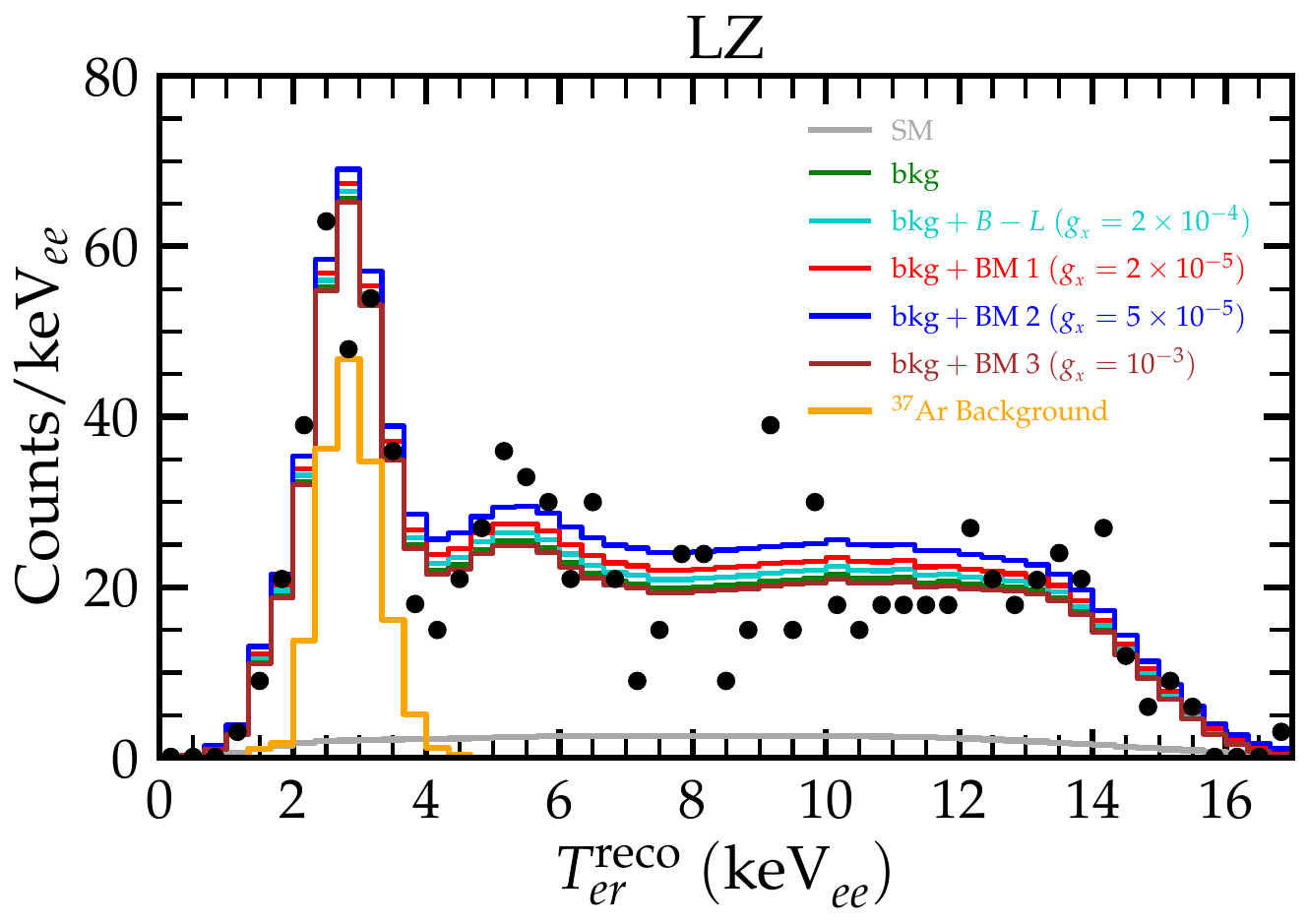}
     \includegraphics[width=0.49\textwidth]{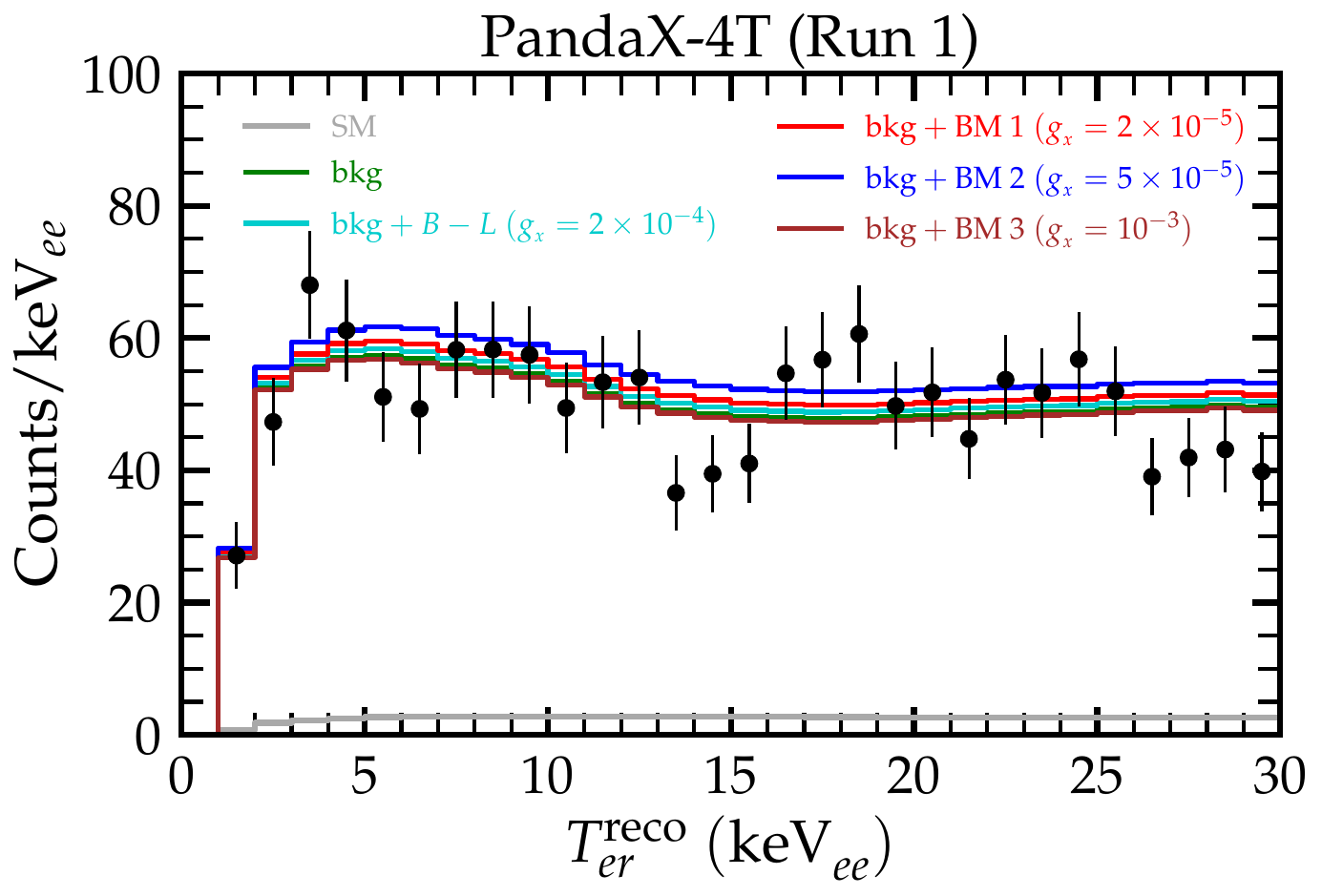}
   \includegraphics[width=0.49\textwidth]{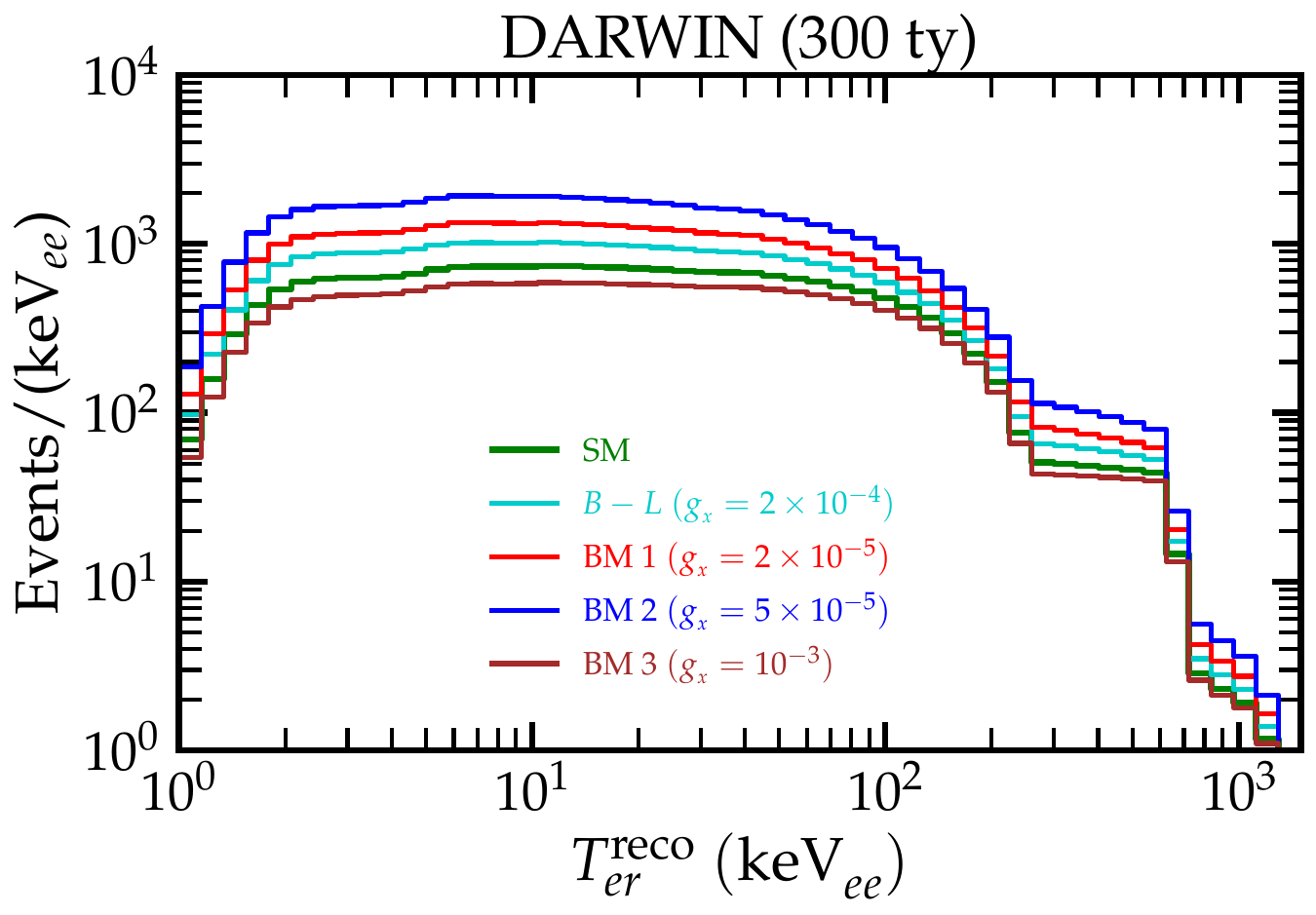}
  \caption{Simulated signals (colored histograms) and experimental data (black points with error bars) as a function of the recoil energy, $T_{er}^\mathrm{reco}$. The panels represent the XENONnT (top left), LZ (top right), PandaX-4T (bottom left), and DARWIN (bottom right) experiments. The BSM events are calculated for a benchmark mediator mass $M_{Z'} = 100$ MeV and various couplings $\textsl{g}_{_x}$, corresponding to different $U(1)_X$ models, as described in the legend. Notably, for XENONnT, LZ, and PandaX-4T, the displayed events include both signal and background contributions as reported by the respective collaborations, whereas the DARWIN events consist only the predicted signal.}\label{Fig:DM_Detector_Events}
\end{figure}

For the spectral analysis of the LZ data, the following Poissonian $\chi^2$ function has been employed
\begin{equation}
\chi^2(\mathcal{S})=2\sum_{k}\sum_{i=1}^{51} \left[R_\mathrm{th}^i(\mathcal{S};\theta_k)- R_\mathrm{exp}^i+  R_\mathrm{exp}^i\ln \left(\frac{R_\mathrm{exp}^i}{R_\mathrm{th}^i(\mathcal{S}; \theta_k)}\right) \right]+\left(\frac{\theta_k}{\sigma_{\theta_k}}\right)^2 \, ,
\label{Eq.:P_chi_2_Func}
\end{equation}
where $R_\mathrm{exp}^i$ denotes the observed events in the $i$th energy bin as reported in~\cite{LZ:2022lsv}, and $R_\mathrm{th}^i(\mathcal{S}; \theta_k)$ represents the predicted events, including signal and background contributions, parametrized by the BSM parameters $\mathcal{S}= \{M_{Z'}, \textsl{g}_{_x}\}$ as well as the  nuisance parameters $\theta_k = \{\alpha, \beta_\mathrm{pp}, \beta_\mathrm{Be}, \delta\}$. The nuisance parameters  account for uncertainties in total background, solar neutrino flux normalizations for pp and $^7\mathrm{Be}_{0.861}$ channels, and $^{37}\text{Ar}$ background component, respectively, with their corresponding uncertainties $\sigma_\alpha=13\%$, $\sigma_{\beta_\mathrm{pp}}=0.6\%$, $\sigma_{\beta_\mathrm{Be}}=3\%$ and $\sigma_\delta=100\%$ respectively~\cite{LZ:2022lsv,AtzoriCorona:2022jeb, DeRomeri:2024dbv, Baxter:2021pqo}. The predicted spectrum is modeled as
\begin{equation}
    R_\mathrm{th}^i(\mathcal{S}; \theta_j)=(1+\alpha)R_\mathrm{bkg}^i+(1+\beta_\mathrm{pp})R_{\mathrm{E}\nu\mathrm{ES}}^{i,\mathrm{pp}}(\mathcal{S})+(1+\beta_\mathrm{Be})R_{\mathrm{E}\nu\mathrm{ES}}^{i,\mathrm{Be}}(\mathcal{S})+(1+\delta)R_{^{37}\text{Ar}}^i\, .
\end{equation}
Notably, the oscillation  uncertainties are tiny compared  to the large uncertainty on the flux normalization, and hence neglected. Following the Ref.~\cite{AtzoriCorona:2022jeb} the background rate $R_\mathrm{bkg}^i$ is derived by subtracting the SM \eves and $^{37}\text{\text{Ar}}$ contributions from the total background reported in~\cite{LZ:2022lsv}.

For XENONnT and PandaX-4T, a Gaussian $\chi^2$ test statistic has been employed
\begin{equation}
\chi^2(\mathcal{S})= \sum_{i=1}^{30} \left( \frac{R_\text{th}^i(\mathcal{S};\alpha,\beta)-R_\text{exp}^i}{\sigma^i_\mathrm{stat}} \right)^2 + \left( \frac{\beta_\mathrm{pp}}{\sigma_{\beta_\mathrm{pp}}}\right)^2 + \left( \frac{\beta_\mathrm{Be}}{\sigma_{\beta_\mathrm{Be}}}\right)^2 + \left( \frac{\alpha}{\sigma_\alpha}\right)^2\, ,
\label{Eq.:G_chi_2_Func}
\end{equation}
where the observed events and statistical uncertainty in the $i$th energy bin, $R_\mathrm{exp}^i$ and $\sigma^i_\mathrm{stat}$, for XENONnT are taken from Ref.~\cite{XENON:2022ltv}, while for PandaX-4T (run 1) it is taken from Ref.~\cite{PandaX:2024cic}. The corresponding predicted events are given by 
\begin{equation}
R_\text{th}^i(\mathcal{S};\alpha,\beta)=(1+\beta_\mathrm{pp})R_{\mathrm{E}\nu\mathrm{ES}}^{i,\mathrm{pp}}(\mathcal{S})+(1+\beta_\mathrm{Be})R_{\mathrm{E}\nu\mathrm{ES}}^{i,\mathrm{Be}}(\mathcal{S}) + (1+\alpha) R_\mathrm{bkg}^i\, .
\end{equation}
 For XENONnT, \( R_\mathrm{bkg} \) is based on the modeled background \( B_0 \) as detailed in Ref.~\cite{XENON:2022ltv}. Since the XENONnT Collaboration has already provided a fitted background \( B_0 \), we have omitted the nuisance parameter \( \alpha \) on the background during the analysis of XENONnT data. For PandaX-4T, we utilize the background reported by the collaboration in Ref.~\cite{PandaX:2024cic} and retain the nuisance parameter \( \alpha \) to account for background uncertainty, with \( \sigma_\alpha=2.5\% \). Similar to the LZ approach, the SM \eves contribution is also deducted from the total background for both XENONnT and LZ.

Finally, for the analysis of DARWIN  a Poissonian $\chi^2$ function has been considered as shown in Eq.~\eqref{Eq.:P_chi_2_Func}, by substituting $R_\mathrm{exp}^i$ with $R_\mathrm{bkg}^i$. Let us  notice that in this case the background spectra are taken from Ref.~\cite{DARWIN:2020bnc}, also accounting for appropriate normalization of the exposure.

\subsection{\label{subsec:TEXONO}Reactor neutrino experiment: TEXONO}
The TEXONO (Taiwan EXperiment On NeutrinO) collaboration has released data of their measurements of reactor $\bar{\nu}_e$--$e$ scattering events, using an array of CsI(Tl) scintillating crystal detectors with a total mass of 187 kg, positioned 28 meters from the core of the Kuo-Sheng Nuclear Power Station in Taiwan~\cite{TEXONO:2009knm}. Here, we elaborate on the analysis details of the $\bar{\nu}_e$--$e$ scattering signals at the TEXONO experiment, as adopted in the present study.

The $\bar{\nu}_e$ flux distribution emitted from nuclear reactors can be determined by the following equation
\begin{equation}
\label{Eq:Reactor_Flux}
\frac{d \Phi_{\bar{\nu}_e}(E_\nu)}{d E_\nu} = \frac{P}{4 \pi d^2 \epsilon} \, \sum_{k} \frac{d N^k_{\bar{\nu}_e}(E_\nu)}{d E_\nu}\, .
\end{equation}
These fluxes primarily consist of $\bar{\nu}_e$ produced through two main processes: (a) the decay of fission products from the four dominant fissile isotopes, $\mathrm{^{235}U}$, $\mathrm{^{238}U}$, $\mathrm{^{239}Pu}$, and $\mathrm{^{241}Pu}$; for these flux components the flux distribution, $d N^k_{\bar{\nu}_e}/d E_\nu$, is estimated using the Huber-Muller model~\cite{Huber:2011wv, Mueller:2011nm} for $E_\nu > 2$ MeV and from Ref.~\cite{Vogel:1989iv} for $E_\nu < 2$ MeV; and (b) neutron capture by $\mathrm{^{238}U}$ fuel ($\mathrm{^{238}U(n, \gamma)^{239}U}$), with the flux distribution taken from Ref.~\cite{TEXONO:2006xds}. Following the procedure described in Ref.~\cite{TEXONO:2006xds}, Eq.~\eqref{Eq:Reactor_Flux} is then summed over the contributions from all $\bar{\nu}_e$ production processes, specifically $k = \{\mathrm{^{235}U}, \mathrm{^{238}U}, \mathrm{^{239}Pu}, \mathrm{^{241}Pu}, \mathrm{^{238}U(n, \gamma)^{239}U}\}$, according to their relative contributions per fission. Given the experimental baseline of $d = 28$ m, a nominal reactor thermal power output of $P = 2.9$ GW, and the average energy release per fission of $\epsilon = 205.24$ MeV, the normalization of the neutrino flux reaching the TEXONO detector is estimated to be $\mathscr{N} = 6.4 \times 10^{12} \mathrm{~cm}^{-2} \mathrm{s}^{-1}$~\cite{TEXONO:2009knm}.

The expected events in the $i$th bin, resulting from $\bar{\nu}_e$--$e$ scattering for the  different interaction channels $\kappa \equiv \{\mathrm{SM}, \mathrm{~BSM}\}$, are  estimated as
\begin{equation}
\label{Eq:Events_Rate}
R^i_{\bar{\nu}_e\text{--}e} = t_\mathrm{run} N_T \int_{T_{er}^i}^{T_{er}^{i+1}} \hspace{-0.3cm} dT_{er}\mathrm{~} \mathbb{Z}_\text{eff}(T_{er})\int_{E_\nu^\mathrm{min}}^{E_{\nu}^\mathrm{max}}dE_\nu \frac{d \Phi_{\bar{\nu}_e}(E_\nu)}{d E_\nu} \left[\frac{d\sigma_{\bar{\nu}_e}}{dT_{er}}\right]_\kappa^{\nu e},
\end{equation}
where the effective number of ionized electrons given an energy deposition of $T_{er}$ in Cs and I atoms, $\mathbb{Z}_\text{eff}(T_{er})$ is taken from Ref.~\cite{Thompson2009}.

\begin{figure}[ht]
   \centering
   \captionsetup{justification=raggedright}
   \includegraphics[width=0.49\textwidth]{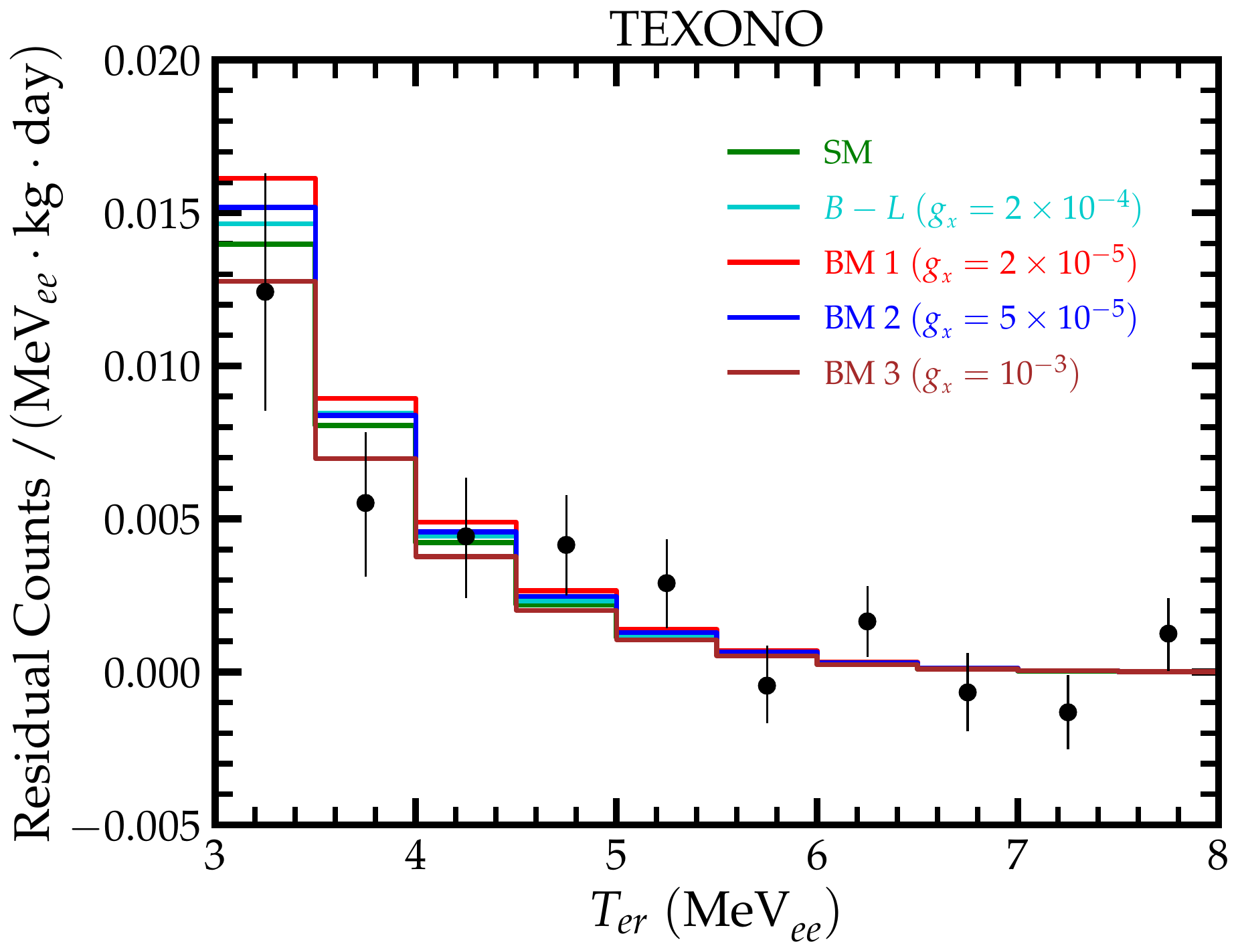}
  \caption{Simulated signals (colored histograms) and background-subtracted reactor-on data (black points with error bars) as a function of the recoil energy, $T_{er}$, at the TEXONO experiment. The BSM events are computed for a benchmark mediator mass $M_{Z'} = 100$ MeV and various couplings $\textsl{g}_{_x}$, corresponding to different $U(1)_X$ models, as indicated in the legend.}\label{Fig:TEXONO_Events}
\end{figure}

In Fig.~\ref{Fig:TEXONO_Events}, we present the simulated signals and background-subtracted reactor-on measured data at  TEXONO experiment as a function of the recoil energy, $T_{er}$. The colored histograms represent the simulated signals for the SM and various $U(1)_X$ models, calculated for a benchmark mediator mass of $M_{Z'} = 100$ MeV and different coupling constants $\textsl{g}_{_x}$. The black points with error bars indicate the background-subtracted reactor-on data, provided by the collaboration \cite{TEXONO:2009knm}.

For the statistical analysis of TEXONO data we have adopted the Gaussian $\chi^2$ statistic, following Ref.~\cite{TEXONO:2009knm}
\begin{equation}
\chi^2(\mathcal{S})= \sum_{i=1}^{10} \left( \frac{(1+\beta)R^i_{\bar{\nu}_e\text{--}e}(\mathcal{S})-R_\text{exp}^i}{\sigma^i_\mathrm{stat}} \right)^2 + \left( \frac{\beta}{\sigma_\beta}\right)^2 \, ,
\label{Eq.:G_chi_2_Func_2}
\end{equation}
where $R_\text{exp}^i$ and $\sigma^i_\mathrm{stat}$ represent the experimentally measured residual counts (background-subtracted reactor-on data) and the corresponding statistical error in each bin $i$, respectively, as reported in Ref.~\cite{TEXONO:2009knm}. Additionally, systematic uncertainties are accounted for through a nuisance parameter $\beta$ with $\sigma_\beta = 20$\%.

\section{Results}
\label{Sec:Results}
We present new constraints on the ($M_{Z'},\textsl{g}_{_{x}}$) plane for all the chiral $U(1)_X$ models (BM 1 -- BM 3) discussed in Sec.~\ref{Sec:Theory}, as well as for the vector $B-L$ model, extracted from the various experiments considered in this study.

\begin{figure}[ht]
   \centering
   \captionsetup{justification=raggedright}
   \includegraphics[width=0.65\textwidth]{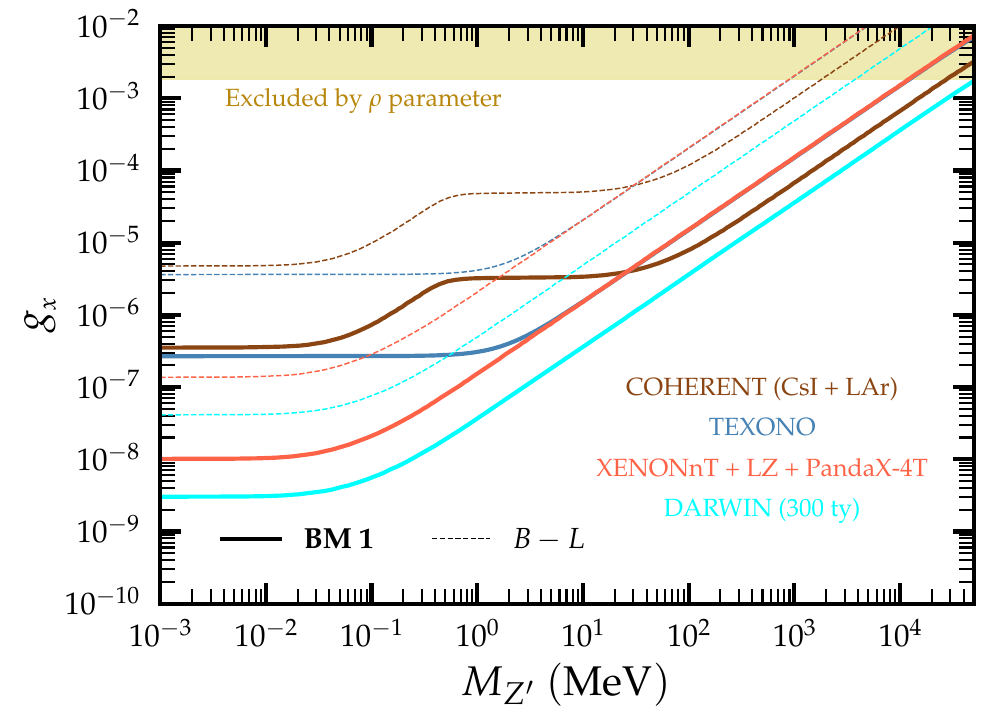}
   \caption{90\% CL exclusion limits for the case of BM 1 model (thick solid lines) in the $(M_{Z'}, \textsl{g}_{_x})$ plane. The depicted limits follow from a combined analysis of COHERENT CsI and LAr data (brown), a combined analysis of all current DM direct detection data, i.e., combining XENONnT, LZ, and PandaX-4T (run 1) data (light red). Also shown are the limits from TEXONO  (blue) and the projected sensitivity of the future DARWIN experiment assuming an exposure of $300 \mathrm{~ton} \times \mathrm{years}$ (cyan). The region of the parameter space shaded in dark yellow is ruled out for BM 1 by the $\rho$ parameter constraint. For comparison, the corresponding $B - L$ model exclusion limits are superimposed with thin dashed lines.}
   \label{Fig:BP1_Results}
\end{figure}

First, focusing on the BM 1 model the 90\% confidence level (CL) limits in the parameter space $(M_{Z'}, \textsl{g}_{_x})$ are depicted with thick solid lines in Fig.~\ref{Fig:BP1_Results}. The brown lines correspond to the combined analysis of COHERENT CsI (\cevns plus E$\nu$ES) and LAr (\cevns only) data. For $M_{Z'} \lesssim 1$~MeV, the COHERENT constraints are driven by \eves events at the CsI detector, while for larger mediator masses the bounds are entirely driven by \cevns events. Notice that, the same feature was also found in previous  searches for light mediators utilizing the COHERENT data~\cite{DeRomeri:2022twg, Candela:2023rvt, Candela:2024ljb}. Indeed, this behavior is well-understood since for low-recoil energies the E$\nu$ES-induced event rates are dominating the corresponding \cevns rates. Turning to the analysis of recent data from DM direct detection experiments, the light red lines show the resulting constraints  from a combined analysis of  XENONnT, LZ, and PandaX-4T (run 1). Additional constraints extracted  from the analysis of TEXONO data using E$\nu$ES, are illustrated with blue lines. As can be seen from the plot, DM direct detection experiments impose stronger constraints for $M_{Z'} \lesssim 30\mathrm{~MeV}$, improving by almost 1.5 order of magnitude compared to COHERENT and TEXONO. In contrast, for heavier mediators, i.e., $M_{Z'} \gtrsim 30\mathrm{~MeV}$, COHERENT places slightly improved constraints compared to DM direct detection experiments. Finally, we demonstrate the attainable sensitivity expected from the future DARWIN experiment assuming a $300 \mathrm{~ton} \times \mathrm{years}$ exposure, with cyan lines. From the latter analysis, we find that the anticipated bounds will surpass all current bounds by at least half an order of magnitude across the entire parameter space. 

For the sake of comparison, the exclusion limits for the vector $B-L$ model are superimposed  with thin dashed lines. 
As can be seen, the constraints placed on  BM 1 are more severe compared to those obtained for the vector $B-L$ model for all the experiments analyzed here. This is mainly due to the fact that the fermion charges within BM 1 are larger and hence lead to higher event rates for the same set of $(M_{Z'}, \textsl{g}_{_x})$. Finally, the shaded area with dark yellow color shows the excluded region from the constraint on the $\rho$ parameter for the case of BM 1 model.

\begin{figure}[t]
   \centering
   \captionsetup{justification=raggedright}
   \includegraphics[width=0.65\textwidth]{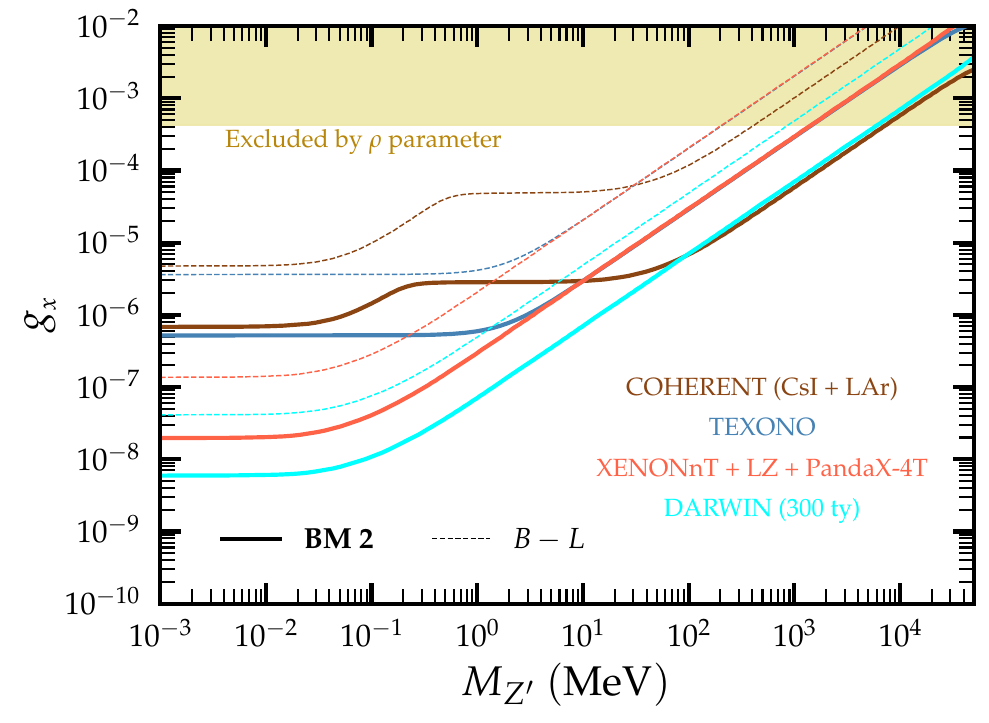}
   \caption{Same as Fig.~\ref{Fig:BP1_Results}, but for the case of BM 2 model.}
   \label{Fig:BP2_Results}
\end{figure}

Similarly, in Fig.~\ref{Fig:BP2_Results}, we present the 90\% CL the respective bounds  focusing on the  BM 2 model (thick solid lines) along with the limits corresponding the vector $B-L$ model (thin dashed lines). The color coding is consistent with that of Fig.~\ref{Fig:BP1_Results}. Although many features and the general behavior of the resulted limits are similar to those found for the case of BM 1 model, it is interesting to notice that for BM 2 the sensitivity of COHERENT  becomes comparable to the projected sensitivity of the future DARWIN experiment at $M_{Z'} \gtrsim 100\mathrm{~MeV}$. It is also noteworthy that for BM 2, the $\rho$ parameter excludes a larger region of the parameter space compared to BM 1, since the SM Higgs ($\phi$) carries a larger charge ($X_\phi$).

\begin{figure}[ht]
   \centering
   \captionsetup{justification=raggedright}
   \includegraphics[width=0.65\textwidth]{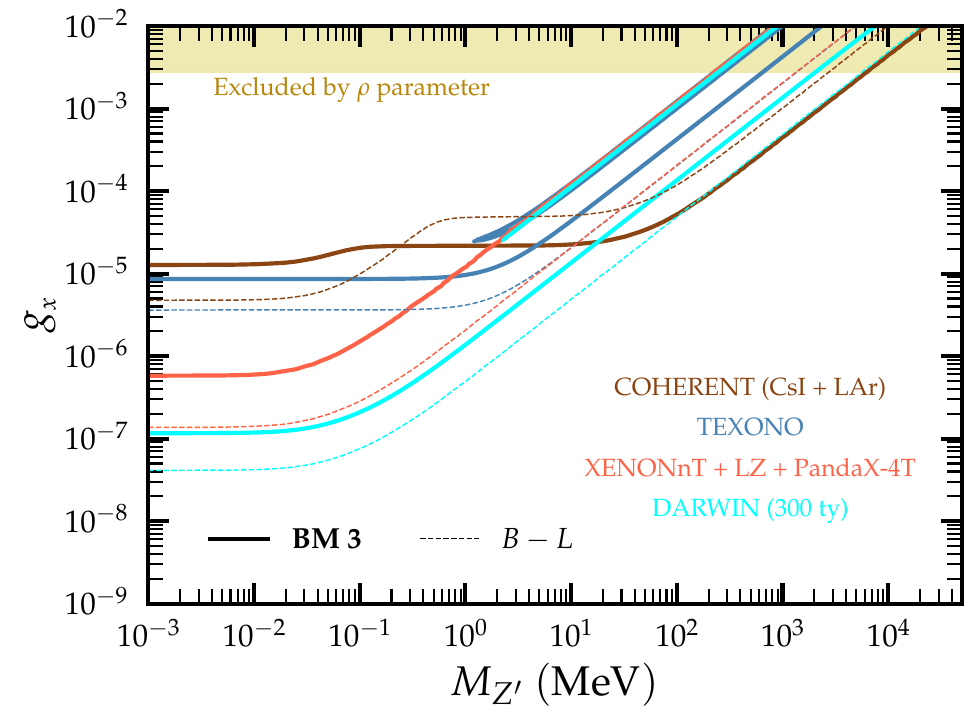}
   \caption{Same as Fig.~\ref{Fig:BP1_Results}, but for the case of BM 3 model.}
   \label{Fig:BP3_Results}
\end{figure}

Finally for the case of BM 3 model,  the 90\% CL exclusion limits on the parameters $(M_{Z'}, \textsl{g}_{_x})$ are presented in Fig.~\ref{Fig:BP3_Results}. Unlike the BM 1 and BM 2 models, the sensitivities for BM 3 are generally weaker compared to those found for the vector $B-L$ model. This holds true across all experiments considered in this study, except for the case of COHERENT. For $M_{Z'} \gtrsim 0.3 \mathrm{~MeV}$, the COHERENT experiment provides more stringent constraints for BM 3 compared to the vector $B-L$ model. In particular, the E$\nu$ES-driven limits corresponding to the BM 3 model are in general weaker in comparison to the $B-L$ case, while for the CE$\nu$NS-related ones the opposite behavior is found. This can be understood from the behavior of the \cevns and \eves cross sections within different BSM scenarios, demonstrated in Appendix \ref{Appendix:2}. Before closing this discussion let us note that, for BM 3 when considering the \eves interaction we identify a region in the parameter space where the BSM contributions interfere destructively with the SM contributions. This interference is particularly evident in the parameter ranges $M_{Z'} \in (1, 10^{3}) \mathrm{~MeV}$ and $\textsl{g}_{_x} \in (10^{-5}, 10^{-2})$. This behavior is notably visible in Fig.~\ref{Fig:BP3_Results} for the TEXONO and DARWIN experiments (see Appendix~\ref{Appendix:2} for details).

\begin{figure}[t]
   \centering
   \captionsetup{justification=raggedright}
   \includegraphics[width=0.65\textwidth]{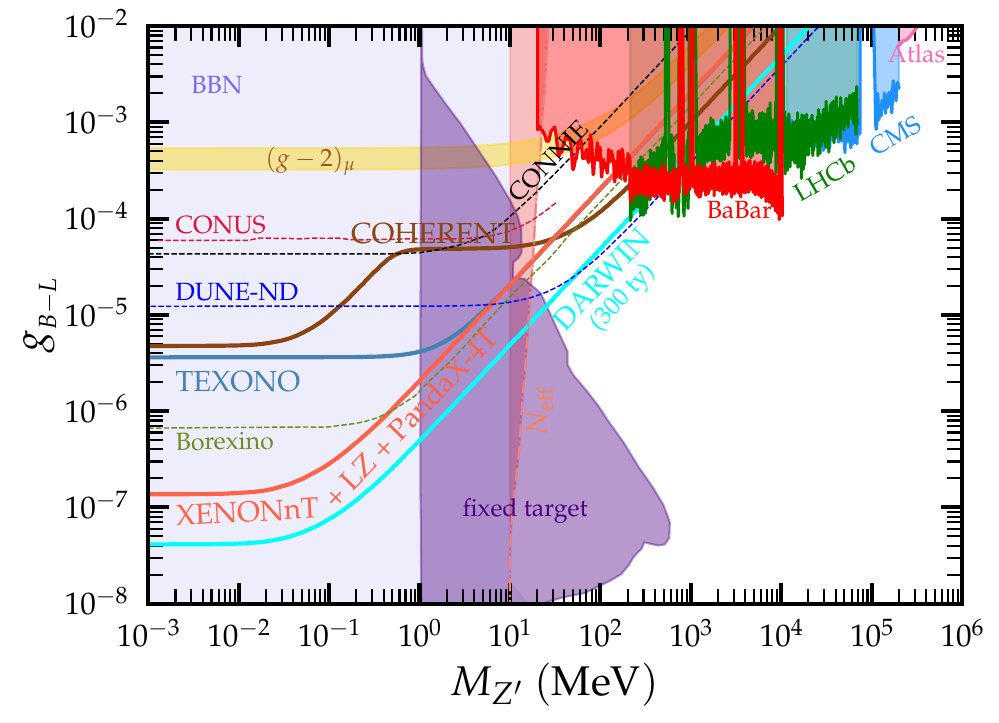}
   \caption{90\% CL constraints on the parameters $(\textsl{g}_{_{B-L}}, M_{Z'})$ of vector $B-L$ model. The limits from the experiments considered in this study are presented along with the existing constraints from relevant literature for comparison (refer to the main text for details).}
   \label{Fig:B_minus_L_Results}
\end{figure}

At this point we are interested to explore the complementarity of the constraints obtained from the analysis of \cevns and \eves data in this work, with additional constraints from further experimental probes. To this purpose, focusing on the $B-L$ model in Fig.~\ref{Fig:B_minus_L_Results}   we reproduce the 90\% CL limits from COHERENT, TEXONO and DM direct detection experiments shown previously~\footnote{We focus on $B-L$ only, since for the  DHC models there are no additional limits available in the literature. However, by comparing $B-L$ with BM 1, BM 2, and BM 3, in Figs.~\ref{Fig:BP1_Results},~\ref{Fig:BP2_Results},~\ref{Fig:BP3_Results}, respectively, serves indirectly this scope.}.  Specifically, we superimpose the limits from other dedicated \cevns experiments, such as CONUS~\cite{CONUS:2021dwh} and CONNIE~\cite{CONNIE:2024pwt}\footnote{The limits on the vector mediator from the CONUS and CONNIE experiments, as provided by the individual collaborations, correspond to the anomalous universal vector model. For the illustration purpose, we have recast these limits for the $B-L$ model by appropriately adjusting the coupling terms.}; solar neutrino experiments, such as Borexino~\cite{Coloma:2022umy}; fixed target electron beam-dump experiments, such as CHARM~\cite{CHARM:1985anb,Gninenko:2012eq}, NA64~\cite{NA64:2016oww,NA64:2019auh}, NOMAD~\cite{NOMAD:2001eyx}, E141~\cite{Riordan:1987aw,Bjorken:2009mm}, E137~\cite{Bjorken:1988as,Andreas:2012mt}, E774~\cite{Bross:1989mp}, KEK~\cite{Konaka:1986cb}, Orsay~\cite{Andreas:2012mt}, U70/$\nu$-CAL~I~\cite{Blumlein:2011mv,Blumlein:2013cua}, APEX~\cite{APEX:2011dww} etc; high energy collider experiments, such as ATLAS~\cite{ATLAS:2016bps} and CMS~\cite{CMS:2019kiy}. We also show existing $B-L$ limits from BaBar~\cite{BaBar:2014zli,BaBar:2017tiz} and LHCb~\cite{LHCb:2019vmc} dark photon analyses~\footnote{The BaBar ($e^+e^- \rightarrow \gamma A' \rightarrow \ell^+ \ell^-$) and LHCb ($pp \rightarrow A' \rightarrow \mu^+ \mu^-$) collaborations provided constraints on the kinetic mixing parameter, $\epsilon$, and the dark photon mass, $m_{A'}$, through their respective dark photon searches. These limits have been recast into the relevant parameters for the vector $B-L$ model, $(\textsl{g}_{_{B-L}}, M_{Z'})$, using the \texttt{darkcast}~\cite{darkcast} software package, as detailed in Ref.~\cite{Ilten:2018crw}. For further information, see Ref.~\cite{Bauer:2018onh}.}, as well as limits from the ${(g-2)}_\mu$ measurement~\cite{AtzoriCorona:2022moj}, and the projected sensitivity of the DUNE Near Detector~\cite{Melas:2023olz}. Finally, we include astrophysical limits from big bang nucleosynthesis (BBN)~\cite{Blinov:2019gcj,Suliga:2020jfa}, as well as from the determination of $N_\mathrm{eff}$~\cite{Esseili:2023ldf, Ghosh:2024cxi}. It is also important to note that for the vector $B-L$ model, the $\rho$ parameter does not impose any constraints on the BSM parameters $(M_{Z'}, \textsl{g}_{_x})$. This is because, in this model the $Z'$ boson does not mix with the SM $Z$ boson, and hence  $\rho$ remains independent of the $Z'$ mass at the tree level (see Eq.~\eqref{Eq.:rho_param_BSM} and \eqref{Eq.Rho_Param} for further details). 
Let us also note  that the XENONnT and PandaX-4T Collaborations have recently reported the observation of \cevns induced by $^8\mathrm{B}$ solar neutrinos~\cite{XENON:2024ijk, PandaX:2024muv}. These datasets also provide constraints on mediators of DHC models as well as the vector $B-L$ model. However, as shown in Ref.~\cite{DeRomeri:2024iaw} for the vector $B-L$ scenario, the resulting constraints lie in a parameter space that is already excluded by other experiments considered in our analysis. Hence, we do not incorporate these datasets in our study. Moreover, after the completion of our present  work, the CONUS+ Collaboration reported its measured reactor antineutrino-induced \cevns data~\cite{Ackermann:2025obx}. This dataset has recently been analyzed to extract limits on the $B-L$ mediator, see e.g., Refs.~\cite{Chattaraj:2025fvx, DeRomeri:2025csu}.

\begin{figure}[t]
   \centering
   \captionsetup{justification=raggedright}
  \includegraphics[width=0.49\textwidth]{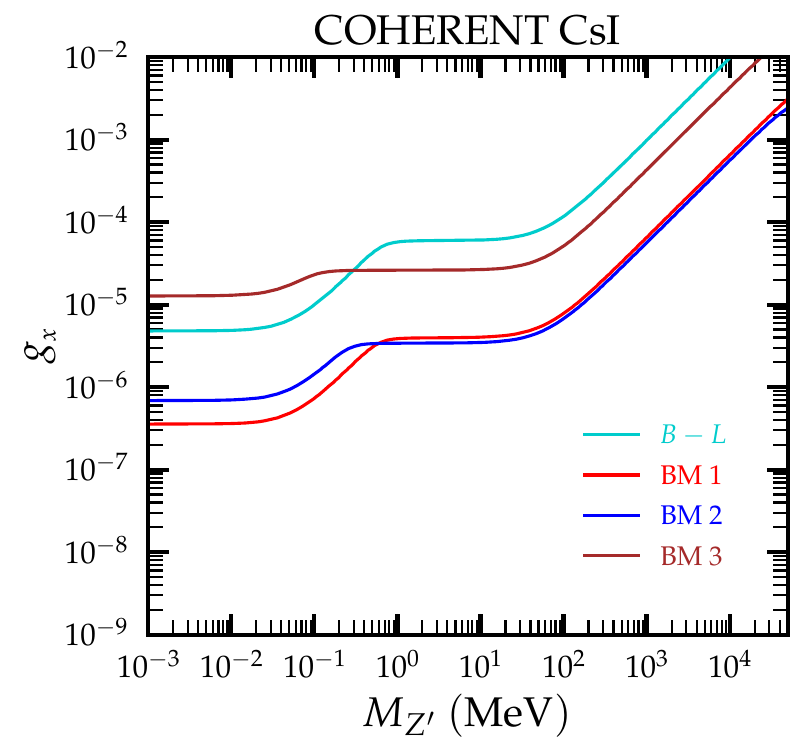}
  \includegraphics[width=0.49\textwidth]{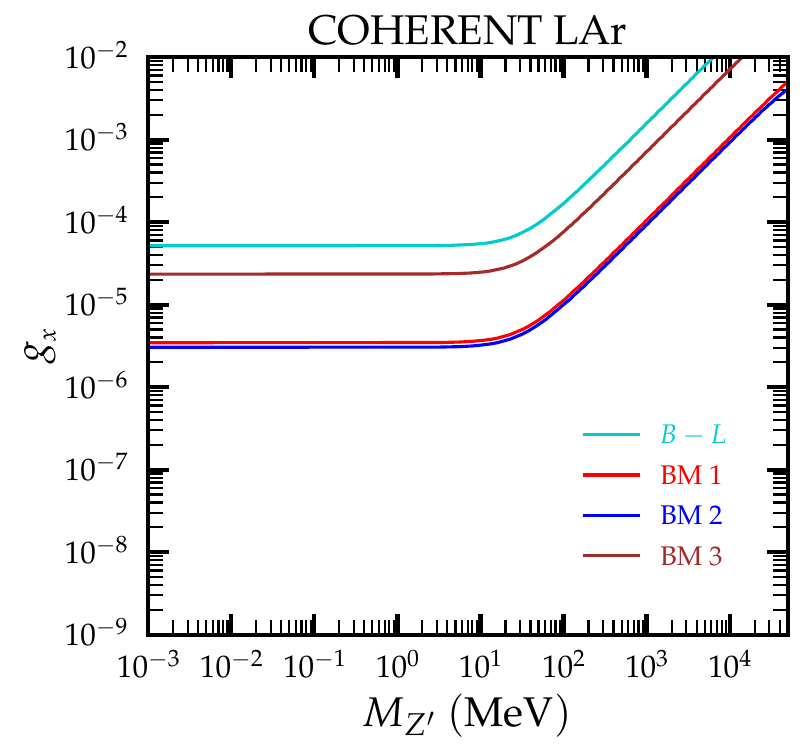}
  \caption{90\% CL constraints on  $(M_{Z'}, \textsl{g}_{_x})$ space obtained from the analysis of COHERENT CsI  (left) and  LAr  (right) data. The  bounds  are depicted for the vector $B-L$ (\textcolor{cyan}{cyan}), BM 1 (\textcolor{red}{red}), BM 2 (\textcolor{blue}{blue}), and BM 3 (\textcolor{matplotlibBrown}{brown}) models.}\label{Fig:COHERENT_Constraints}
\end{figure}

\begin{figure}[ht]
   \centering
   \captionsetup{justification=raggedright}
  \includegraphics[width=0.49\textwidth]{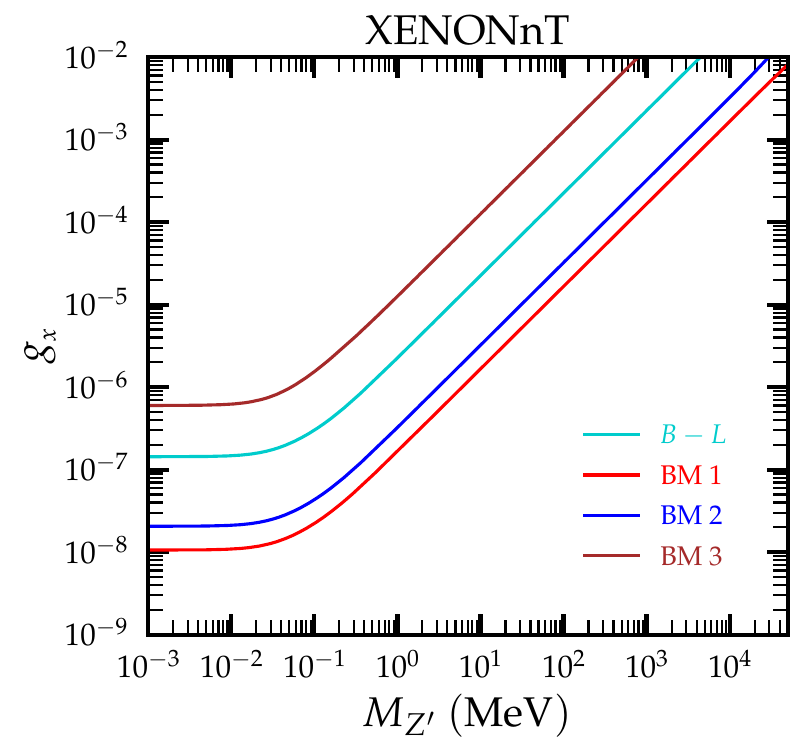}
  \includegraphics[width=0.49\textwidth]{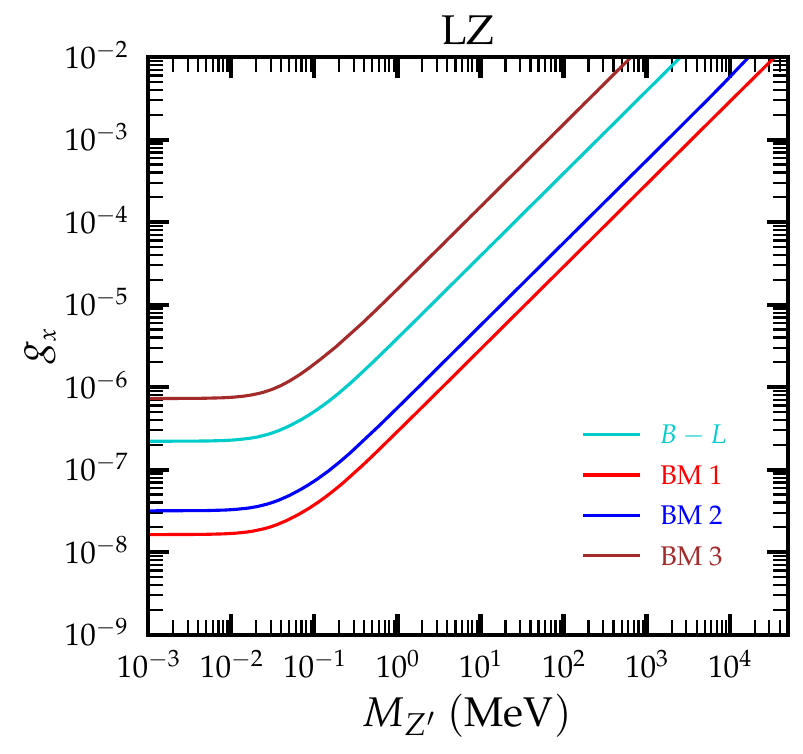}
  \includegraphics[width=0.49\textwidth]{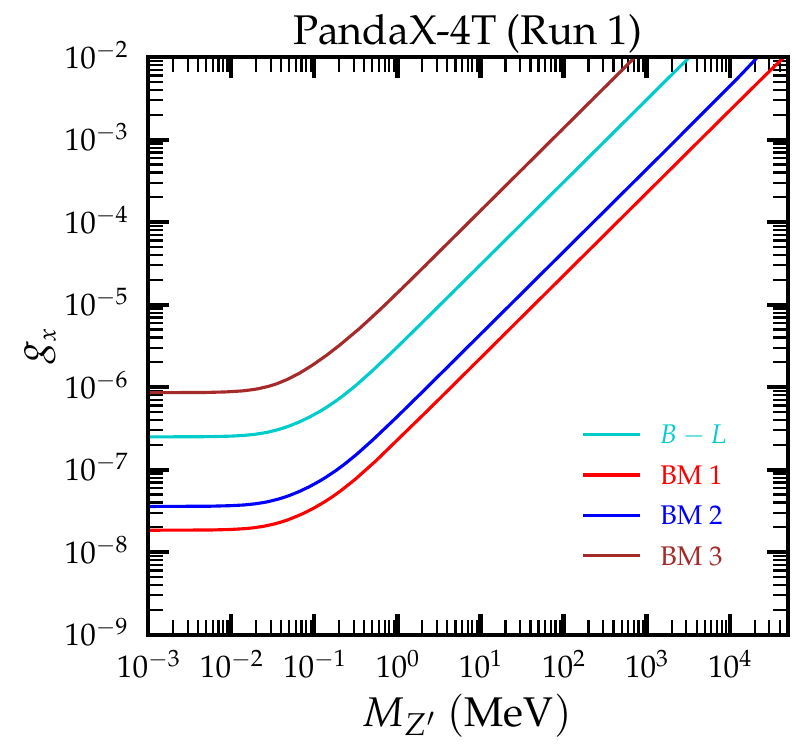}
  \includegraphics[width=0.49\textwidth]{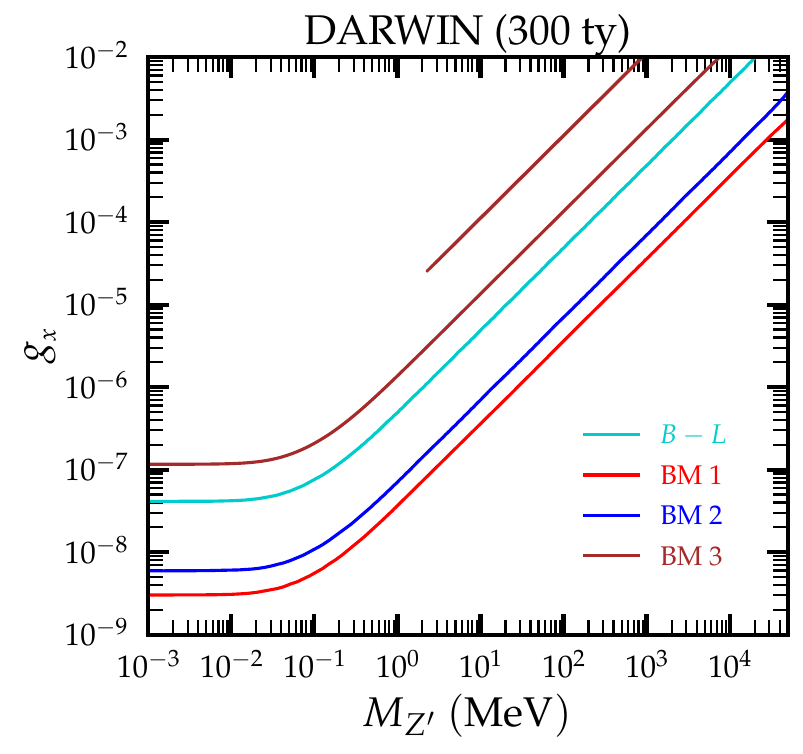}
\caption{90\% CL constraints on $(M_{Z'}, \textsl{g}_{_x})$ space obtained from various DM direct detection experiments: XENONnT (top left), LZ (top right), PandaX-4T (bottom left), and DARWIN (bottom right). The color coding for the different models is consistent with Fig.~\ref{Fig:COHERENT_Constraints}.}\label{Fig:DM_Detection_Constraints}
\end{figure}

For completeness, we intend to perform direct comparisons on the relative strength of the obtained constraints among the different chiral models as well as $B-L$, coming out from the analysis of the individual experiments. We begin our discussion focusing on COHERENT experiment. For the various models under consideration, the left and right panels of Fig.~\ref{Fig:COHERENT_Constraints} showcase the constraints at  90\% CL obtained from the COHERENT CsI and LAr data, respectively. Let us note that the limits corresponding to the vector $B-L$ model are in excellent agreement with a previous analysis performed in Ref.~\cite{DeRomeri:2022twg}.  Evidently,  the COHERENT CsI data impose constraints on a slightly larger region of the parameter space compared to the COHERENT LAr data for all analyzed models. Notice that, the bounds extracted from the analysis of the CsI data exhibit a dip at $M_{Z'} \lesssim 1 \, \text{MeV}$ due to the incorporation of \eves events in the analysis, in addition to \cevns (as discussed in Sec.~\ref{subsec:COHERENT}). In this regime, \eves events dominate over \cevns events, whereas for the analysis of the LAr data only \cevns events are considered, resulting in less stringent constraints in that region. Therefore, the combined CsI+ LAr limits presented previously are mainly driven by the CsI data.  When comparing the different models, we observe that the vector $B-L$ model generally yields the least stringent constraints, while the BM 2 model  presents the most severe limits, except in the region where $M_{Z'} \lesssim 1 \, \text{MeV}$ for the COHERENT CsI limit. In this particular range, where \eves events dominate, the BM 3 model  exhibits the weakest constraints, whereas the BM 1 model becomes the most constrained.

We now turn our attention to the DM direct detection constraints from the leading experiments: XENONnT, LZ, and PandaX-4T, along with the projected sensitivity of the upcoming DARWIN experiment. The corresponding 90\% CL limits are illustrated in Fig.~\ref{Fig:DM_Detection_Constraints}. The limits for the vector $B-L$ model are consistent with previous studies~\cite{A:2022acy, DeRomeri:2024dbv}. As can be seen, XENONnT places a tad stronger  constraints among the different DM direct detection experiments, for all the models considered here.  Looking ahead, the anticipated sensitivity of the DARWIN experiment, with a projected data-taking capability of 300 ton-years, is expected to exceed the current constraints by at least half an order of magnitude. This indicates that future DARWIN experiment could dramatically reshape the exclusion landscape for various $U(1)'$ models~\cite{DeRomeri:2024dbv}.  We also find that the BM 3 (BM 1) model exhibits the weakest (strongest) constraints. Furthermore, as discussed earlier and also in Appendix \ref{Appendix:2}, the constraints from the DARWIN experiment reveal a distinctive feature for BM 3, characterized by the destructive interference between the SM and BSM contributions.

\begin{figure}[t]
   \centering
   \captionsetup{justification=raggedright}
  \includegraphics[width=0.49\textwidth]{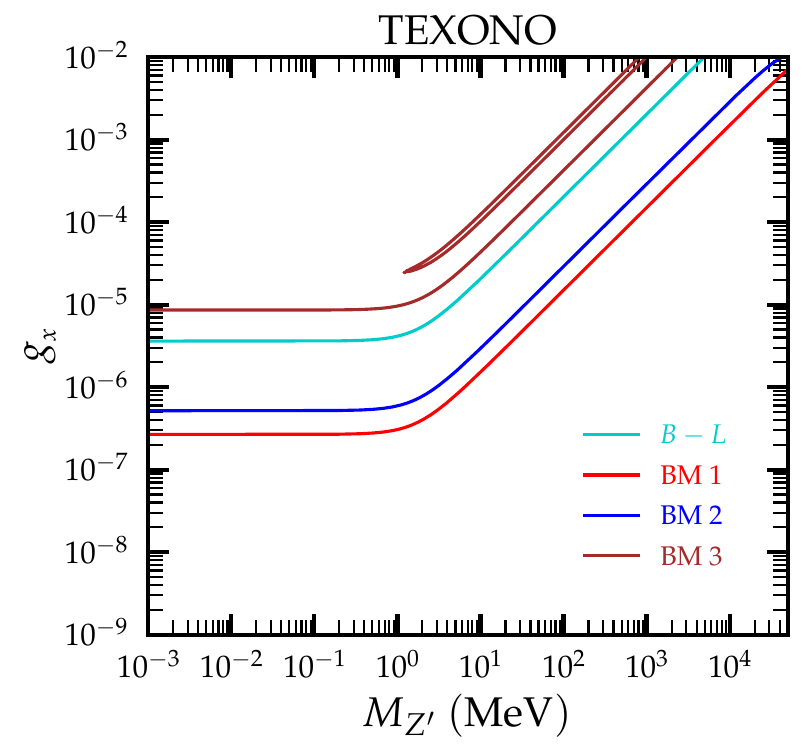}
  \caption{90\% CL constraints on  $(M_{Z'}, \textsl{g}_{_x})$ space derived from the analysis TEXONO experiment. The color coding for the exclusion limits is consistent with  Fig.~\ref{Fig:COHERENT_Constraints}.}\label{Fig:TEXONO_Constraints}
\end{figure}

Finally, the 90\% CL limits obtained from the analysis of the TEXONO experiment are visualized in Fig.~\ref{Fig:TEXONO_Constraints}. Similarly to the limits observed in DM direct detection experiments  shown in Fig.~\ref{Fig:DM_Detection_Constraints}, the BM 3 (BM 1) model exhibits the least (most) stringent constraints, for the different models analyzed here. Likewise DARWIN experiment, for the BM 3 model a destructive interference between the SM and BSM contributions is found. 

\section{Conclusions}
\label{Sec:Conclusions}
We explored a new class of gauge $U(1)_X$ models predicted within the dark hypercharge symmetry (DHC). DHC is phenomenologically interesting as it predicts large nontrivial charges under the new symmetry for both SM and BSM fermions. In this work, we  mainly focused on three benchmark scenarios, namely BM 1, BM 2 and BM 3, each having different charges for quarks and leptons. First, we have shown that within DHC, QED and charged-current interactions remain essentially unchanged, unlike the case of neutral currents. We have furthermore highlighted that stringent upper limits can be inferred from SM precision measurements of the electroweak parameter $\rho$, which were taken into account in the present calculations.  Then, we focused our attention on constraining DHC via   neutral-current interactions, by exploring its implications  on \cevns and \eves processes. At this scope, we analyzed available \cevns data from the COHERENT experiment as well as solar-induced \eves data from dark matter direct detection experiments such as XENONnT, LUX-ZEPLIN, PandaX-4T and the future DARWIN experiment. For all the aforementioned experiments, both individual and combined analyses were carried out. Existing \eves data coming from the reactor TEXONO experiment were also analyzed. The performed statistical analyses resulted in new constraints for the BM 1 -- BM 3 scenarios. Specifically, we arrived at the conclusion  that solar neutrino-induced \eves  bounds in the light of new data from DM direct detection experiments, dominate for $M_{Z'}<30$~MeV. Instead, for larger mediator masses the CE$\nu$NS-driven bounds placed by the COHERENT data become the most relevant. Finally, the constraints from TEXONO were found to be always subdominant, while the anticipated sensitivity of the future DARWIN experiment will offer an improvement of up to one order of magnitude. Moreover, we explored the relative significance of the resulting constraints on different chiral models studied in this work. To this purpose,  comparisons were made with the constraints placed on the widely studied $B-L$ model which served as a reference model. In particular, we concluded that the BM 1 and 2 models are probed with enhanced sensitivity compared to $B-L$, while the case of BM 3 leads to relatively weaker constraints. Finally, we discussed the complementarity of  our present constraints with those coming from astrophysical observations, beam-dump experiment and high energy colliders.

\acknowledgments
The authors are grateful to Sk Jeesun for useful correspondence.
We acknowledge the use of the high-performance computing facilities offered by the Bhaskara Cluster at IISER Bhopal, which significantly facilitated the completion of this research.
A.M. and H.P. appreciate the financial support provided through the Prime Minister Research Fellowship (PMRF), funded by the Government of India (PMRF ID: 0401970 and 0401969).
The work of D.K.P. is   supported by the Spanish grants PID2023-147306NB-I00 and CEX2023-001292-S (MCIU/AEI/10.13039/501100011033), as well as CIDEXG/2022/20  (Generalitat Valenciana) and CNS2023-144124 (MCIN/AEI/10.13039/ 501100011033 and “Next Generation EU”/PRTR).

\appendix

 \section{DERIVATION OF GAUGE INTERACTION VERTICES WITHIN CHIRAL $U(1)_X$ MODELS}\label{Appendix:1}
In this Appendix we discuss all relevant gauge rotations and the interaction strength of gauge bosons with fermions. The interactions of fermions and bosons could be obtained by expanding the kinetic term of fermion 
\begin{equation}
    \mathscr{L}_{K} = i \overline{\psi}\gamma^{\rho}D_{\rho}\psi \, ,
\end{equation}
with the relevant covariant derivative, $D_{\rho}$ being
\begin{equation}\label{codrivative_1}
D_{\rho}= \partial_{\rho} +igT^{a}W^{a}_{\rho} + ig'\frac{Y}{2}B_{\rho}+iX\textsl{g}_{_{x}}C_{\rho}\,.
\end{equation}
All, the relevant coupling and charges are defined in Sec.~\ref{Sub_Sec:Mass_spectrum_of_gauge_mediators}, and Table~\ref{Tab:BP_Charges}. The  $SU(2)$ generator, $T^{a}$, is defined as
\begin{equation}
    T^{1} = \frac{1}{2} \begin{pmatrix}
        0 & 1 \\
        1 & 0
    \end{pmatrix}\,, \quad  T^{2} = \frac{1}{2} \begin{pmatrix}
        0 & -i \\
        i &  0
    \end{pmatrix}\, , \quad  T^{3} = \frac{1}{2} \quad \begin{pmatrix}
        1 & 0 \\
        0 & -1
    \end{pmatrix}\, .
\end{equation}

We  first discuss charged current interactions. In this case it is useful to define a complex field
\begin{equation}
    W^{\pm}_{\rho} = \frac{W^{1}_{\rho}\mp  i W^{2}_{\rho}}{\sqrt{2}}\, .
\end{equation}
Note that flipping the sign on the right-hand side is essential for the electric charge of this field to become \(\pm 1\), while the corresponding generators are defined as $T^{\pm} = (T^{1} \pm i T^{2})/ \sqrt{2}$. With this redefinition, the covariant derivative could be written as
\begin{equation}\label{Eq:codrivative2}
D_{\rho}= \partial_{\rho} +ig( T^{+}W_{\rho}^{+} + T^{-}W_{\rho}^{-}) + ig  T^{3}W_{\rho}^{3}+ ig'\frac{Y}{2}B_{\rho}+iX\textsl{g}_{_{x}}C_{\rho}\,.
\end{equation}
From Eq.~\eqref{Eq:codrivative2} it can be deduced that the charged current interaction remains the same as SM. Focusing now on neutral current interactions, the relevant rotation defined in Eq.~\eqref{unitary matrix} yields
\begin{equation}\label{Eq:neutral_field_rotation}
\begin{aligned}
    &B_{\rho} = \cos \theta_{W}A_{\rho} - \cos \alpha \sin \theta_{W}Z_{\rho} -  \sin \alpha \sin \theta_{W}Z'_{\rho}\,,\\
    &W^{3}_{\rho} =  \sin \theta_{W} A_{\rho} + \cos \alpha \cos \theta_{W} Z_{\rho} + \sin \alpha \cos \theta_{W}Z'_{\rho}\,,\\
    &C_{\rho} =  -\sin \alpha Z_{\rho} + \cos \alpha Z'_{\rho}\,.
    \end{aligned}
\end{equation}
By solving Eq.~\eqref{Eq:codrivative2} using Eq.~\eqref{Eq:neutral_field_rotation} gives
\begin{equation}
\begin{aligned}
   D_{\rho}= \partial_{\rho} & +  ig( T^{+}W_{\rho}^{+} + T^{-}W_{\rho}^{-}) + ieQA_{\rho}\\ & + i \frac{g}{\cos \theta_{W}} \left[ ( T^{3} - Q \sin^{2} \theta_{W} )\cos \alpha ~~-~~ X\textsl{g}_{_{x}} \sin \alpha ~\frac{\cos \theta_{W}}{g}  \right]Z_{\rho}\\
& + i \left[  \left(  g T^{3} \cos \theta_{W} - \frac{g'Y}{2} \sin \theta_{W}     \right) \sin \alpha + X\textsl{g}_{_{x}} \cos \alpha          \right]Z'_{\rho}.
    \end{aligned}
\end{equation}
Again, it is clear that QED interactions remain the same as in the SM, while the weak neutral current mediated by the SM \( Z \) boson is modified.  However, as discussed in Sec.~\ref{SubSec:SM_BSM_interaction}, in the limit that satisfies the \(\rho\) parameter constraint, these terms can be neglected.

\section{\cevns AND \eves CROSS SECTIONS WITHIN SM AND DIFFERENT CHIRAL $U(1)_X$ MODELS}\label{Appendix:2}

In this Appendix, we present the cross sections for the \cevns and \eves processes, as predicted within the SM, the vector $B-L$, and the different chiral $U(1)_X$ models (BM 1, BM 2 and BM 3). The behavior of the \cevns and \eves cross sections dictate the different contour patterns  shown in Figs.~\ref{Fig:COHERENT_Constraints},~\ref{Fig:DM_Detection_Constraints} and~\ref{Fig:TEXONO_Constraints}. Further details and examples are given below.

In Fig.~\ref{Fig:CEvNS_xSec} we show the total \cevns cross section for the Cs nucleus as a function of the incoming neutrino energy assuming SM interactions only as well as the different $U(1)_X$ models considered in this work. For illustration purposes, the chosen benchmark parameter values are $\textsl{g}_{_x} = 5 \times 10^{-5}$ and $M_{Z^\prime} = 1 \, \mathrm{MeV}$. It must be stressed that  the $B-L$ as well as the three chiral models interfere always constructively with the SM. Since in the \cevns cross section the axial vector coupling is negligible, the relative strength of the individual cross sections is controlled by the size of $Z-Z'$ mixing and the different charge assignments for the vector coupling corresponding to the various chiral models. As can be seen from the plot, the smallest (largest) \cevns cross section corresponds to the SM (BM 2) case. In increasing order the cross sections are found to be: SM, $B-L$, BM 3, BM 1, and BM 2 cases, which is in accordance with the constraints shown in~Fig.~\ref{Fig:COHERENT_Constraints}.

\begin{figure}[t]
   \centering
   \captionsetup{justification=raggedright}
   \includegraphics[width=0.49\textwidth]{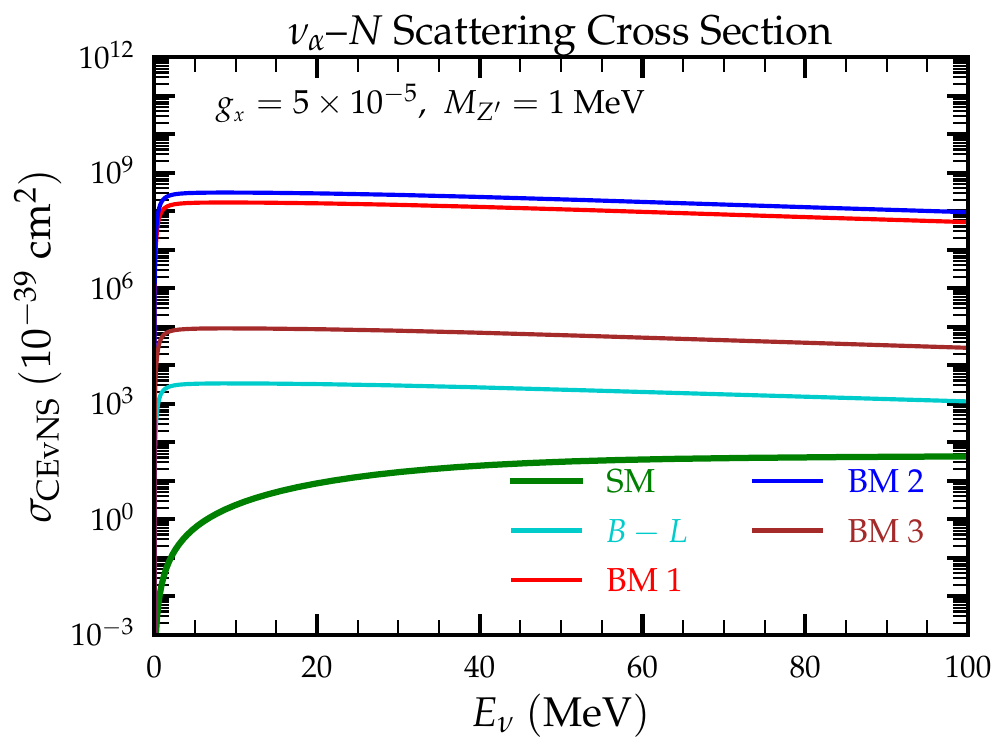}
   \caption{Total \cevns cross section for the Cs nucleus as a function of $E_\nu$ for $\textsl{g}_{_x} = 5 \times 10^{-5}$ and $M_{Z^\prime} = 1 \, \mathrm{MeV}$. The color coding is: SM (green), $B-L$ (cyan), BM 1 (red), BM 2 (blue), and BM 3 (brown).}
   \label{Fig:CEvNS_xSec}
\end{figure}

\begin{figure}[t]
   \centering
   \captionsetup{justification=raggedright}
   \includegraphics[width=0.49\textwidth]{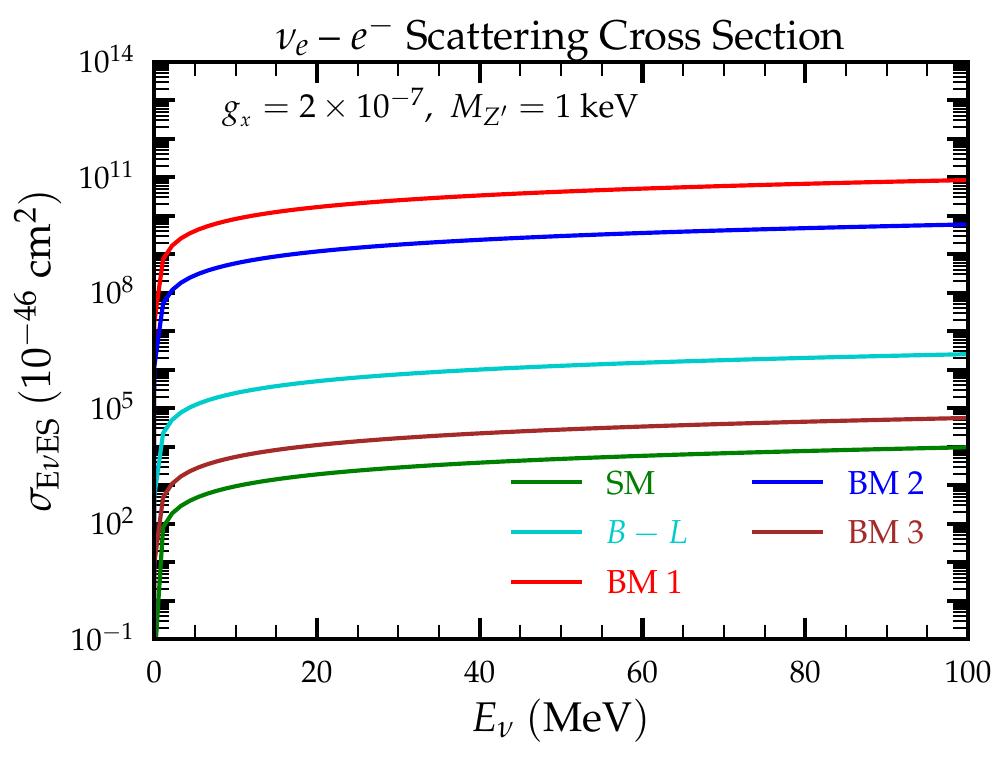}
   \includegraphics[width=0.49\textwidth]{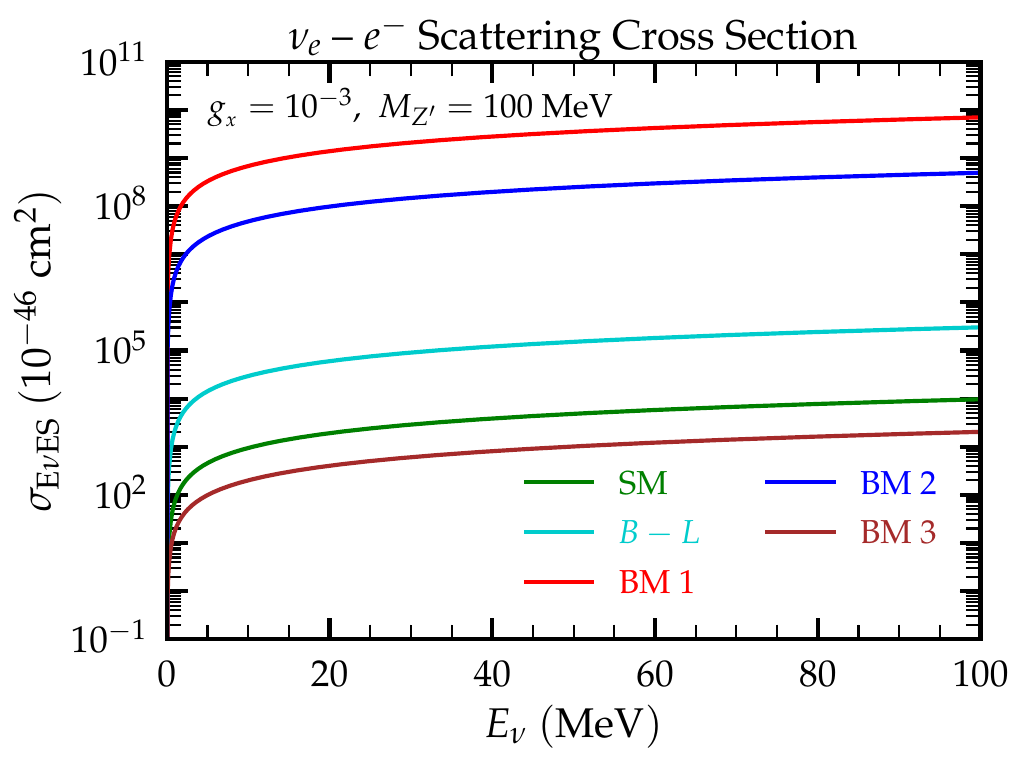}
   \caption{Total \eves cross sections for $\nu_e$--$e$ scattering as a function of $E_\nu$. The left panel shows the case for $\textsl{g}_{_x} = 2 \times 10^{-7}$ and $M_{Z^\prime} = 1 \, \mathrm{keV}$, while the right panel shows the cross section for $\textsl{g}_{_x} = 10^{-3}$ and $M_{Z^\prime} = 100 \, \mathrm{MeV}$. The color coding is the same as in Fig.~\ref{Fig:CEvNS_xSec}.}
   \label{Fig:EvES_xSec}
\end{figure}

Similarly, in Fig.~\ref{Fig:EvES_xSec} we present the \eves cross sections ($\nu_e$--$e$ scattering) predicted in the SM or within the different $U(1)_X$ models. The values of $g_{_x}$ and $M_{Z'}$ are intentionally chosen to differ largely for reasons that will become obvious in the following. The left (right) panel shows the cross section assuming $\textsl{g}_{_x} = 2 \times 10^{-7}$ and $M_{Z^\prime} = 1 \, \mathrm{keV}$  ($\textsl{g}_{_x} = 10^{-3}$ and $M_{Z^\prime} = 100 \, \mathrm{MeV}$). For most of the parameter space, the cross sections in increasing order are: SM, BM 3, $B-L$, BM 2, and BM 1. Notice that the order is not exactly the same as for the \cevns case, since for \eves both vector and axial vector contributions to the cross section are of comparable size. This, combined with the fact that the kinematic terms --which involve different signs themselves-- in the \eves cross section do not factorize from the couplings, renders the problem too complicated to be understood  analytically as in the \cevns case. Numerically, we find that for certain regions in the parameter space (see, e.g., the right panel of Fig.~\ref{Fig:EvES_xSec}), the BM 3 cross section becomes smaller than the SM one due to destructive interference.  Numerically we found that, for  TEXONO (DARWIN) the destructive interference in BM 3 arises due to the vector coupling term, $\textsl{g}_V$, in the $\bar{\nu}_e$--$e^-$ ($\nu_e$--$e^-$) cross section. This is reflected in the subdominant contour regions found for the case of DARWIN and TEXONO experiments shown in Fig.~\ref{Fig:DM_Detection_Constraints} and Fig.~\ref{Fig:TEXONO_Constraints}. Finally, in the case of DARWIN the event rate also includes subleading contributions from the $\nu_{\mu, \tau}$--$e^-$ cross section. However, the latter  do not induce any destructive interference. 

\bibliographystyle{utphys}
\bibliography{bibliography}

\providecommand{\href}[2]{#2}\begingroup\raggedright\begin{thebibliography}{100}

\bibitem{Super-Kamiokande:1998kpq}
{\bfseries Super-Kamiokande} Collaboration, Y.~Fukuda {\em et~al.}, ``{Evidence
  for oscillation of atmospheric neutrinos},''
  \href{http://dx.doi.org/10.1103/PhysRevLett.81.1562}{{\em Phys. Rev. Lett.}
  {\bfseries 81} (1998) 1562--1567},
  \href{http://arxiv.org/abs/hep-ex/9807003}{{\ttfamily arXiv:hep-ex/9807003}}.

\bibitem{SNO:2002ziz}
{\bfseries SNO} Collaboration, A.~B. McDonald {\em et~al.}, ``{Direct evidence
  for neutrino flavor transformation from neutral-current interactions in
  SNO},'' \href{http://dx.doi.org/10.1063/1.1524553}{{\em AIP Conf. Proc.}
  {\bfseries 646} no.~1, (2002) 43--58}.

\bibitem{Zwicky:1933gu}
F.~Zwicky, ``{Die Rotverschiebung von extragalaktischen Nebeln},''
  \href{http://dx.doi.org/10.1007/s10714-008-0707-4}{{\em Helv. Phys. Acta}
  {\bfseries 6} (1933) 110--127}.

\bibitem{Sofue:2000jx}
Y.~Sofue and V.~Rubin, ``{Rotation curves of spiral galaxies},''
  \href{http://dx.doi.org/10.1146/annurev.astro.39.1.137}{{\em Ann. Rev.
  Astron. Astrophys.} {\bfseries 39} (2001) 137--174},
  \href{http://arxiv.org/abs/astro-ph/0010594}{{\ttfamily
  arXiv:astro-ph/0010594}}.

\bibitem{Bertone:2004pz}
G.~Bertone, D.~Hooper, and J.~Silk, ``{Particle dark matter: Evidence,
  candidates and constraints},''
  \href{http://dx.doi.org/10.1016/j.physrep.2004.08.031}{{\em Phys. Rept.}
  {\bfseries 405} (2005) 279--390},
  \href{http://arxiv.org/abs/hep-ph/0404175}{{\ttfamily arXiv:hep-ph/0404175}}.

\bibitem{Planck:2018vyg}
{\bfseries Planck} Collaboration, N.~Aghanim {\em et~al.}, ``{Planck 2018
  results. VI. Cosmological parameters},''
  \href{http://dx.doi.org/10.1051/0004-6361/201833910}{{\em Astron. Astrophys.}
  {\bfseries 641} (2020) A6}, \href{http://arxiv.org/abs/1807.06209}{{\ttfamily
  arXiv:1807.06209 [astro-ph.CO]}}. [Erratum: Astron.Astrophys. 652, C4
  (2021)].

\bibitem{Langacker:1981hs}
P.~Langacker, ``{Grand Unified Theories},'' {\em eConf} {\bfseries C810824}
  (1981) 823. Contribution to the 10th International Symposium on Lepton and
  Photon Interactions at High Energy, Bonn, Germany, August 24–29, 1981,
  \url{http://lss.fnal.gov/conf/C810824/p823.pdf}.

\bibitem{ATLAS:2019erb}
{\bfseries ATLAS} Collaboration, G.~Aad {\em et~al.}, ``{Search for high-mass
  dilepton resonances using 139 fb$^{-1}$ of $pp$ collision data collected at
  $\sqrt{s}=$13 TeV with the ATLAS detector},''
  \href{http://dx.doi.org/10.1016/j.physletb.2019.07.016}{{\em Phys. Lett. B}
  {\bfseries 796} (2019) 68--87},
  \href{http://arxiv.org/abs/1903.06248}{{\ttfamily arXiv:1903.06248
  [hep-ex]}}.

\bibitem{CMS:2021ctt}
{\bfseries CMS} Collaboration, A.~M. Sirunyan {\em et~al.}, ``{Search for
  resonant and nonresonant new phenomena in high-mass dilepton final states at
  $ \sqrt{s} $ = 13 TeV},''
  \href{http://dx.doi.org/10.1007/JHEP07(2021)208}{{\em JHEP} {\bfseries 07}
  (2021) 208}, \href{http://arxiv.org/abs/2103.02708}{{\ttfamily
  arXiv:2103.02708 [hep-ex]}}.

\bibitem{Electroweak:2003ram}
{\bfseries LEP, ALEPH, DELPHI, L3, OPAL, LEP Electroweak Working Group, SLD
  Electroweak Group, SLD Heavy Flavor Group} Collaboration, t.~S. Electroweak,
  ``{A Combination of preliminary electroweak measurements and constraints on
  the standard model},'' \href{http://arxiv.org/abs/hep-ex/0312023}{{\ttfamily
  arXiv:hep-ex/0312023}}.

\bibitem{Essig:2009nc}
R.~Essig, P.~Schuster, and N.~Toro, ``{Probing Dark Forces and Light Hidden
  Sectors at Low-Energy e+e- Colliders},''
  \href{http://dx.doi.org/10.1103/PhysRevD.80.015003}{{\em Phys. Rev. D}
  {\bfseries 80} (2009) 015003},
  \href{http://arxiv.org/abs/0903.3941}{{\ttfamily arXiv:0903.3941 [hep-ph]}}.

\bibitem{Bross:1989mp}
A.~Bross, M.~Crisler, S.~H. Pordes, J.~Volk, S.~Errede, and J.~Wrbanek, ``{A
  Search for Shortlived Particles Produced in an Electron Beam Dump},''
  \href{http://dx.doi.org/10.1103/PhysRevLett.67.2942}{{\em Phys. Rev. Lett.}
  {\bfseries 67} (1991) 2942--2945}.

\bibitem{CHARM:1985anb}
{\bfseries CHARM} Collaboration, F.~Bergsma {\em et~al.}, ``{Search for Axion
  Like Particle Production in 400-{GeV} Proton - Copper Interactions},''
  \href{http://dx.doi.org/10.1016/0370-2693(85)90400-9}{{\em Phys. Lett. B}
  {\bfseries 157} (1985) 458--462}.

\bibitem{Gninenko:2012eq}
S.~N. Gninenko, ``{Constraints on sub-GeV hidden sector gauge bosons from a
  search for heavy neutrino decays},''
  \href{http://dx.doi.org/10.1016/j.physletb.2012.06.002}{{\em Phys. Lett. B}
  {\bfseries 713} (2012) 244--248},
  \href{http://arxiv.org/abs/1204.3583}{{\ttfamily arXiv:1204.3583 [hep-ph]}}.

\bibitem{NA64:2016oww}
{\bfseries NA64} Collaboration, D.~Banerjee {\em et~al.}, ``{Search for
  invisible decays of sub-GeV dark photons in missing-energy events at the CERN
  SPS},'' \href{http://dx.doi.org/10.1103/PhysRevLett.118.011802}{{\em Phys.
  Rev. Lett.} {\bfseries 118} no.~1, (2017) 011802},
  \href{http://arxiv.org/abs/1610.02988}{{\ttfamily arXiv:1610.02988
  [hep-ex]}}.

\bibitem{NA64:2019auh}
{\bfseries NA64} Collaboration, D.~Banerjee {\em et~al.}, ``{Improved limits on
  a hypothetical X(16.7) boson and a dark photon decaying into $e^+e^-$
  pairs},'' \href{http://dx.doi.org/10.1103/PhysRevD.101.071101}{{\em Phys.
  Rev. D} {\bfseries 101} no.~7, (2020) 071101},
  \href{http://arxiv.org/abs/1912.11389}{{\ttfamily arXiv:1912.11389
  [hep-ex]}}.

\bibitem{Spergel:1999mh}
D.~N. Spergel and P.~J. Steinhardt, ``{Observational evidence for
  selfinteracting cold dark matter},''
  \href{http://dx.doi.org/10.1103/PhysRevLett.84.3760}{{\em Phys. Rev. Lett.}
  {\bfseries 84} (2000) 3760--3763},
  \href{http://arxiv.org/abs/astro-ph/9909386}{{\ttfamily
  arXiv:astro-ph/9909386}}.

\bibitem{Duerr:2018mbd}
M.~Duerr, K.~Schmidt-Hoberg, and S.~Wild, ``{Self-interacting dark matter with
  a stable vector mediator},''
  \href{http://dx.doi.org/10.1088/1475-7516/2018/09/033}{{\em JCAP} {\bfseries
  09} (2018) 033}, \href{http://arxiv.org/abs/1804.10385}{{\ttfamily
  arXiv:1804.10385 [hep-ph]}}.

\bibitem{AtzoriCorona:2022moj}
M.~Atzori~Corona, M.~Cadeddu, N.~Cargioli, F.~Dordei, C.~Giunti, Y.~F. Li,
  E.~Picciau, C.~A. Ternes, and Y.~Y. Zhang, ``{Probing light mediators and (g
  \ensuremath{-} 2)$_{\mu}$ through detection of coherent elastic neutrino
  nucleus scattering at COHERENT},''
  \href{http://dx.doi.org/10.1007/JHEP05(2022)109}{{\em JHEP} {\bfseries 05}
  (2022) 109}, \href{http://arxiv.org/abs/2202.11002}{{\ttfamily
  arXiv:2202.11002 [hep-ph]}}.

\bibitem{AristizabalSierra:2020edu}
D.~Aristizabal~Sierra, V.~De~Romeri, L.~J. Flores, and D.~K. Papoulias,
  ``{Light vector mediators facing XENON1T data},''
  \href{http://dx.doi.org/10.1016/j.physletb.2020.135681}{{\em Phys. Lett. B}
  {\bfseries 809} (2020) 135681},
  \href{http://arxiv.org/abs/2006.12457}{{\ttfamily arXiv:2006.12457
  [hep-ph]}}.

\bibitem{Majumdar:2021vdw}
A.~Majumdar, D.~K. Papoulias, and R.~Srivastava, ``{Dark matter detectors as a
  novel probe for light new physics},''
  \href{http://dx.doi.org/10.1103/PhysRevD.106.013001}{{\em Phys. Rev. D}
  {\bfseries 106} no.~1, (2022) 013001},
  \href{http://arxiv.org/abs/2112.03309}{{\ttfamily arXiv:2112.03309
  [hep-ph]}}.

\bibitem{A:2022acy}
S.~K. A., A.~Majumdar, D.~K. Papoulias, H.~Prajapati, and R.~Srivastava,
  ``{Implications of first LZ and XENONnT results: A comparative study of
  neutrino properties and light mediators},''
  \href{http://dx.doi.org/10.1016/j.physletb.2023.137742}{{\em Phys. Lett. B}
  {\bfseries 839} (2023) 137742},
  \href{http://arxiv.org/abs/2208.06415}{{\ttfamily arXiv:2208.06415
  [hep-ph]}}.

\bibitem{DeRomeri:2024dbv}
V.~De~Romeri, D.~K. Papoulias, and C.~A. Ternes, ``{Light vector mediators at
  direct detection experiments},''
  \href{http://dx.doi.org/10.1007/JHEP05(2024)165}{{\em JHEP} {\bfseries 05}
  (2024) 165}, \href{http://arxiv.org/abs/2402.05506}{{\ttfamily
  arXiv:2402.05506 [hep-ph]}}.

\bibitem{Lindner:2018kjo}
M.~Lindner, F.~S. Queiroz, W.~Rodejohann, and X.-J. Xu, ``{Neutrino-electron
  scattering: general constraints on Z$^{\prime}$ and dark photon models},''
  \href{http://dx.doi.org/10.1007/JHEP05(2018)098}{{\em JHEP} {\bfseries 05}
  (2018) 098}, \href{http://arxiv.org/abs/1803.00060}{{\ttfamily
  arXiv:1803.00060 [hep-ph]}}.

\bibitem{AristizabalSierra:2022axl}
D.~Aristizabal~Sierra, V.~De~Romeri, and D.~K. Papoulias, ``{Consequences of
  the Dresden-II reactor data for the weak mixing angle and new physics},''
  \href{http://dx.doi.org/10.1007/JHEP09(2022)076}{{\em JHEP} {\bfseries 09}
  (2022) 076}, \href{http://arxiv.org/abs/2203.02414}{{\ttfamily
  arXiv:2203.02414 [hep-ph]}}.

\bibitem{Coloma:2022umy}
P.~Coloma, P.~Coloma, M.~C. Gonzalez-Garcia, M.~C. Gonzalez-Garcia, M.~Maltoni,
  M.~Maltoni, J.~a.~P. Pinheiro, J.~a.~P. Pinheiro, S.~Urrea, and S.~Urrea,
  ``{Constraining new physics with Borexino Phase-II spectral data},''
  \href{http://dx.doi.org/10.1007/JHEP07(2022)138}{{\em JHEP} {\bfseries 07}
  (2022) 138}, \href{http://arxiv.org/abs/2204.03011}{{\ttfamily
  arXiv:2204.03011 [hep-ph]}}. [Erratum: JHEP 11, 138 (2022)].

\bibitem{DeRomeri:2022twg}
V.~De~Romeri, O.~G. Miranda, D.~K. Papoulias, G.~Sanchez~Garcia, M.~T\'ortola,
  and J.~W.~F. Valle, ``{Physics implications of a combined analysis of
  COHERENT CsI and LAr data},''
  \href{http://dx.doi.org/10.1007/JHEP04(2023)035}{{\em JHEP} {\bfseries 04}
  (2023) 035}, \href{http://arxiv.org/abs/2211.11905}{{\ttfamily
  arXiv:2211.11905 [hep-ph]}}.

\bibitem{Melas:2023olz}
P.~Melas, D.~K. Papoulias, and N.~Saoulidou, ``{Probing generalized neutrino
  interactions with the DUNE Near Detector},''
  \href{http://dx.doi.org/10.1007/JHEP07(2023)190}{{\em JHEP} {\bfseries 07}
  (2023) 190}, \href{http://arxiv.org/abs/2303.07094}{{\ttfamily
  arXiv:2303.07094 [hep-ph]}}.

\bibitem{ATLAS:2016bps}
{\bfseries ATLAS} Collaboration, M.~Aaboud {\em et~al.}, ``{Search for
  high-mass new phenomena in the dilepton final state using proton-proton
  collisions at $\sqrt{s}=13$ TeV with the ATLAS detector},''
  \href{http://dx.doi.org/10.1016/j.physletb.2016.08.055}{{\em Phys. Lett. B}
  {\bfseries 761} (2016) 372--392},
  \href{http://arxiv.org/abs/1607.03669}{{\ttfamily arXiv:1607.03669
  [hep-ex]}}.

\bibitem{Cerdeno:2021cdz}
D.~G. Cerde\~no, M.~Cerme\~no, M.~A. P\'erez-Garc\'\i{}a, and E.~Reid,
  ``{Medium effects in supernovae constraints on light mediators},''
  \href{http://dx.doi.org/10.1103/PhysRevD.104.063013}{{\em Phys. Rev. D}
  {\bfseries 104} no.~6, (2021) 063013},
  \href{http://arxiv.org/abs/2106.11660}{{\ttfamily arXiv:2106.11660
  [hep-ph]}}.

\bibitem{Li:2023puz}
S.-P. Li and X.-J. Xu, ``{N$_{eff}$ constraints on light mediators coupled to
  neutrinos: the dilution-resistant effect},''
  \href{http://dx.doi.org/10.1007/JHEP10(2023)012}{{\em JHEP} {\bfseries 10}
  (2023) 012}, \href{http://arxiv.org/abs/2307.13967}{{\ttfamily
  arXiv:2307.13967 [hep-ph]}}.

\bibitem{Esseili:2023ldf}
H.~Esseili and G.~D. Kribs, ``{Cosmological implications of gauged U(1)$_{B-L}$
  on \ensuremath{\Delta}N $_{eff}$ in the CMB and BBN},''
  \href{http://dx.doi.org/10.1088/1475-7516/2024/05/110}{{\em JCAP} {\bfseries
  05} (2024) 110}, \href{http://arxiv.org/abs/2308.07955}{{\ttfamily
  arXiv:2308.07955 [hep-ph]}}.

\bibitem{Ghosh:2023ilw}
D.~K. Ghosh, P.~Ghosh, S.~Jeesun, and R.~Srivastava, ``{Hubble tension and
  cosmological imprints of $U(1)_X$ gauge symmetry: $U(1)_{B_3-3 L_i}$ as a
  case study},'' \href{http://dx.doi.org/10.1140/epjc/s10052-024-13220-8}{{\em
  Eur. Phys. J. C} {\bfseries 84} no.~8, (2024) 853},
  \href{http://arxiv.org/abs/2312.16304}{{\ttfamily arXiv:2312.16304
  [hep-ph]}}.

\bibitem{Ghosh:2024cxi}
D.~K. Ghosh, P.~Ghosh, S.~Jeesun, and R.~Srivastava, ``{Neff at CMB challenges
  U(1)X light gauge boson scenarios},''
  \href{http://dx.doi.org/10.1103/PhysRevD.110.075032}{{\em Phys. Rev. D}
  {\bfseries 110} no.~7, (2024) 075032},
  \href{http://arxiv.org/abs/2404.10077}{{\ttfamily arXiv:2404.10077
  [hep-ph]}}.

\bibitem{He:1990pn}
X.~G. He, G.~C. Joshi, H.~Lew, and R.~R. Volkas, ``{NEW Z-prime
  PHENOMENOLOGY},'' \href{http://dx.doi.org/10.1103/PhysRevD.43.R22}{{\em Phys.
  Rev. D} {\bfseries 43} (1991) 22--24}.

\bibitem{Ma:1997nq}
E.~Ma, ``{Gauged B - 3L(tau) and radiative neutrino masses},''
  \href{http://dx.doi.org/10.1016/S0370-2693(98)00599-1}{{\em Phys. Lett. B}
  {\bfseries 433} (1998) 74--81},
  \href{http://arxiv.org/abs/hep-ph/9709474}{{\ttfamily arXiv:hep-ph/9709474}}.

\bibitem{Lee:2010hf}
H.-S. Lee and E.~Ma, ``{Gauged $B-x_i L$ origin of $R$ Parity and its
  implications},'' \href{http://dx.doi.org/10.1016/j.physletb.2010.04.032}{{\em
  Phys. Lett. B} {\bfseries 688} (2010) 319--322},
  \href{http://arxiv.org/abs/1001.0768}{{\ttfamily arXiv:1001.0768 [hep-ph]}}.

\bibitem{Bonilla:2017lsq}
C.~Bonilla, T.~Modak, R.~Srivastava, and J.~W.~F. Valle, ``{$U(1)_{B_3-3L_\mu}$
  gauge symmetry as a simple description of $b\to s$ anomalies},''
  \href{http://dx.doi.org/10.1103/PhysRevD.98.095002}{{\em Phys. Rev. D}
  {\bfseries 98} no.~9, (2018) 095002},
  \href{http://arxiv.org/abs/1705.00915}{{\ttfamily arXiv:1705.00915
  [hep-ph]}}.

\bibitem{Alonso:2017uky}
R.~Alonso, P.~Cox, C.~Han, and T.~T. Yanagida, ``{Flavoured $B-L$ local
  symmetry and anomalous rare $B$ decays},''
  \href{http://dx.doi.org/10.1016/j.physletb.2017.10.027}{{\em Phys. Lett. B}
  {\bfseries 774} (2017) 643--648},
  \href{http://arxiv.org/abs/1705.03858}{{\ttfamily arXiv:1705.03858
  [hep-ph]}}.

\bibitem{Prajapati:2024wuu}
H.~Prajapati and R.~Srivastava, ``{The Dark HyperCharge Symmetry},''
  \href{http://arxiv.org/abs/2411.02512}{{\ttfamily arXiv:2411.02512
  [hep-ph]}}.

\bibitem{COHERENT:2017ipa}
{\bfseries COHERENT} Collaboration, D.~Akimov {\em et~al.}, ``{Observation of
  Coherent Elastic Neutrino-Nucleus Scattering},''
  \href{http://dx.doi.org/10.1126/science.aao0990}{{\em Science} {\bfseries
  357} no.~6356, (2017) 1123--1126},
  \href{http://arxiv.org/abs/1708.01294}{{\ttfamily arXiv:1708.01294
  [nucl-ex]}}.

\bibitem{COHERENT:2021xmm}
{\bfseries COHERENT} Collaboration, D.~Akimov {\em et~al.}, ``{Measurement of
  the Coherent Elastic Neutrino-Nucleus Scattering Cross Section on CsI by
  COHERENT},'' \href{http://dx.doi.org/10.1103/PhysRevLett.129.081801}{{\em
  Phys. Rev. Lett.} {\bfseries 129} no.~8, (2022) 081801},
  \href{http://arxiv.org/abs/2110.07730}{{\ttfamily arXiv:2110.07730
  [hep-ex]}}.

\bibitem{COHERENT:2020iec}
{\bfseries COHERENT} Collaboration, D.~Akimov {\em et~al.}, ``{First
  Measurement of Coherent Elastic Neutrino-Nucleus Scattering on Argon},''
  \href{http://dx.doi.org/10.1103/PhysRevLett.126.012002}{{\em Phys. Rev.
  Lett.} {\bfseries 126} no.~1, (2021) 012002},
  \href{http://arxiv.org/abs/2003.10630}{{\ttfamily arXiv:2003.10630
  [nucl-ex]}}.

\bibitem{Abdullah:2022zue}
M.~Abdullah {\em et~al.}, ``{Coherent elastic neutrino-nucleus scattering:
  Terrestrial and astrophysical applications},''
  \href{http://arxiv.org/abs/2203.07361}{{\ttfamily arXiv:2203.07361
  [hep-ph]}}.

\bibitem{Dent:2016wcr}
J.~B. Dent, B.~Dutta, S.~Liao, J.~L. Newstead, L.~E. Strigari, and J.~W.
  Walker, ``{Probing light mediators at ultralow threshold energies with
  coherent elastic neutrino-nucleus scattering},''
  \href{http://dx.doi.org/10.1103/PhysRevD.96.095007}{{\em Phys. Rev. D}
  {\bfseries 96} no.~9, (2017) 095007},
  \href{http://arxiv.org/abs/1612.06350}{{\ttfamily arXiv:1612.06350
  [hep-ph]}}.

\bibitem{Papoulias:2017qdn}
D.~K. Papoulias and T.~S. Kosmas, ``{COHERENT constraints to conventional and
  exotic neutrino physics},''
  \href{http://dx.doi.org/10.1103/PhysRevD.97.033003}{{\em Phys. Rev. D}
  {\bfseries 97} no.~3, (2018) 033003},
  \href{http://arxiv.org/abs/1711.09773}{{\ttfamily arXiv:1711.09773
  [hep-ph]}}.

\bibitem{Khan:2019cvi}
A.~N. Khan and W.~Rodejohann, ``{New physics from COHERENT data with an
  improved quenching factor},''
  \href{http://dx.doi.org/10.1103/PhysRevD.100.113003}{{\em Phys. Rev. D}
  {\bfseries 100} no.~11, (2019) 113003},
  \href{http://arxiv.org/abs/1907.12444}{{\ttfamily arXiv:1907.12444
  [hep-ph]}}.

\bibitem{AristizabalSierra:2019ykk}
D.~Aristizabal~Sierra, B.~Dutta, S.~Liao, and L.~E. Strigari, ``{Coherent
  elastic neutrino-nucleus scattering in multi-ton scale dark matter
  experiments: Classification of vector and scalar interactions new physics
  signals},'' \href{http://dx.doi.org/10.1007/JHEP12(2019)124}{{\em JHEP}
  {\bfseries 12} (2019) 124}, \href{http://arxiv.org/abs/1910.12437}{{\ttfamily
  arXiv:1910.12437 [hep-ph]}}.

\bibitem{Miranda:2020tif}
O.~G. Miranda, D.~K. Papoulias, G.~Sanchez~Garcia, O.~Sanders, M.~T\'ortola,
  and J.~W.~F. Valle, ``{Implications of the first detection of coherent
  elastic neutrino-nucleus scattering (CEvNS) with Liquid Argon},''
  \href{http://dx.doi.org/10.1007/JHEP05(2020)130}{{\em JHEP} {\bfseries 05}
  (2020) 130}, \href{http://arxiv.org/abs/2003.12050}{{\ttfamily
  arXiv:2003.12050 [hep-ph]}}. [Erratum: JHEP 01, 067 (2021)].

\bibitem{Cadeddu:2020nbr}
M.~Cadeddu, N.~Cargioli, F.~Dordei, C.~Giunti, Y.~F. Li, E.~Picciau, and Y.~Y.
  Zhang, ``{Constraints on light vector mediators through coherent elastic
  neutrino nucleus scattering data from COHERENT},''
  \href{http://dx.doi.org/10.1007/JHEP01(2021)116}{{\em JHEP} {\bfseries 01}
  (2021) 116}, \href{http://arxiv.org/abs/2008.05022}{{\ttfamily
  arXiv:2008.05022 [hep-ph]}}.

\bibitem{Abdullah:2018ykz}
M.~Abdullah, J.~B. Dent, B.~Dutta, G.~L. Kane, S.~Liao, and L.~E. Strigari,
  ``{Coherent elastic neutrino nucleus scattering as a probe of a Z' through
  kinetic and mass mixing effects},''
  \href{http://dx.doi.org/10.1103/PhysRevD.98.015005}{{\em Phys. Rev. D}
  {\bfseries 98} no.~1, (2018) 015005},
  \href{http://arxiv.org/abs/1803.01224}{{\ttfamily arXiv:1803.01224
  [hep-ph]}}.

\bibitem{Flores:2020lji}
L.~J. Flores, N.~Nath, and E.~Peinado, ``{Non-standard neutrino interactions in
  U(1)' model after COHERENT data},''
  \href{http://dx.doi.org/10.1007/JHEP06(2020)045}{{\em JHEP} {\bfseries 06}
  (2020) 045}, \href{http://arxiv.org/abs/2002.12342}{{\ttfamily
  arXiv:2002.12342 [hep-ph]}}.

\bibitem{Banerjee:2020zvi}
H.~Banerjee, B.~Dutta, and S.~Roy, ``{Supersymmetric gauged
  $\mathrm{U}{(1)}_{L_{\mu }-{L}_{\tau }}$ model for electron and muon $(g -2)$
  anomaly},'' \href{http://dx.doi.org/10.1007/JHEP03(2021)211}{{\em JHEP}
  {\bfseries 03} (2021) 211}, \href{http://arxiv.org/abs/2011.05083}{{\ttfamily
  arXiv:2011.05083 [hep-ph]}}.

\bibitem{delaVega:2021wpx}
L.~M.~G. de~la Vega, L.~J. Flores, N.~Nath, and E.~Peinado, ``{Complementarity
  between dark matter direct searches and CE\ensuremath{\nu}NS experiments in
  U(1)' models},'' \href{http://dx.doi.org/10.1007/JHEP09(2021)146}{{\em JHEP}
  {\bfseries 09} (2021) 146}, \href{http://arxiv.org/abs/2107.04037}{{\ttfamily
  arXiv:2107.04037 [hep-ph]}}.

\bibitem{Coloma:2022avw}
P.~Coloma, I.~Esteban, M.~C. Gonzalez-Garcia, L.~Larizgoitia, F.~Monrabal, and
  S.~Palomares-Ruiz, ``{Bounds on new physics with data of the Dresden-II
  reactor experiment and COHERENT},''
  \href{http://dx.doi.org/10.1007/JHEP05(2022)037}{{\em JHEP} {\bfseries 05}
  (2022) 037}, \href{http://arxiv.org/abs/2202.10829}{{\ttfamily
  arXiv:2202.10829 [hep-ph]}}.

\bibitem{DeRomeri:2023cjt}
V.~De~Romeri, V.~M. Lozano, and G.~Sanchez~Garcia, ``{Neutrino window to scalar
  leptoquarks: From low energy to colliders},''
  \href{http://dx.doi.org/10.1103/PhysRevD.109.055014}{{\em Phys. Rev. D}
  {\bfseries 109} no.~5, (2024) 055014},
  \href{http://arxiv.org/abs/2307.13790}{{\ttfamily arXiv:2307.13790
  [hep-ph]}}.

\bibitem{Bernal:2022qba}
N.~Bernal and Y.~Farzan, ``{Neutrino nonstandard interactions with arbitrary
  couplings to u and d quarks},''
  \href{http://dx.doi.org/10.1103/PhysRevD.107.035007}{{\em Phys. Rev. D}
  {\bfseries 107} no.~3, (2023) 035007},
  \href{http://arxiv.org/abs/2211.15686}{{\ttfamily arXiv:2211.15686
  [hep-ph]}}.

\bibitem{AristizabalSierra:2018eqm}
D.~Aristizabal~Sierra, V.~De~Romeri, and N.~Rojas, ``{COHERENT analysis of
  neutrino generalized interactions},''
  \href{http://dx.doi.org/10.1103/PhysRevD.98.075018}{{\em Phys. Rev. D}
  {\bfseries 98} (2018) 075018},
  \href{http://arxiv.org/abs/1806.07424}{{\ttfamily arXiv:1806.07424
  [hep-ph]}}.

\bibitem{Flores:2021kzl}
L.~J. Flores, N.~Nath, and E.~Peinado, ``{CE\ensuremath{\nu}NS as a probe of
  flavored generalized neutrino interactions},''
  \href{http://dx.doi.org/10.1103/PhysRevD.105.055010}{{\em Phys. Rev. D}
  {\bfseries 105} no.~5, (2022) 055010},
  \href{http://arxiv.org/abs/2112.05103}{{\ttfamily arXiv:2112.05103
  [hep-ph]}}.

\bibitem{Brdar:2018qqj}
V.~Brdar, W.~Rodejohann, and X.-J. Xu, ``{Producing a new Fermion in Coherent
  Elastic Neutrino-Nucleus Scattering: from Neutrino Mass to Dark Matter},''
  \href{http://dx.doi.org/10.1007/JHEP12(2018)024}{{\em JHEP} {\bfseries 12}
  (2018) 024}, \href{http://arxiv.org/abs/1810.03626}{{\ttfamily
  arXiv:1810.03626 [hep-ph]}}.

\bibitem{Candela:2023rvt}
P.~M. Candela, V.~De~Romeri, and D.~K. Papoulias, ``{COHERENT production of a
  dark fermion},'' \href{http://dx.doi.org/10.1103/PhysRevD.108.055001}{{\em
  Phys. Rev. D} {\bfseries 108} no.~5, (2023) 055001},
  \href{http://arxiv.org/abs/2305.03341}{{\ttfamily arXiv:2305.03341
  [hep-ph]}}.

\bibitem{Candela:2024ljb}
P.~M. Candela, V.~De~Romeri, P.~Melas, D.~K. Papoulias, and N.~Saoulidou,
  ``{Up-scattering production of a sterile fermion at DUNE: complementarity
  with spallation source and direct detection experiments},''
  \href{http://dx.doi.org/10.1007/JHEP10(2024)032}{{\em JHEP} {\bfseries 10}
  (2024) 032}, \href{http://arxiv.org/abs/2404.12476}{{\ttfamily
  arXiv:2404.12476 [hep-ph]}}.

\bibitem{Baxter:2019mcx}
D.~Baxter {\em et~al.}, ``{Coherent Elastic Neutrino-Nucleus Scattering at the
  European Spallation Source},''
  \href{http://dx.doi.org/10.1007/JHEP02(2020)123}{{\em JHEP} {\bfseries 02}
  (2020) 123}, \href{http://arxiv.org/abs/1911.00762}{{\ttfamily
  arXiv:1911.00762 [physics.ins-det]}}.

\bibitem{Chatterjee:2022mmu}
S.~S. Chatterjee, S.~Lavignac, O.~G. Miranda, and G.~Sanchez~Garcia,
  ``{Constraining nonstandard interactions with coherent elastic
  neutrino-nucleus scattering at the European Spallation Source},''
  \href{http://dx.doi.org/10.1103/PhysRevD.107.055019}{{\em Phys. Rev. D}
  {\bfseries 107} no.~5, (2023) 055019},
  \href{http://arxiv.org/abs/2208.11771}{{\ttfamily arXiv:2208.11771
  [hep-ph]}}.

\bibitem{CONUS:2021dwh}
{\bfseries CONUS} Collaboration, H.~Bonet {\em et~al.}, ``{Novel constraints on
  neutrino physics beyond the standard model from the CONUS experiment},''
  \href{http://dx.doi.org/10.1007/JHEP05(2022)085}{{\em JHEP} {\bfseries 05}
  (2022) 085}, \href{http://arxiv.org/abs/2110.02174}{{\ttfamily
  arXiv:2110.02174 [hep-ph]}}.

\bibitem{CONNIE:2019xid}
{\bfseries CONNIE} Collaboration, A.~Aguilar-Arevalo {\em et~al.}, ``{Search
  for light mediators in the low-energy data of the CONNIE reactor neutrino
  experiment},'' \href{http://dx.doi.org/10.1007/JHEP04(2020)054}{{\em JHEP}
  {\bfseries 04} (2020) 054}, \href{http://arxiv.org/abs/1910.04951}{{\ttfamily
  arXiv:1910.04951 [hep-ex]}}.

\bibitem{Alfonso-Pita:2022eli}
E.~Alfonso-Pita, L.~J. Flores, E.~Peinado, and E.~V\'azquez-J\'auregui, ``{New
  physics searches in a low threshold scintillating argon bubble chamber
  measuring coherent elastic neutrino-nucleus scattering in reactors},''
  \href{http://dx.doi.org/10.1103/PhysRevD.105.113005}{{\em Phys. Rev. D}
  {\bfseries 105} no.~11, (2022) 113005},
  \href{http://arxiv.org/abs/2203.05982}{{\ttfamily arXiv:2203.05982
  [hep-ph]}}.

\bibitem{Lindner:2024eng}
M.~Lindner, T.~Rink, and M.~Sen, ``{Light vector bosons and the weak mixing
  angle in the light of future germanium-based reactor CE\ensuremath{\nu}NS
  experiments},'' \href{http://dx.doi.org/10.1007/JHEP08(2024)171}{{\em JHEP}
  {\bfseries 08} (2024) 171}, \href{http://arxiv.org/abs/2401.13025}{{\ttfamily
  arXiv:2401.13025 [hep-ph]}}.

\bibitem{XENON:2022ltv}
{\bfseries XENON} Collaboration, E.~Aprile {\em et~al.}, ``{Search for New
  Physics in Electronic Recoil Data from XENONnT},''
  \href{http://dx.doi.org/10.1103/PhysRevLett.129.161805}{{\em Phys. Rev.
  Lett.} {\bfseries 129} no.~16, (2022) 161805},
  \href{http://arxiv.org/abs/2207.11330}{{\ttfamily arXiv:2207.11330
  [hep-ex]}}.

\bibitem{LZ:2022lsv}
{\bfseries LZ} Collaboration, J.~Aalbers {\em et~al.}, ``{First Dark Matter
  Search Results from the LUX-ZEPLIN (LZ) Experiment},''
  \href{http://dx.doi.org/10.1103/PhysRevLett.131.041002}{{\em Phys. Rev.
  Lett.} {\bfseries 131} no.~4, (2023) 041002},
  \href{http://arxiv.org/abs/2207.03764}{{\ttfamily arXiv:2207.03764
  [hep-ex]}}.

\bibitem{PandaX:2024cic}
{\bfseries PandaX} Collaboration, X.~Zeng {\em et~al.}, ``{Exploring New
  Physics with PandaX-4T Low Energy Electronic Recoil Data},''
  \href{http://dx.doi.org/10.1103/PhysRevLett.134.041001}{{\em Phys. Rev.
  Lett.} {\bfseries 134} no.~4, (2025) 041001},
  \href{http://arxiv.org/abs/2408.07641}{{\ttfamily arXiv:2408.07641
  [hep-ex]}}.

\bibitem{Cerdeno:2016sfi}
D.~G. Cerde\~no, M.~Fairbairn, T.~Jubb, P.~A.~N. Machado, A.~C. Vincent, and
  C.~B\oe{}hm, ``{Physics from solar neutrinos in dark matter direct detection
  experiments},'' \href{http://dx.doi.org/10.1007/JHEP09(2016)048}{{\em JHEP}
  {\bfseries 05} (2016) 118}, \href{http://arxiv.org/abs/1604.01025}{{\ttfamily
  arXiv:1604.01025 [hep-ph]}}. [Erratum: JHEP 09, 048 (2016)].

\bibitem{Schwemberger:2022fjl}
T.~Schwemberger and T.-T. Yu, ``{Detecting beyond the standard model
  interactions of solar neutrinos in low-threshold dark matter detectors},''
  \href{http://dx.doi.org/10.1103/PhysRevD.106.015002}{{\em Phys. Rev. D}
  {\bfseries 106} no.~1, (2022) 015002},
  \href{http://arxiv.org/abs/2202.01254}{{\ttfamily arXiv:2202.01254
  [hep-ph]}}.

\bibitem{Khan:2022bel}
A.~N. Khan, ``{Light new physics and neutrino electromagnetic interactions in
  XENONnT},'' \href{http://dx.doi.org/10.1016/j.physletb.2022.137650}{{\em
  Phys. Lett. B} {\bfseries 837} (2023) 137650},
  \href{http://arxiv.org/abs/2208.02144}{{\ttfamily arXiv:2208.02144
  [hep-ph]}}.

\bibitem{DARWIN:2020bnc}
{\bfseries DARWIN} Collaboration, J.~Aalbers {\em et~al.}, ``{Solar neutrino
  detection sensitivity in DARWIN via electron scattering},''
  \href{http://dx.doi.org/10.1140/epjc/s10052-020-08602-7}{{\em Eur. Phys. J.
  C} {\bfseries 80} no.~12, (2020) 1133},
  \href{http://arxiv.org/abs/2006.03114}{{\ttfamily arXiv:2006.03114
  [physics.ins-det]}}.

\bibitem{TEXONO:2009knm}
{\bfseries TEXONO} Collaboration, M.~Deniz {\em et~al.}, ``{Measurement of
  Nu(e)-bar -Electron Scattering Cross-Section with a CsI(Tl) Scintillating
  Crystal Array at the Kuo-Sheng Nuclear Power Reactor},''
  \href{http://dx.doi.org/10.1103/PhysRevD.81.072001}{{\em Phys. Rev. D}
  {\bfseries 81} (2010) 072001},
  \href{http://arxiv.org/abs/0911.1597}{{\ttfamily arXiv:0911.1597 [hep-ex]}}.

\bibitem{Witten:1982fp}
E.~Witten, ``{An SU(2) Anomaly},''
  \href{http://dx.doi.org/10.1016/0370-2693(82)90728-6}{{\em Phys. Lett. B}
  {\bfseries 117} (1982) 324--328}.

\bibitem{Adler:1969gk}
S.~L. Adler, ``{Axial vector vertex in spinor electrodynamics},''
  \href{http://dx.doi.org/10.1103/PhysRev.177.2426}{{\em Phys. Rev.} {\bfseries
  177} (1969) 2426--2438}.

\bibitem{Bell:1969ts}
J.~S. Bell and R.~Jackiw, ``{A PCAC puzzle: $\pi^0 \to \gamma \gamma$ in the
  $\sigma$ model},'' \href{http://dx.doi.org/10.1007/BF02823296}{{\em Nuovo
  Cim. A} {\bfseries 60} (1969) 47--61}.

\bibitem{Alvarez-Gaume:1983ihn}
L.~Alvarez-Gaume and E.~Witten, ``{Gravitational Anomalies},''
  \href{http://dx.doi.org/10.1016/0550-3213(84)90066-X}{{\em Nucl. Phys. B}
  {\bfseries 234} (1984) 269}.

\bibitem{Delbourgo:1972xb}
R.~Delbourgo and A.~Salam, ``{The gravitational correction to pcac},''
  \href{http://dx.doi.org/10.1016/0370-2693(72)90825-8}{{\em Phys. Lett. B}
  {\bfseries 40} (1972) 381--382}.

\bibitem{Geng:1989tcu}
C.~Q. Geng and R.~E. Marshak, ``{Uniqueness of Quark and Lepton Representations
  in the Standard Model From the Anomalies Viewpoint},''
  \href{http://dx.doi.org/10.1103/PhysRevD.39.693}{{\em Phys. Rev. D}
  {\bfseries 39} (1989) 693}.

\bibitem{Minahan:1989vd}
J.~A. Minahan, P.~Ramond, and R.~C. Warner, ``{A Comment on Anomaly
  Cancellation in the Standard Model},''
  \href{http://dx.doi.org/10.1103/PhysRevD.41.715}{{\em Phys. Rev. D}
  {\bfseries 41} (1990) 715}.

\bibitem{ROSS1975135}
D.~Ross and M.~Veltman, ``Neutral currents and the higgs mechanism,''
  \href{http://dx.doi.org/https://doi.org/10.1016/0550-3213(75)90485-X}{{\em
  Nuclear Physics B} {\bfseries 95} no.~1, (1975) 135--147}.

\bibitem{Bento:2023weq}
M.~P. Bento, H.~E. Haber, and J.~a.~P. Silva, ``{Tree-level Unitarity in $
  \textrm{SU}{(2)}_L\times \textrm{U}{(1)}_Y\times \textrm{U}{(1)}_{Y^{\prime
  }} $ Models},'' \href{http://dx.doi.org/10.1007/JHEP10(2023)083}{{\em JHEP}
  {\bfseries 10} (2023) 083}, \href{http://arxiv.org/abs/2306.01836}{{\ttfamily
  arXiv:2306.01836 [hep-ph]}}.

\bibitem{Bento:2023flt}
M.~P. Bento, H.~E. Haber, and J.~a.~P. Silva, ``{Classes of complete dark
  photon models constrained by Z-physics},''
  \href{http://dx.doi.org/10.1016/j.physletb.2024.138501}{{\em Phys. Lett. B}
  {\bfseries 850} (2024) 138501},
  \href{http://arxiv.org/abs/2311.04976}{{\ttfamily arXiv:2311.04976
  [hep-ph]}}.

\bibitem{ParticleDataGroup:2020ssz}
{\bfseries Particle Data Group} Collaboration, P.~A. Zyla {\em et~al.},
  ``{Review of Particle Physics},''
  \href{http://dx.doi.org/10.1093/ptep/ptaa104}{{\em PTEP} {\bfseries 2020}
  no.~8, (2020) 083C01}.

\bibitem{Giunti:2007ry}
C.~Giunti and C.~W. Kim,
  \href{http://dx.doi.org/10.1093/acprof:oso/9780198508717.001.0001}{{\em
  {Fundamentals of Neutrino Physics and Astrophysics}}}.
\newblock 2007.

\bibitem{Kayser:1979mj}
B.~Kayser, E.~Fischbach, S.~P. Rosen, and H.~Spivack, ``{Charged and Neutral
  Current Interference in $\nu_e e$ Scattering},''
  \href{http://dx.doi.org/10.1103/PhysRevD.20.87}{{\em Phys. Rev. D} {\bfseries
  20} (1979) 87}.

\bibitem{Barranco:2005yy}
J.~Barranco, O.~G. Miranda, and T.~I. Rashba, ``{Probing new physics with
  coherent neutrino scattering off nuclei},''
  \href{http://dx.doi.org/10.1088/1126-6708/2005/12/021}{{\em JHEP} {\bfseries
  12} (2005) 021}, \href{http://arxiv.org/abs/hep-ph/0508299}{{\ttfamily
  arXiv:hep-ph/0508299}}.

\bibitem{Freedman:1973yd}
D.~Z. Freedman, ``{Coherent Neutrino Nucleus Scattering as a Probe of the Weak
  Neutral Current},'' \href{http://dx.doi.org/10.1103/PhysRevD.9.1389}{{\em
  Phys. Rev. D} {\bfseries 9} (1974) 1389--1392}.

\bibitem{Papoulias:2018uzy}
D.~K. Papoulias, R.~Sahu, T.~S. Kosmas, V.~K.~B. Kota, and B.~Nayak, ``{Novel
  neutrino-floor and dark matter searches with deformed shell model
  calculations},'' \href{http://dx.doi.org/10.1155/2018/6031362}{{\em Adv. High
  Energy Phys.} {\bfseries 2018} (2018) 6031362},
  \href{http://arxiv.org/abs/1804.11319}{{\ttfamily arXiv:1804.11319
  [hep-ph]}}.

\bibitem{Papoulias:2015vxa}
D.~K. Papoulias and T.~S. Kosmas, ``{Standard and Nonstandard Neutrino-Nucleus
  Reactions Cross Sections and Event Rates to Neutrino Detection
  Experiments},'' \href{http://dx.doi.org/10.1155/2015/763648}{{\em Adv. High
  Energy Phys.} {\bfseries 2015} (2015) 763648},
  \href{http://arxiv.org/abs/1502.02928}{{\ttfamily arXiv:1502.02928
  [nucl-th]}}.

\bibitem{Klein:1999qj}
S.~Klein and J.~Nystrand, ``{Exclusive vector meson production in relativistic
  heavy ion collisions},''
  \href{http://dx.doi.org/10.1103/PhysRevC.60.014903}{{\em Phys. Rev. C}
  {\bfseries 60} (1999) 014903},
  \href{http://arxiv.org/abs/hep-ph/9902259}{{\ttfamily arXiv:hep-ph/9902259}}.

\bibitem{Adamski:2024yqt}
S.~Adamski {\em et~al.}, ``{First detection of coherent elastic
  neutrino-nucleus scattering on germanium},''
  \href{http://arxiv.org/abs/2406.13806}{{\ttfamily arXiv:2406.13806
  [hep-ex]}}.

\bibitem{Michel:1949qe}
L.~Michel, ``{Interaction between four half spin particles and the decay of the
  $\mu$ meson},'' \href{http://dx.doi.org/10.1088/0370-1298/63/5/311}{{\em
  Proc. Phys. Soc. A} {\bfseries 63} (1950) 514--531}.

\bibitem{Bouchiat:1957zz}
C.~Bouchiat and L.~Michel, ``{Theory of $\mu$-Meson Decay with the Hypothesis
  of Nonconservation of Parity},''
  \href{http://dx.doi.org/10.1103/PhysRev.106.170}{{\em Phys. Rev.} {\bfseries
  106} (1957) 170--172}.

\bibitem{COHERENT:2021pcd}
{\bfseries COHERENT} Collaboration, D.~Akimov {\em et~al.}, ``{Measurement of
  scintillation response of CsI[Na] to low-energy nuclear recoils by
  COHERENT},'' \href{http://dx.doi.org/10.1088/1748-0221/17/10/P10034}{{\em
  JINST} {\bfseries 17} no.~10, (2022) P10034},
  \href{http://arxiv.org/abs/2111.02477}{{\ttfamily arXiv:2111.02477
  [physics.ins-det]}}.

\bibitem{Thompson2009}
A.~Thompson {\em et~al.}, ``X-ray data booklet,'' 2009.
\newblock \url{https://xdb.lbl.gov/}.

\bibitem{Picciau:2022xzi}
E.~Picciau, {\em {Low-energy signatures in DarkSide-50 experiment and neutrino
  scattering processes}}.
\newblock PhD thesis, Cagliari U., 2022.

\bibitem{COHERENT:2020ybo}
{\bfseries COHERENT} Collaboration, D.~Akimov {\em et~al.}, ``{COHERENT
  Collaboration data release from the first detection of coherent elastic
  neutrino-nucleus scattering on argon},''
  \href{http://arxiv.org/abs/2006.12659}{{\ttfamily arXiv:2006.12659
  [nucl-ex]}}.

\bibitem{Chen:2016eab}
J.-W. Chen, H.-C. Chi, C.~P. Liu, and C.-P. Wu, ``{Low-energy electronic recoil
  in xenon detectors by solar neutrinos},''
  \href{http://dx.doi.org/10.1016/j.physletb.2017.10.029}{{\em Phys. Lett. B}
  {\bfseries 774} (2017) 656--661},
  \href{http://arxiv.org/abs/1610.04177}{{\ttfamily arXiv:1610.04177
  [hep-ex]}}.

\bibitem{BahcallSNData}
J.~Bahcall.
\newblock \url{http://www.sns.ias.edu/~jnb/SNdata/sndata.html}.

\bibitem{Bahcall:1987jc}
J.~N. Bahcall and R.~K. Ulrich, ``{Solar Models, Neutrino Experiments and
  Helioseismology},'' \href{http://dx.doi.org/10.1103/RevModPhys.60.297}{{\em
  Rev. Mod. Phys.} {\bfseries 60} (1988) 297--372}.

\bibitem{Bahcall:1994cf}
J.~N. Bahcall, ``{The Be-7 solar neutrino line: A Reflection of the central
  temperature distribution of the sun},''
  \href{http://dx.doi.org/10.1103/PhysRevD.49.3923}{{\em Phys. Rev. D}
  {\bfseries 49} (1994) 3923--3945},
  \href{http://arxiv.org/abs/astro-ph/9401024}{{\ttfamily
  arXiv:astro-ph/9401024}}.

\bibitem{Bahcall:1995bt}
J.~N. Bahcall and M.~H. Pinsonneault, ``{Solar models with helium and heavy
  element diffusion},'' \href{http://dx.doi.org/10.1103/RevModPhys.67.781}{{\em
  Rev. Mod. Phys.} {\bfseries 67} (1995) 781--808},
  \href{http://arxiv.org/abs/hep-ph/9505425}{{\ttfamily arXiv:hep-ph/9505425}}.

\bibitem{Bahcall:1996qv}
J.~N. Bahcall, E.~Lisi, D.~E. Alburger, L.~De~Braeckeleer, S.~J. Freedman, and
  J.~Napolitano, ``{Standard neutrino spectrum from B-8 decay},''
  \href{http://dx.doi.org/10.1103/PhysRevC.54.411}{{\em Phys. Rev. C}
  {\bfseries 54} (1996) 411--422},
  \href{http://arxiv.org/abs/nucl-th/9601044}{{\ttfamily
  arXiv:nucl-th/9601044}}.

\bibitem{Haxton:2012wfz}
W.~C. Haxton, R.~G. Hamish~Robertson, and A.~M. Serenelli, ``{Solar Neutrinos:
  Status and Prospects},''
  \href{http://dx.doi.org/10.1146/annurev-astro-081811-125539}{{\em Ann. Rev.
  Astron. Astrophys.} {\bfseries 51} (2013) 21--61},
  \href{http://arxiv.org/abs/1208.5723}{{\ttfamily arXiv:1208.5723
  [astro-ph.SR]}}.

\bibitem{Baxter:2021pqo}
D.~Baxter {\em et~al.}, ``{Recommended conventions for reporting results from
  direct dark matter searches},''
  \href{http://dx.doi.org/10.1140/epjc/s10052-021-09655-y}{{\em Eur. Phys. J.
  C} {\bfseries 81} no.~10, (2021) 907},
  \href{http://arxiv.org/abs/2105.00599}{{\ttfamily arXiv:2105.00599
  [hep-ex]}}.

\bibitem{Escrihuela:2009up}
F.~J. Escrihuela, O.~G. Miranda, M.~A. Tortola, and J.~W.~F. Valle,
  ``{Constraining nonstandard neutrino-quark interactions with solar, reactor
  and accelerator data},''
  \href{http://dx.doi.org/10.1103/PhysRevD.80.129908}{{\em Phys. Rev. D}
  {\bfseries 80} (2009) 105009},
  \href{http://arxiv.org/abs/0907.2630}{{\ttfamily arXiv:0907.2630 [hep-ph]}}.
  [Erratum: Phys.Rev.D 80, 129908 (2009)].

\bibitem{LZ:2023poo}
{\bfseries LZ} Collaboration, J.~Aalbers {\em et~al.}, ``{Search for new
  physics in low-energy electron recoils from the first LZ exposure},''
  \href{http://dx.doi.org/10.1103/PhysRevD.108.072006}{{\em Phys. Rev. D}
  {\bfseries 108} no.~7, (2023) 072006},
  \href{http://arxiv.org/abs/2307.15753}{{\ttfamily arXiv:2307.15753
  [hep-ex]}}.

\bibitem{XENON:2020iwh}
{\bfseries XENON} Collaboration, E.~Aprile {\em et~al.}, ``{Energy resolution
  and linearity of XENON1T in the MeV energy range},''
  \href{http://dx.doi.org/10.1140/epjc/s10052-020-8284-0}{{\em Eur. Phys. J. C}
  {\bfseries 80} no.~8, (2020) 785},
  \href{http://arxiv.org/abs/2003.03825}{{\ttfamily arXiv:2003.03825
  [physics.ins-det]}}.

\bibitem{XENON:2020rca}
{\bfseries XENON} Collaboration, E.~Aprile {\em et~al.}, ``{Excess electronic
  recoil events in XENON1T},''
  \href{http://dx.doi.org/10.1103/PhysRevD.102.072004}{{\em Phys. Rev. D}
  {\bfseries 102} no.~7, (2020) 072004},
  \href{http://arxiv.org/abs/2006.09721}{{\ttfamily arXiv:2006.09721
  [hep-ex]}}.

\bibitem{Pereira:2023rte}
{\bfseries LZ} Collaboration, G.~Pereira, C.~Silva, and V.~N. Solovov,
  ``{Energy resolution of the LZ detector for high-energy electronic
  recoils},'' \href{http://dx.doi.org/10.1088/1748-0221/18/04/C04007}{{\em
  JINST} {\bfseries 18} no.~04, (2023) C04007}.

\bibitem{PandaX:2022ood}
{\bfseries PandaX} Collaboration, D.~Zhang {\em et~al.}, ``{Search for Light
  Fermionic Dark Matter Absorption on Electrons in PandaX-4T},''
  \href{http://dx.doi.org/10.1103/PhysRevLett.129.161804}{{\em Phys. Rev.
  Lett.} {\bfseries 129} no.~16, (2022) 161804},
  \href{http://arxiv.org/abs/2206.02339}{{\ttfamily arXiv:2206.02339
  [hep-ex]}}.

\bibitem{AtzoriCorona:2022jeb}
M.~Atzori~Corona, W.~M. Bonivento, M.~Cadeddu, N.~Cargioli, and F.~Dordei,
  ``{New constraint on neutrino magnetic moment and neutrino millicharge from
  LUX-ZEPLIN dark matter search results},''
  \href{http://dx.doi.org/10.1103/PhysRevD.107.053001}{{\em Phys. Rev. D}
  {\bfseries 107} no.~5, (2023) 053001},
  \href{http://arxiv.org/abs/2207.05036}{{\ttfamily arXiv:2207.05036
  [hep-ph]}}.

\bibitem{Huber:2011wv}
P.~Huber, ``{On the determination of anti-neutrino spectra from nuclear
  reactors},'' \href{http://dx.doi.org/10.1103/PhysRevC.85.029901}{{\em Phys.
  Rev. C} {\bfseries 84} (2011) 024617},
  \href{http://arxiv.org/abs/1106.0687}{{\ttfamily arXiv:1106.0687 [hep-ph]}}.
  [Erratum: Phys.Rev.C 85, 029901 (2012)].

\bibitem{Mueller:2011nm}
T.~A. Mueller {\em et~al.}, ``{Improved Predictions of Reactor Antineutrino
  Spectra},'' \href{http://dx.doi.org/10.1103/PhysRevC.83.054615}{{\em Phys.
  Rev. C} {\bfseries 83} (2011) 054615},
  \href{http://arxiv.org/abs/1101.2663}{{\ttfamily arXiv:1101.2663 [hep-ex]}}.

\bibitem{Vogel:1989iv}
P.~Vogel and J.~Engel, ``{Neutrino Electromagnetic Form-Factors},''
  \href{http://dx.doi.org/10.1103/PhysRevD.39.3378}{{\em Phys. Rev. D}
  {\bfseries 39} (1989) 3378}.

\bibitem{TEXONO:2006xds}
{\bfseries TEXONO} Collaboration, H.~T. Wong {\em et~al.}, ``{A Search of
  Neutrino Magnetic Moments with a High-Purity Germanium Detector at the
  Kuo-Sheng Nuclear Power Station},''
  \href{http://dx.doi.org/10.1103/PhysRevD.75.012001}{{\em Phys. Rev. D}
  {\bfseries 75} (2007) 012001},
  \href{http://arxiv.org/abs/hep-ex/0605006}{{\ttfamily arXiv:hep-ex/0605006}}.

\bibitem{CONNIE:2024pwt}
{\bfseries CONNIE} Collaboration, A.~A. Aguilar-Arevalo {\em et~al.},
  ``{Searches for CE\ensuremath{\nu}NS and Physics beyond the Standard Model
  using Skipper-CCDs at CONNIE},''
  \href{http://arxiv.org/abs/2403.15976}{{\ttfamily arXiv:2403.15976
  [hep-ex]}}.

\bibitem{NOMAD:2001eyx}
{\bfseries NOMAD} Collaboration, P.~Astier {\em et~al.}, ``{Search for heavy
  neutrinos mixing with tau neutrinos},''
  \href{http://dx.doi.org/10.1016/S0370-2693(01)00362-8}{{\em Phys. Lett. B}
  {\bfseries 506} (2001) 27--38},
  \href{http://arxiv.org/abs/hep-ex/0101041}{{\ttfamily arXiv:hep-ex/0101041}}.

\bibitem{Riordan:1987aw}
E.~M. Riordan {\em et~al.}, ``{A Search for Short Lived Axions in an Electron
  Beam Dump Experiment},''
  \href{http://dx.doi.org/10.1103/PhysRevLett.59.755}{{\em Phys. Rev. Lett.}
  {\bfseries 59} (1987) 755}.

\bibitem{Bjorken:2009mm}
J.~D. Bjorken, R.~Essig, P.~Schuster, and N.~Toro, ``{New Fixed-Target
  Experiments to Search for Dark Gauge Forces},''
  \href{http://dx.doi.org/10.1103/PhysRevD.80.075018}{{\em Phys. Rev. D}
  {\bfseries 80} (2009) 075018},
  \href{http://arxiv.org/abs/0906.0580}{{\ttfamily arXiv:0906.0580 [hep-ph]}}.

\bibitem{Bjorken:1988as}
J.~D. Bjorken, S.~Ecklund, W.~R. Nelson, A.~Abashian, C.~Church, B.~Lu, L.~W.
  Mo, T.~A. Nunamaker, and P.~Rassmann, ``{Search for Neutral Metastable
  Penetrating Particles Produced in the SLAC Beam Dump},''
  \href{http://dx.doi.org/10.1103/PhysRevD.38.3375}{{\em Phys. Rev. D}
  {\bfseries 38} (1988) 3375}.

\bibitem{Andreas:2012mt}
S.~Andreas, C.~Niebuhr, and A.~Ringwald, ``{New Limits on Hidden Photons from
  Past Electron Beam Dumps},''
  \href{http://dx.doi.org/10.1103/PhysRevD.86.095019}{{\em Phys. Rev. D}
  {\bfseries 86} (2012) 095019},
  \href{http://arxiv.org/abs/1209.6083}{{\ttfamily arXiv:1209.6083 [hep-ph]}}.

\bibitem{Konaka:1986cb}
A.~Konaka {\em et~al.}, ``{Search for Neutral Particles in Electron Beam Dump
  Experiment},'' \href{http://dx.doi.org/10.1103/PhysRevLett.57.659}{{\em Phys.
  Rev. Lett.} {\bfseries 57} (1986) 659}.

\bibitem{Blumlein:2011mv}
J.~Blumlein and J.~Brunner, ``{New Exclusion Limits for Dark Gauge Forces from
  Beam-Dump Data},''
  \href{http://dx.doi.org/10.1016/j.physletb.2011.05.046}{{\em Phys. Lett. B}
  {\bfseries 701} (2011) 155--159},
  \href{http://arxiv.org/abs/1104.2747}{{\ttfamily arXiv:1104.2747 [hep-ex]}}.

\bibitem{Blumlein:2013cua}
J.~Bl\"umlein and J.~Brunner, ``{New Exclusion Limits on Dark Gauge Forces from
  Proton Bremsstrahlung in Beam-Dump Data},''
  \href{http://dx.doi.org/10.1016/j.physletb.2014.02.029}{{\em Phys. Lett. B}
  {\bfseries 731} (2014) 320--326},
  \href{http://arxiv.org/abs/1311.3870}{{\ttfamily arXiv:1311.3870 [hep-ph]}}.

\bibitem{APEX:2011dww}
{\bfseries APEX} Collaboration, S.~Abrahamyan {\em et~al.}, ``{Search for a New
  Gauge Boson in Electron-Nucleus Fixed-Target Scattering by the APEX
  Experiment},'' \href{http://dx.doi.org/10.1103/PhysRevLett.107.191804}{{\em
  Phys. Rev. Lett.} {\bfseries 107} (2011) 191804},
  \href{http://arxiv.org/abs/1108.2750}{{\ttfamily arXiv:1108.2750 [hep-ex]}}.

\bibitem{CMS:2019kiy}
{\bfseries CMS} Collaboration, A.~M. Sirunyan {\em et~al.}, ``{Search for a
  Narrow Resonance Lighter than 200 GeV Decaying to a Pair of Muons in
  Proton-Proton Collisions at $\sqrt{s} =$ TeV},''
  \href{http://dx.doi.org/10.1103/PhysRevLett.124.131802}{{\em Phys. Rev.
  Lett.} {\bfseries 124} no.~13, (2020) 131802},
  \href{http://arxiv.org/abs/1912.04776}{{\ttfamily arXiv:1912.04776
  [hep-ex]}}.

\bibitem{BaBar:2014zli}
{\bfseries BaBar} Collaboration, J.~P. Lees {\em et~al.}, ``{Search for a Dark
  Photon in $e^+e^-$ Collisions at BaBar},''
  \href{http://dx.doi.org/10.1103/PhysRevLett.113.201801}{{\em Phys. Rev.
  Lett.} {\bfseries 113} no.~20, (2014) 201801},
  \href{http://arxiv.org/abs/1406.2980}{{\ttfamily arXiv:1406.2980 [hep-ex]}}.

\bibitem{BaBar:2017tiz}
{\bfseries BaBar} Collaboration, J.~P. Lees {\em et~al.}, ``{Search for
  Invisible Decays of a Dark Photon Produced in ${e}^{+}{e}^{-}$ Collisions at
  BaBar},'' \href{http://dx.doi.org/10.1103/PhysRevLett.119.131804}{{\em Phys.
  Rev. Lett.} {\bfseries 119} no.~13, (2017) 131804},
  \href{http://arxiv.org/abs/1702.03327}{{\ttfamily arXiv:1702.03327
  [hep-ex]}}.

\bibitem{LHCb:2019vmc}
{\bfseries LHCb} Collaboration, R.~Aaij {\em et~al.}, ``{Search for
  $A'\to\mu^+\mu^-$ Decays},''
  \href{http://dx.doi.org/10.1103/PhysRevLett.124.041801}{{\em Phys. Rev.
  Lett.} {\bfseries 124} no.~4, (2020) 041801},
  \href{http://arxiv.org/abs/1910.06926}{{\ttfamily arXiv:1910.06926
  [hep-ex]}}.

\bibitem{darkcast}
\url{https://gitlab.com/philten/darkcast}.

\bibitem{Ilten:2018crw}
P.~Ilten, Y.~Soreq, M.~Williams, and W.~Xue, ``{Serendipity in dark photon
  searches},'' \href{http://dx.doi.org/10.1007/JHEP06(2018)004}{{\em JHEP}
  {\bfseries 06} (2018) 004}, \href{http://arxiv.org/abs/1801.04847}{{\ttfamily
  arXiv:1801.04847 [hep-ph]}}.

\bibitem{Bauer:2018onh}
M.~Bauer, P.~Foldenauer, and J.~Jaeckel, ``{Hunting All the Hidden Photons},''
  \href{http://dx.doi.org/10.1007/JHEP07(2018)094}{{\em JHEP} {\bfseries 07}
  (2018) 094}, \href{http://arxiv.org/abs/1803.05466}{{\ttfamily
  arXiv:1803.05466 [hep-ph]}}.

\bibitem{Blinov:2019gcj}
N.~Blinov, K.~J. Kelly, G.~Z. Krnjaic, and S.~D. McDermott, ``{Constraining the
  Self-Interacting Neutrino Interpretation of the Hubble Tension},''
  \href{http://dx.doi.org/10.1103/PhysRevLett.123.191102}{{\em Phys. Rev.
  Lett.} {\bfseries 123} no.~19, (2019) 191102},
  \href{http://arxiv.org/abs/1905.02727}{{\ttfamily arXiv:1905.02727
  [astro-ph.CO]}}.

\bibitem{Suliga:2020jfa}
A.~M. Suliga and I.~Tamborra, ``{Astrophysical constraints on nonstandard
  coherent neutrino-nucleus scattering},''
  \href{http://dx.doi.org/10.1103/PhysRevD.103.083002}{{\em Phys. Rev. D}
  {\bfseries 103} no.~8, (2021) 083002},
  \href{http://arxiv.org/abs/2010.14545}{{\ttfamily arXiv:2010.14545
  [hep-ph]}}.

\bibitem{XENON:2024ijk}
{\bfseries XENON} Collaboration, E.~Aprile {\em et~al.}, ``{First Indication of
  Solar B8 Neutrinos via Coherent Elastic Neutrino-Nucleus Scattering with
  XENONnT},'' \href{http://dx.doi.org/10.1103/PhysRevLett.133.191002}{{\em
  Phys. Rev. Lett.} {\bfseries 133} no.~19, (2024) 191002},
  \href{http://arxiv.org/abs/2408.02877}{{\ttfamily arXiv:2408.02877
  [nucl-ex]}}.

\bibitem{PandaX:2024muv}
{\bfseries PandaX} Collaboration, Z.~Bo {\em et~al.}, ``{First Indication of
  Solar B8 Neutrinos through Coherent Elastic Neutrino-Nucleus Scattering in
  PandaX-4T},'' \href{http://dx.doi.org/10.1103/PhysRevLett.133.191001}{{\em
  Phys. Rev. Lett.} {\bfseries 133} no.~19, (2024) 191001},
  \href{http://arxiv.org/abs/2407.10892}{{\ttfamily arXiv:2407.10892
  [hep-ex]}}.

\bibitem{DeRomeri:2024iaw}
V.~De~Romeri, D.~K. Papoulias, and C.~A. Ternes, ``{Bounds on new neutrino
  interactions from the first CE$\nu$NS data at direct detection
  experiments},'' \href{http://arxiv.org/abs/2411.11749}{{\ttfamily
  arXiv:2411.11749 [hep-ph]}}.

\bibitem{Ackermann:2025obx}
N.~Ackermann {\em et~al.}, ``{First observation of reactor antineutrinos by
  coherent scattering},'' \href{http://arxiv.org/abs/2501.05206}{{\ttfamily
  arXiv:2501.05206 [hep-ex]}}.

\bibitem{Chattaraj:2025fvx}
A.~Chattaraj, A.~Majumdar, and R.~Srivastava, ``{Probing standard model and
  beyond with reactor CE\ensuremath{\nu}NS data of CONUS+ experiment},''
  \href{http://dx.doi.org/10.1016/j.physletb.2025.139438}{{\em Phys. Lett. B}
  {\bfseries 864} (2025) 139438},
  \href{http://arxiv.org/abs/2501.12441}{{\ttfamily arXiv:2501.12441
  [hep-ph]}}.

\bibitem{DeRomeri:2025csu}
V.~De~Romeri, D.~K. Papoulias, and G.~Sanchez~Garcia, ``{Implications of the
  first CONUS+ measurement of coherent elastic neutrino-nucleus scattering},''
  \href{http://arxiv.org/abs/2501.17843}{{\ttfamily arXiv:2501.17843
  [hep-ph]}}.

\end{thebibliography}\endgroup

\end{document}